\documentclass{llncs}

\usepackage[ruled, noend, linesnumbered]{algorithm2e}

\usepackage{mathtools}

\usepackage{newtxmath}
\usepackage[capitalise, noabbrev]{cleveref}

\usepackage{amsfonts}
\usepackage{graphicx, caption}
\usepackage{subcaption}
\usepackage{adjustbox}

\usepackage{array,multirow}
\usepackage{booktabs}
\usepackage[misc]{ifsym}

\usepackage{xcolor}

\newcommand{\jakir}[1]{{\color{blue}#1}}

\newboolean{extended} 
\setboolean{extended}{true}

\newcommand\longversion[1]{\ifthenelse{\boolean{extended}}{#1}
{

}}

\newcommand\inappendix[1]{\ifthenelse{\boolean{extended}}{\cref{#1} in Appendix}{\cite{extended}}}

\begin{document}

\title{Quantifying Node-based Core Resilience}

\author{Jakir Hossain \inst{1}\textsuperscript{(\Letter)} 
\and
Sucheta Soundarajan\inst{2} \and
Ahmet Erdem Sar{\i}y\"{u}ce\inst{1}}

\institute{University at Buffalo, Buffalo, NY 14260, USA\\
\email{\{mh267,erdem\}@buffalo.edu} \and
Syracuse University, Syracuse, NY 13244, USA\\
\email{susounda@syr.edu}}

\authorrunning{J. Hossain et al.}
\toctitle{Quantifying Node-based Core Resilience}
\tocauthor{Jakir Hossain, Sucheta Soundarajan, and Ahmet Erdem Sar{\i}y\"{u}ce}

\maketitle
 \vspace{-5ex}
\begin{abstract}
Core decomposition is an efficient building block for various graph analysis tasks such as dense subgraph discovery and identifying influential nodes.
One crucial weakness of the core decomposition is its sensitivity to changes in the graph: inserting or removing a few edges can drastically change the core structure of a graph.
Hence, it is essential to characterize, quantify, and, if possible, improve the resilience of the core structure of a given graph in global and local levels.
Previous works mostly considered the core resilience of the entire graph or important subgraphs in it.
In this work, we study node-based core resilience measures upon edge removals and insertions.
We first show that a previously proposed measure, Core Strength, does not correctly capture the core resilience of a node upon edge removals.
Next, we introduce the concept of dependency graph to capture the impact of neighbor nodes (for edge removal) and probable future neighbor nodes (for edge insertion) on the core number of a given node.
Accordingly, we define Removal Strength and Insertion Strength measures to capture the resilience of an individual node upon removing and inserting an edge, respectively.
As naive computation of those measures is costly, we provide efficient heuristics built on key observations about the core structure.
We consider two key applications, finding critical edges and identifying influential spreaders, to demonstrate the usefulness of our new measures on various real-world networks and against several baselines. 
We also show that our heuristic algorithms are more efficient than the naive approaches.
\end{abstract}

\section{Introduction}
\vspace{-1ex}

$k$-cores are proposed as the seedbeds in which cohesive subsets of nodes can be found~\cite{k-core_definition}.
$k$-core is defined as the maximal connected subgraph in which every vertex has at least $k$ neighbors in the subgraph.
Each node is assigned a core number which denotes the maximum $k$ for which the node is a part of $k$-core.
Thanks to its linear time complexity, $k$-cores are used as a standard tool in various applications at downstream graph analytics.
Examples include the analysis of internet topology~\cite{carmi2007model_internet_topology}, predicting protein interactions~\cite{altaf2003prediction_protein_protein}, identifying influential spreaders~\cite{kitsak2010identification_influential_kshell}, and community detection~\cite{giatsidis2013d_community_detection,andersen2009finding_dense_find_1,balasundaram2011clique_dense_find_2,kortsarz1994generating_dense_find_3}. 

Despite its widespread use, $k$-cores are known to have a weak resilience against a few changes in the graph~\cite{adiga2013robust_k_core,www18}.
Inserting or removing a few edges can drastically change the core structure of a graph.
In applications where noise is common or the studied graph has uncertain parts, core decomposition is not reliable.
For example, networks are constructed as a result of indirect measurements in various applications, such as the Internet router/AS level graphs by traceroutes~\cite{faloutsos1999power}, biological networks by experimental correlations~\cite{schwab2010rhythmogenic}, and social media based networks by limited samples via the APIs~\cite{bakshy2011everyone}.
It is essential to characterize, quantify, and, if possible, improve the resilience of the core structure of a given graph in such applications at global and local levels.

In previous works, the resilience of $k$-core is studied under node or edge removal to improve users' involvement in social networks~\cite{zhang2017finding_collapsed_kcore,k-core_maximization,medya2020game}, bolster connections to protect a social network from unraveling~\cite{bhawalkar2015preventing_unraveling} and determining the edges that should be monitored for attacks on technological networks~\cite{www18}. 
Those studies only consider the core structure of the entire graph or a few important subgraphs (e.g., maximum $k$-cores).
{\bf There is no holistic study to quantify the node-based core resilience for any given node in the graph upon edge removals and edge insertions.}
Considering the query-driven scenarios in uncertain or noisy networks where the properties of the nodes are important, such as identifying influential nodes in spreading processes~\cite{lin2014identifying_disease_spreading} or information diffusion~\cite{liu2014information_advertise_2} and finding critical nodes/edges~\cite{purevsuren2019efficient_social_network_analysis}, it is crucial to measure the resilience of core numbers against edge removals as well as insertions.

In this work, we study node-based core resilience measures upon edge removals and insertions.
We first demonstrate that a previously proposed node-based measure, Core Strength~\cite{www18}, is inaccurate at capturing the changes in the core number upon edge removals.
Next, we propose the concept of dependency graph which captures the impact of neighbor nodes (for removal case) and probable future neighbor nodes (for insertion case) on the core number of a given node.
In the dependency graph for removal, one-way dependency relationships between neighboring nodes help to identify the resilience of core numbers.
Likewise, in the insertion case, we discover the one-way dependency relationships to quantify the likelihood of a change in the core number.
Using the dependency graphs, we define a pair of Removal Strength and Insertion Strength measures for each node.
Calculating those node strengths for big graphs in a naive way is computationally intensive. For edge removal, we use the equal edges~\cite{k-core_minimization} and $k$-corona \cite{corona_percolation} properties to design RSC algorithm. For insertion, we design ISC algorithm based on the number of connections a node has with the same or higher core number.
As node-level aspects of a graph are important in many real-world applications, we consider two applications to demonstrate the benefit of our new measures: finding the most critical edges to remove/insert such that the number of nodes that changes their initial core numbers is maximized~\cite{dey2020network_global_kcore,k-core_maximization,k-core_minimization} and identifying influential spreaders~\cite{zareie2018hierarchical,medo2009adaptive_news_recomendation,chen2012identifying_influential}. For both applications we compare our node-based measures against several state-of-the-art baselines.
 
Our contributions can be summarized as follows:
\vspace{-2ex}
\begin{itemize}
    \item We point out that the Core Strength definition (by~\cite{www18}) is incorrect and provide counterexamples as well as empirical results to show its unreliability.
    \item To quantify the node resilience upon edge removal and edge insertion, we use the concept of dependency graphs. Accordingly, we introduce a pair of Removal Strength and Insertion Strength measures.
    \item We design RSC and ISC algorithms to compute the new node resilience measures for removal and insertion.
    \item We consider two motivating applications to examine the effectiveness of those metrics: finding critical edges and identifying influential spreaders.
    \item We evaluate our measures and algorithms on real-world networks. We demonstrate the efficiency and effectiveness of our techniques against several baselines on the two applications mentioned above.    
\end{itemize}

\section{Background}
\vspace{-1ex}

In this work, we consider $G = (V, E)$ as an undirected and unweighted graph, where $V$ and $E$ represent the set of nodes and edges in $G$, respectively.
We use $\bar{E}$ to denote the complement of $E$, i.e., $\bar{E}=\{(u, v) | u \in V, v \in V, (u, v) \notin E\}$.
We use $N(u, G)$ to represent the set of neighbors of $u$ in $G$ and $\Gamma(u, G)$ to denote the distance-2 neighbors of $u$.
Let $S\subseteq G$ be a subgraph of $G$.  We use $deg(u, S)$ to denote the degree of $u$ in $S$. 
In some cases we consider a directed graph $G'$ in which $deg^{-}(u, G')$ and $deg^{+}(u, G')$ denotes the in-degree and out-degree of $u$ in $G'$, respectively. 
In our notations, we omit $G$ when it is obvious.

The $k$-core, denoted by $C_k(G)$, is the maximal connected subgraph $S\subseteq G$ where every vertex has at least $k$ connections in $S$, i.e., $deg(u, S) \ge k~\forall u \in G$.
The core number of a vertex is the largest $k$ value for which a $k$-core contains the vertex.
Here, $K(u, G)$ denotes the core number of $u$ in $G$, and $K_.(G)$ is the core vector, which is the core numbers of all vertices in $G$. 
The maximum $k$-core(s) of a graph are the (non-empty) $k$-cores with largest value of $k$. 
The $k$-shell of a graph is the set of nodes with core number $k$~\cite{carmi2007model_internet_topology} and a subcore is a connected subgraph of nodes with core number $k$~\cite{sariyuce2013streaming}. 
The $k$-cores (for all $k$) are computed by recursively removing vertices with degree less than $k$ and their adjacent edges, while assigning core numbers during the process, which takes $O(\lvert E\lvert )$ time~\cite{batagelj2011fast_k-shell_decomposition}.

We define the subset of neighbors of a node $u$ based on the relative core numbers: $\Delta_<(u, G)$ denotes the neighbors with smaller core numbers, i.e., $\{v : v \in N(u,G) \wedge K(v,G) < K(u,G)\}$ and $\Delta_=(u, G)$ is the neighbors with equal core numbers.
Similarly, $\Delta_>(u, G)$ and $\Delta_\geq(u, G)$ are defined.

\longversion{

Lastly, we recall the $h$-index based core number definition from earlier work.

\vspace{-1ex}
\begin{lemma}\label{def:core_number_h_index}
The core number of a node is equal to the $h$-index of the core numbers of its neighbors~\cite{h-index_k_core}. Formally, $K(u) = \mathcal{H}\Big(K(v_1), K(v_2),..., K(v_{d_u})\Big)$,
where $N(u,G) = \{v_1, v_2,..., v_{d_u}\}$ and $\mathcal{H}$ is a function that returns the maximum integer $h > 0$ such that there exist at least $h$ elements in $(K(v_1), K(v_2),..., K(v_{d_u})$ whose value is at least $h$. Here, $v_{d_u}=deg(u, G)$. 
\end{lemma}
}

\section{Related Work}
\vspace{-1ex}

Network resilience is the capability of a network to maintain or restore its function under faults.
Characterizing the resilience of a network is important for critical systems such as power grids and transportation systems~\cite{lewis2022many}.
Characterization of the resilience is made with respect to various graph characteristics, such as components and paths~\cite{survey}.
One interesting direction in this context is the resilience of the core structure.
Core decomposition is one of the most widely used graph algorithms thanks to its linear complexity~\cite{k-core_definition}.
However, it is quite sensitive to changes in the graph and there are a few studies to characterize and improve its robustness~\cite{adiga2013robust_k_core,www18,dey2020network_global_kcore,burleson2020k,Bo22}.
Here we summarize the literature on core resilience and explore its significance in two motivating applications: finding critical edges and identifying influential spreaders.

\noindent{\bf Core resilience.} Adiga and Vullikanti found that the stability of maximum $k$-cores under noise and sampling perturbations does not degrade in a monotonic way~\cite{adiga2013robust_k_core}.
Laishram et al. defined the core resilience of a graph as the correlation between the core number rankings of the top r\% nodes before and after p\% edges or nodes are removed at random~\cite{www18}.
As computing this is costly, they proposed Core Strength and Core Influence measures as proxy to quantify the resilience of a node's core number upon node or edge deletions.
Burleson-Lesser et al. modeled network robustness by using the histogram of core numbers~\cite{burleson2020k} and found that ecological and financial networks with U-shaped histograms are resilient to node deletion attacks.
Dey et al. defined a graph's stability based on changes in each node's core number upon node removals and studied identifying critical nodes to delete to maximize the number of nodes falling from their initial cores~\cite{dey2020network_global_kcore}.
More recently, Zhou et al. studied attack strategies to change the core numbers of the nodes by rewiring edges~\cite{Bo22}. 
Unlike those studies, we focus on node-based core resilience and consider both removal and insertion. For a given node, we quantify the resilience of its core number upon edge insertion and removals.~\cref{table:core_resilience} compares our work and previous studies on core resilience.
\begin{table*}[!t]
\centering
\resizebox{\textwidth}{!}{
\begin{tabular}{|c|c|c|c|c|c|c|}
\hline
\textbf{\begin{tabular}[c]{@{}c@{}}\end{tabular}} & 
\begin{tabular}[c]{@{}c@{}}\textbf{\cite{adiga2013robust_k_core}}  \\ \end{tabular} & \begin{tabular}[c]{@{}c@{}}\textbf{\cite{www18}}  \\ \end{tabular} & \begin{tabular}[c]{@{}c@{}}\textbf{\cite{burleson2020k}}  \\ \end{tabular} &  \begin{tabular}[c]{@{}c@{}}\textbf{\cite{dey2020network_global_kcore}}  \\ \end{tabular} &  \begin{tabular}[c]{@{}c@{}}\textbf{\cite{Bo22}}  \\ \end{tabular} & \textbf{Our work} \\ \hline
\begin{tabular}[c]{@{}c@{}}Graph structure\end{tabular}& Max cores & Entire graph & Entire graph & Entire graph & $k$-shells & \begin{tabular}[c]{@{}c@{}}Core number \end{tabular} \\ \hline
Edge insertion & Yes & No & No & No & Yes & Yes \\ \hline
Edge removal & Yes & Yes & No & No & Yes & Yes \\ \hline
    \end{tabular}
    }
\vspace{1ex}
\caption{\footnotesize {\bf Comparison of previous works on core resilience and our work.}}
\label{table:core_resilience}
\vspace{-7ex}
\end{table*}

\noindent{\bf Finding Critical Edges.} A related line of work has proposed problems to minimize and maximize the size of a $k$-core by inserting/removing nodes/edges.
For the removal, the motivation is often to find critical nodes/edges that should be kept in the graph to avoid unraveling in social networks or be watched against targeted attacks in infrastructure networks~\cite{k-core_minimization,medya2020game,zhang2017finding_collapsed_kcore,zhao2021finding_critical_convolution}.
In the context of core resilience, such nodes/edges are the weak structures with low resilience against removal and are suitable for targeted attacks.
Regarding the insertion, the objective is to find new edges that can increase the user engagement~\cite{k-core_maximization,sun2022fast} or incentivize existing nodes to stay engaged so that other nodes are kept engaged as well~\cite{bhawalkar2015preventing_unraveling,zhang2017olak,laishram2020residual_core_maximize}.
In the scope of core resilience, such nodes/edges are the critical graph structures that are most vulnerable to increases in core numbers or core sizes.
All those works consider a specific $k$-core and study targeted attacks to change the core structure with a minimal number of edge/node changes. 
In this work, we focus on the core numbers and use our new node-based core resilience measures to select a limited number of edges so as to maximize the number of nodes whose core numbers change (see~\cref{sec:criticalEdges}).

\noindent{\bf Identifying Influential Spreaders.} Another application that core numbers are heavily used is identifying influential spreaders~\cite{kitsak2010identification_influential_kshell}.
Influential spreaders are the nodes that determine how information spreads over the network or how a virus is propagated~\cite{anderson1992infectious_disease,guille2013information_diffusion_survey}.
SIR (Susceptible-Infected-Recovered) model is a classical tool to measure the influence of a given set of nodes \cite{sharkey2011deterministic_SIR_model}.  In the SIR model, a set of initially infected nodes are chosen which will spread the disease at each time step, $t$. The fraction of infected nodes, denoted by $S(t)$, is used to measure the spread after $t$ iterations.
Kitsak et al. demonstrated that the most efficient spreaders are located in the highest $k$-shells~\cite{kitsak2010identification_influential_kshell}.
Wang et al. discovered that greedily choosing multiple spreaders may result in some of them being too close to each other and hence their influence overlaps~\cite{wang2020identifying_improved_kshell}.
They proposed the $IKS$ algorithm to select nodes from different $k$-shells based on the highest node information entropy, which outperforms the other centrality or core-number based measures.
Considering the success of core-based measures, we use node-based core resilience as a proxy to identify influential spreaders (see~\cref{sec:infspreader}).

\section{Node-based Core Resilience}\label{sec:noderes}
\vspace{-1ex}

Earlier studies mostly defined core resilience measures for the entire graph. 
One exception is the core strength ($CS$) definition by Laishram et al.~\cite{www18}, which aims to measure the resilience of a node's core number upon edge removals and is defined as $CS(u, G) = |\Delta_\geq(u,G)| - K(u,G) + 1$.
They claim that in order to decrease $K(u,G)$, at least $CS(u,G)$ connections from $u$ must be removed.
Here we show that this claim is not true by a simple counterexample and give empirical evidence to show how frequently it fails in practice.

\begin{figure}[!t]
    \centering
  \begin{subfigure}[t]{.48\linewidth}
    \centering
    \includegraphics[width=\linewidth]{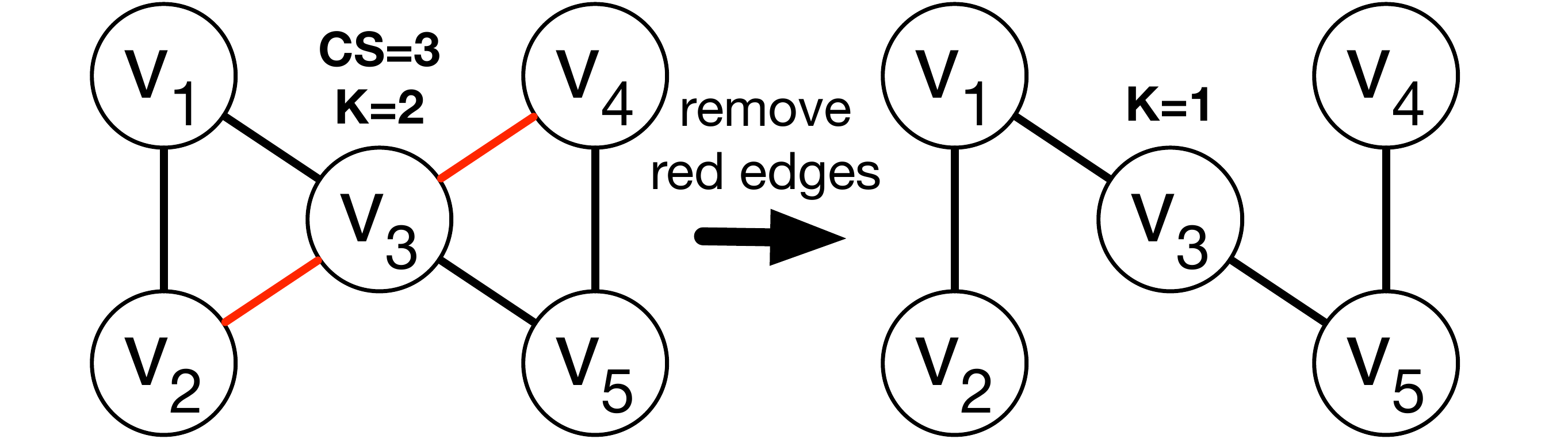}
\caption{\footnotesize A counterexample: removing less than $CS(v_3)$ edges decreases $K(v_3)$.}
\label{fig:CS_wrong}
  \end{subfigure}
   \hspace{1mm}
  \begin{subfigure}[t]{.48\linewidth}
    \centering\includegraphics[width=\linewidth]{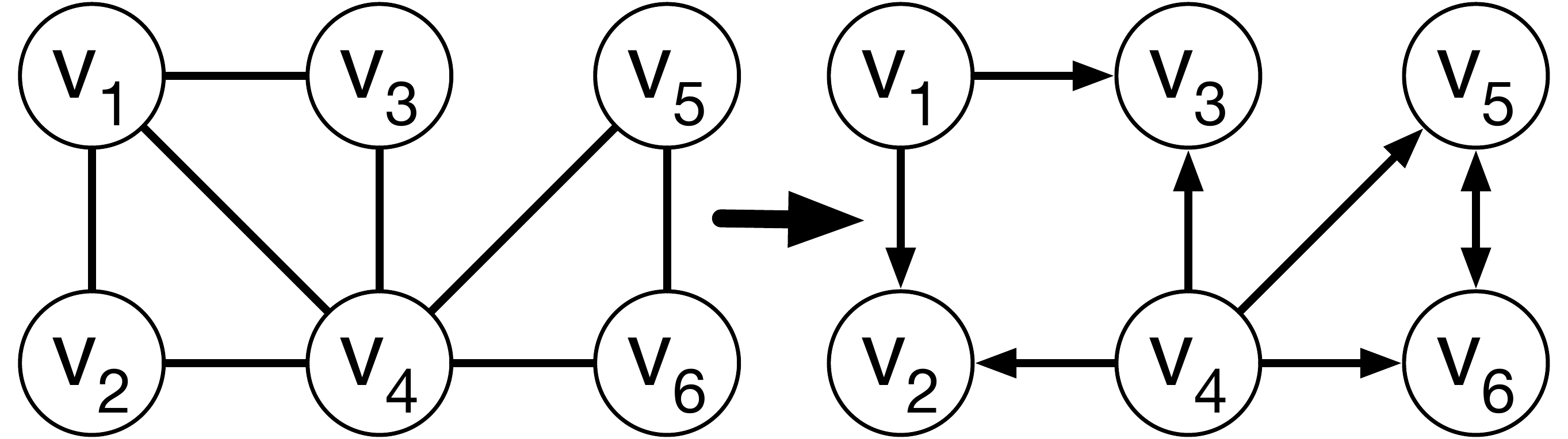}
    \caption{\footnotesize A toy graph (left) and its removal dependency graph (right)}
        \label{fig:removal_dependency}
  \end{subfigure}
      \vspace{-1ex}
    \caption{\footnotesize {\bf Illustrative examples}} 
\vspace{-5ex}
\end{figure}

Consider the toy graph in~\cref{fig:CS_wrong}. 
$CS(v_3)$ is 3, which means that at least three edges of $v_3$ should be removed to decrease its core number, according to Laishram et al.~\cite{www18}.
However, if we remove only $(v_3, v_2)$ and $(v_3, v_4)$, $K(v_3)$ decreases to 1.
Note that removing two edges does not always decrease $K(v_3)$, e.g,. deleting $(v_3, v_1)$ and $(v_3, v_2)$ does not affect $K(v_3)$.
Depending on the edges being removed, other nodes may have their core numbers changed too, and this cascading effect may result in decreasing the core number of the vertex of interest.
Hence, not only the count but also the position of the removed edges matters in quantifying the node-based core resilience.

One question is how likely to see such structures, where removing less than $CS(u)$ edges decreases $K(u)$, in real-world networks.
We perform a simple experiment to check this.
We consider the nodes in the maximum $k$-cores with a $CS$ of at least two. For each node, we remove one of its edges and observe its core number changes. We repeat this for each edge of a node.
Removing even a single edge is sufficient around $10\%$ of the time to decrease the core number (as shown in \inappendix{fig:appendix_CS_wrong}).
Hence, the $CS$ definition also fails frequently in practice.

As the $CS$ definition is inaccurate in capturing the core resilience of a node upon edge removals, we define a new measure, Removal Strength, to compute the likelihood of a node's core number change (\cref{sec:removal_strength}).
Moreover, we propose a new measure, Insertion Strength, to assess the stability of a node's core number after an edge insertion~(\cref{sec:insertion_strength}).

\subsection{Resilience Against Edge Removal}  \label{sec:removal_strength}

We capture a node's core resilience against edge removals by analyzing its dependency on neighbor nodes. 
We focus on single edge removal, with a conjecture that multiple edge removals can be approximated by considering multiple single edge removals.

We define that the node $u$ is dependent on node $v$, denoted as a relationship $v \rightarrow u$, if $K(u)$ decrements after removing the edge $(u, v)$.
For a given graph $G=(V, E)$, we define the {\it removal dependency graph} , denoted by $G^{rd}=(V, E^{rd})$, as a directed graph such that an edge $(u, v) \in E^{rd}$ if $(u, v) \in E$ and $K(v, G \setminus (u, v)) < K (v, G)$.
We give an example in~\cref{fig:removal_dependency}.
For the toy graph on the left, the corresponding removal dependency graph ($G^{rd}$) is given on the right.
In the $G^{rd}$, $v_2$ has two in-neighbors ($v_1$ and $v_4$) which means it is dependent on $v_1$ and $v_4$. 
For an  edge in $G$, if neither node is dependent on the other, then no edge will appear in the $G^{rd}$, such as ($v_1, v_4$).
If each node is dependent on another, then there are two edges in both directions, as for ($v_5, v_6$).

In-degree and out-degree of a node in the removal dependency graph give important insights about its core resilience.
A node with a large in-degree is dependent on many of its neighbors, hence removing a nearby edge could reduce its core number, implying a lower core resilience.
We define {\it In-Degree Removal Strength} of a node $u$, $RS_{ID}(u)$, to quantify the resilience of $u$ to retain its core number upon edge removal(s): $RS_{ID}(u)= \frac{1}{deg^{-}(u, G^{rd})}$.
The higher a node's out-degree in the dependency graph, the more strength it has to change the other nodes' core numbers. 
We define {\it Out-Degree Removal Strength} of a node $u$, $RS_{OD}(u)$, to quantify the strength of $u$ to change the core number of other nodes: $RS_{OD}(u)= {deg^{+}(u, G^{rd})}$.

\subsubsection{Removal Strength Computation} \label{sec:removal}
A naive way to compute the removal dependency graph is to run incremental core decomposition algorithm for every single edge removal~\cite{sariyuce2013streaming,sariyuce2016incremental}, which will be costly.
Here we propose efficient heuristics by using key observations about the core  structure. 

We define node $u \in G$ as {\bf vulnerable} if $K(u, G)=\lvert\Delta_\geq(u,G)\lvert$. For a vulnerable node $u$, the set of edges $(u,v)$ where $v \in \Delta_\geq(u,G)$ is called the {\bf sensitive edges of $u$} (also called as equal edges in~\cite{k-core_minimization}).

\begin{lemma}\label{lem:sensitive}
If a sensitive edge $(u,v)$ of a vulnerable node $u$ is removed, then $K(u, G)$ will decrease.
\end{lemma}
\ifthenelse{\boolean{extended}}{
\begin{proof}
Let us consider that node $u$ in a $k$-core subgraph where $K(u, G)=\lvert\Delta_\geq(u,G)\lvert = k$. Which states that node $u$ has exactly $k$ neighbors whose core numbers are at least $k$. Now, if we delete any sensitive edge $(u, v)$ of $u$ where $v \in \Delta_\geq(u,G)$, then node $u$ will have less than $k$ neighbors whose core numbers are at least $k$. As a result, from \cref{def:core_number_h_index}, $K(u, G)$ will be less than $k$ in the graph $G\setminus\{(u,v)\}$, and the theorem holds.
\end{proof}
}
{
\begin{proof}
Proofs of Lemmas~\ref{lem:sensitive}-\ref{lem:insert_diff_shell} are available in the extended version~\cite{extended}.

\end{proof}
}

Sensitive edges of a vulnerable node provide a way to group certain edges  whose removal yields the same core vector, as first shown in~\cite{k-core_minimization}.

\begin{lemma} \label{lem:same_core_vector}
For a vulnerable node $u$, removing any sensitive edge yields the same core vector, i.e., $K_.(G\setminus\{(u,v_1)\})=K_.(G\setminus\{(u,v_2)\})$ where both edges are sensitive.
\end{lemma}
\longversion{
\begin{proof}
 Any two sensitive edges for a vulnerable node $u$ are referred to as equal edges by \cite{k-core_minimization}; Theorem 4 in~\cite{k-core_minimization} proves that removing equal edges would result in the same core vector, which also proves that removing any sensitive edge would yield the same core vector.
\end{proof}}

According to \cite{corona_percolation}, {\bf $k$-corona} is a maximal connected subgraph of vulnerable vertices with the same core number, $k$. Formally, $S \subseteq G$ is a $k$-corona if $\forall u \in S$, $K(u, G)=k$ and $u$ is a vulnerable vertex.
We define {\bf $k$-corona adjacent edge set}, $KAES(S)$, as the union of the sensitive edges of the vulnerable nodes in a $k$-corona $S$, i.e., $\bigcup_{u \in S}{\{(u, v) \lvert v \in \Delta_\geq(u,G)\}}$.

\begin{lemma} \label{lem:change_only_kaes}
When an edge $(u, v)$ is removed from the graph, there will be a change in the core numbers if and only if the removed edge $(u, v)$ is part of a $KAES$.
\end{lemma}

\longversion{
\begin{proof}
($\rightarrow$) We know that the $KAES$ only contains the sensitive edges, and, by~\cref{lem:sensitive}, the core number will change after deleting any sensitive edges. Hence, there will be a change in the core number if we remove any edge $(u, v) \in KAES$. \\

($\leftarrow$) Now, we prove that removing any edge $(u, v) \notin KAES$ will not change the core number of any node. Assume w.l.o.g that $K(u)\leq K(v)$.  Now, to determine  whether an edge is part of a $KAES$ or not, we have to consider the following four conditions:

\begin{enumerate}
    \item Both node $u$ and $v$ are vulnerable.
    \item Only node $u$ is vulnerable.
    \item Only node $v$ is vulnerable.
    \item Neither node $u$ nor $v$ is vulnerable.
\end{enumerate}

For (1), since both nodes are vulnerable, then $K(u, G)=\lvert\Delta_\geq(u,G)\lvert$ and $K(v, G)=\lvert\Delta_\geq(v,G)\lvert$. Hence, the edge $(u,v)$ will be part of a $KAES$ and is not need to be considered.

For (2), as the node $u$ is vulnerable and $K(u) <= K(v)$, the edge $(u,v)$ is a sensitive edge of $u$. Hence, the edge $(u,v)$ will be part of a $KAES$ and is not need to be considered. 

For (3), when $K(u) = K(v)$, the edge $(u,v) \in KAES$. So, in order to have $(u,v) \notin KAES$, we have to consider only the case $K(u) < K(v)$. Now, when $K(u) < K(v)$, deleting $(u, v)$ has no effect on $K(v)$ (by~\cite{sariyuce2013streaming}). Besides, since $u$ is not a vulnerable node, it has at least $K(u) + 1$ neighbors in its $k$-core of $G$. So, there will be at least $K(u)$ neighbors of $u$ in its $k$-core of $G^\prime$ where $G'=G\setminus\{(u, v)\}$. Hence, by \cref{def:core_number_h_index}, $K(u)$ will not also change.

For (4), as both node $u$ and $v$ are not vulnerable, the edge $(u, v) \notin KAES$. Hence, the node $u$ has at least $K(u) + 1$ neighbors and the node $v$ has at least $K(v) + 1$ neighbors in their $k$-core of $G$. So, there will be at least $K(u)$ neighbors of $u$ and at least $K(v)$ neighbors of $v$ in their $k$-core of $G^\prime$ where $G'=G\setminus\{(u, v)\}$. By~\cref{def:core_number_h_index}, $K(u)$ and $K(v)$ will remain unchanged.

Since $K(u)$ and $K(v)$ are unchanged in the last two conditions, there will be no changes in the core number of any other vertex (by~\cite{sariyuce2013streaming}). So, if we remove an edge $(u,v) \notin KAES$, then there will be no changes in the core number, and the theorem holds.

\end{proof}}

\begin{lemma} \label{lem:same_corevector_kaes}
For a $k$-corona $S$, removing any edge in $KAES(S)$ yields the same core vector, i.e., $K_.(G\setminus\{(u,v)\})=K_.(G\setminus\{(x,y)\})$ for $(u,v), (x,y) \in KAES(S)$.
\end{lemma}

\longversion{
\begin{proof}
According to~\cref{lem:same_core_vector}, for a vulnerable node $u$, removing any sensitive edge $(u, v)$ yields the same core vector. When the size of a $k$-corona is 1,~\cref{lem:same_corevector_kaes} is the same as  ~\cref{lem:same_core_vector}. 
Otherwise, the proof follows from the fact that there are (consecutive) pairs of edges that connect $(u, v)$ to $(x, y)$ where~\cref{lem:same_core_vector} holds for each edge pair.
We define two edges to be adjacent if they share a node.
As $k$-corona is a connected subgraph, consider a path of edges between $u$ and $x$ \{$e_1, e_2, \cdots, e_{n-1}, e_n$\} in a $KAES(S)$ where $e_1 = (u, v)$, $e_n = (x,y)$, and $e_1$ is adjacent to only \{$e_2$\}, $e_2$  is adjacent to only \{$e_1$, $e_3$\}, and $e_i$ is adjacent to only \{$e_{i-1}$, $e_{i+1}$\} for $i<n$ .
Here, $e_1$ and $e_2$  are the sensitive edges of a vulnerable node, and, by~\cref{lem:same_core_vector}, $K_.(G\setminus\{e_1\})=K_.(G\setminus\{e_2\})$.
Likewise, edges $e_i$ and $e_{i+1}$ are also sensitive edges of some vulnerable node and, by~\cref{lem:same_core_vector}, $K_.(G\setminus\{e_i\})=K_.(G\setminus\{e_{i+1}\})$.
As $e_1$ and $e_n$ are connected by a series of adjacent edges, we can conclude that $K_.(G\setminus\{e_1\})=K_.(G\setminus\{e_{n}\})$; which implies that for any two edges $(u,v), (x,y) \in KAES(S)$,  $K_.(G\setminus\{(u, v)\})=K_.(G\setminus\{(x, y)\})$. As a result, all of the edges $e \in KAES(S)$ will have the same effect on the core vector if we choose any edge from them to remove, and the theorem holds. \\
\end{proof}}

We define the subset of nodes whose core numbers change after removing a single edge as {\bf Core Changed Nodes ($CCN$)}. According to ~\cref{lem:same_corevector_kaes}, for a $k$-corona $S$, if we choose any edge $(u,v) \in KAES(S)$ to delete, then we will get the same core vector. 
Hence we denote the set of nodes whose core numbers change after removing any edge in a $KAES(S)$ as $CCN_{KAES(S)}$.

Instead of examining every single edge in a graph, we can utilize $CCN_{KAES(S)}$ to efficiently detect the changes in the core numbers of nodes.
Assume w.l.o.g. that $K(u)\leq K(v)$. If $u$ is a vulnerable node, then, by~\cref{lem:sensitive}, deleting an edge $(u, v)$ will decrement the core number of $u$.
If $u$ is not a vulnerable node, we need to look at the properties of both $u$ and $v$.
If $v$ is also not a vulnerable node, then the edge $(u,v) \notin KAES$, and there will be no changes in $K(u)$ or $K(v)$ (by~\cref{lem:change_only_kaes}).
However, if the node $v$ is vulnerable, we need to consider the following two cases to determine the changes in $K(v)$:\\
\indent\textbf{Case 1: $K(u) = K(v)$.} Here, $(u,v)$ is a sensitive edge, and deleting $(u,v)$ will decrement $K(v)$ (\cref{lem:sensitive}). In this case, $K(u)$ will also decrement if it becomes affected by the changes in $K(v)$. This information is actually captured by the $CCN$ set. If two nodes are in a  same $CCN$, a change in the core number of one node affects the core number of the other node. Hence, if $u$ and $v$ are in the same $CCN_{KAES(S)}$ for any $k$-corona $S$, then their core numbers depend on each other.\\
\indent\textbf{Case 2: $K(u) < K(v)$.} In this case, $K(v)$ will not change, as shown by~\cite{sariyuce2013streaming}. Regarding $K(u)$, as $u$ is not a vulnerable node, it has at least $K(u) + 1$ neighbors in its $k$-core. Since $v$ is not in the $k$-core of $u$, there will still be at least $K(u)$ neighbors in $u$'s $k$-core in $G\setminus\{(u, v\}$, and thus $K(u)$ will not change either.

\begin{algorithm}[!t]
\caption{\textbf{RSC: Removal Strength Computation} {$(G(V, E))$} } \label{alg:dependency_graph_removal}  \DontPrintSemicolon
{\bf Input:} $G~(V, E)$: graph\;
{\bf Output:} $RS_{ID}$, $RS_{OD}$: in and out-degree removal strength, respectively\;
$G^{rd}~(V,E^{'}) \leftarrow $ empty graph \tcp*{removal dependency graph}
Compute all the $k$-cores of $G$, $C_k(G)$, and put in $\cal{C}$\;
\ForEach{$k$-core $C_k(G) \in \cal{C}$} {
	Compute all $k$-coronas in $C_k(G)$ and put in $\cal{S}$\;
	\ForEach{$k$-corona $S \in \cal{S}$} {
		Find $KAES(S)$\;
		Delete any single edge $e \in KAES(S)$, compute $CCN_{KAES(S)}$~\tcp*{by~\cite{sariyuce2016incremental}}
	}
}

\ForEach{$u \in V$} {
	\If{$K(u) = |\Delta_{\geq} (u, G)|$} {
		\ForEach{$v \in N_{u}$} {
			\If{$K(u) \leq  K(v)$}{$E^{'}$.push($(v,u)$)\tcp*{$K(u)$ will decrement, by~\cref{lem:sensitive}}}
		}
	}
	\Else {
		\ForEach {$v$ in $N_{u}$} {
			\If {$K(v) = K(u) \And {K(v)=|\Delta_{\geq} (v, G)|} \And$ \hfill\;
			 $u$ and $v$ are in a same $CCN_{KAES(S)}$} {
				$E^{'}$.push($(v,u)$)~~~~\tcp*{$K(u)$ will decrement, by Case 1}
			}			
		}
	}
}
\ForEach {$u$ in $V$} {
	$RS_{ID}(u) \leftarrow  \frac {1}{deg^{-}(u, G^{rd})}$, $RS_{OD}(u) \leftarrow  {deg^{+}(u, G^{rd})}$
}
Return $RS_{ID}$, $RS_{OD}$
\end{algorithm}

\subsubsection{Removal Strength Algorithm}\label{sec:remalg}
Building on the definitions and observations above, we propose RSC algorithm (\cref{alg:dependency_graph_removal}) to compute $RS_{ID}$ and $RS_{OD}$ for each node in a given graph.
At the beginning, we find the $k$-core(s) of a graph using the classical peeling based algorithm proposed by Batagelj et al.~\cite{cores_decomposition_peeling_2} (line 4).
Then using the BFS traversal, we compute the set of $k$-coronas ($\cal{S}$) in every $k$-core subgraph (line 6). 
Since removing an edge $(u, v)\notin KAES$ does not affect the core number of any node (by~\cref{lem:change_only_kaes}), we only consider the edges $(u, v) \in KAES$ in each $k$-core.
For each $k$-corona $S$, we find the $KAES$ (line 8).
Thanks to~\cref{lem:same_corevector_kaes}, we remove only one edge in $KAES(S)$ and compute $CCN_{KAES(S)}$ for a $KAES(S)$ (line 9) by using the incremental core decomposition algorithm from~\cite{sariyuce2016incremental}.
Next, we use~\cref{lem:sensitive} and the two observations (Case 1 and 2) to quickly determine whether the core numbers will change after each edge removal (lines 10-19).
At the end, we calculate and return the in-degree and out-degree removal strengths of all the nodes by using $G^{rd}$ (lines 20-22).\\
\noindent {\bf Time and space complexity.} Line 4, as well as lines 10-19 takes $O(|E|)$ time. 
In the worst case, lines 5-9 takes $O(|V|\cdot|E|)$ time---if each node is a $k$-corona, one edge is removed per node, hence $|V|$ edge removals in total (line 9) and each edge removal takes $O(|E|)$ time per~\cite{sariyuce2016incremental}.
Overall time complexity is $O(|V|\cdot|E|)$, but this is a loose bound as the number of $k$-coronas is significantly less than $|V|$ in real-world networks (even for large networks we observe the number of $k$-coronas to be small, requiring us to remove fewer edges, as shown in \cref{table:runtime} column 4 (\% Gain)).
In addition to graph ($O(|E|)$), we store $|\Delta_{\geq} (u, G)|$ values ($O(|V|)$), component ids to bookkeep $CCN_{KAES(S)}$ for each node ($O(|V|)$), and $RS_{ID}, RS_{OD}$ values ($O(|V|)$). Overall space complexity is $O(|E|)$.

\subsection{Resilience Against Edge Insertion} \label{sec:insertion_strength}
We now characterize the resilience of a node's core number against edge insertions.
We again focus on the impact of a single edge change, consider the changes in a node's core number based on new links it forms with other nodes, and model the resilience accordingly.
Regarding the set of edge insertions, it is impractical and unrealistic to think about all possible new links between any pair of unconnected nodes, namely $\bar{E}$.
It is impractical because real-world networks are sparse, i.e., $|E| << {|V| \choose 2}$, which implies $|\bar{E}| >> |E|$.
It is unrealistic as it is unexpected that a link will form between two nodes if they have no common neighbors, i.e., if they are not distance-2 neighbors~\cite{liben2003link_common_count,newman2001clustering_common_count_collaboration}.
Even the number of non-neighbor node pairs with at least one common neighbor is too large to be considered, reaching up to 100$\cdot|E|$ for some real-world networks.
Furthermore, those node pairs are not located homogeneously in the graph; some nodes (mostly low-degree) have very few (or no) distance-2 neighbors, hence it is not clear how to define insertion core resilience for those (see \inappendix{fig:d2-hist}~for statistics).

To address these issues, we consider a fixed number ($b$) of edge insertions for each node and construct the {\it insertion candidate graph}, $G^{ic}$, accordingly.
Here, we fix $b=5$ as it is close to the average degrees of the networks used in experiments and no significant advantage is observed for larger $b$ values.
For any node $u \in G$, and its distance-2 neighbors $\Gamma(u)$, we consider the below cases to select the edges and add to $G^{ic}$:

\vspace{-2ex}
\begin{itemize}
    \item If $|\Gamma(u)| > b$, choose $b$ random edges $(u, v)$ such that $v \in \Gamma(u)$.
    \item Else, choose all $(u, v)$ edges such that $v \in \Gamma(u)$ and choose $b - |\Gamma(u)|$ random $(u, w)$ edge(s) such that $w \in V$ (and $w \notin \Gamma(u))$.
\end{itemize}
\vspace{-2ex}

\noindent Note that $b$ ensures a lower bound on the degree of a node in $G^{ic}$, there can be nodes with larger degree due to random edges coming from the other nodes.

\begin{figure}[!t]
    \centering
\begin{minipage}{1.0\textwidth}
   \vspace{-1ex}
\begin{subfigure}[t]{.33\linewidth}
    \includegraphics[width=\linewidth]{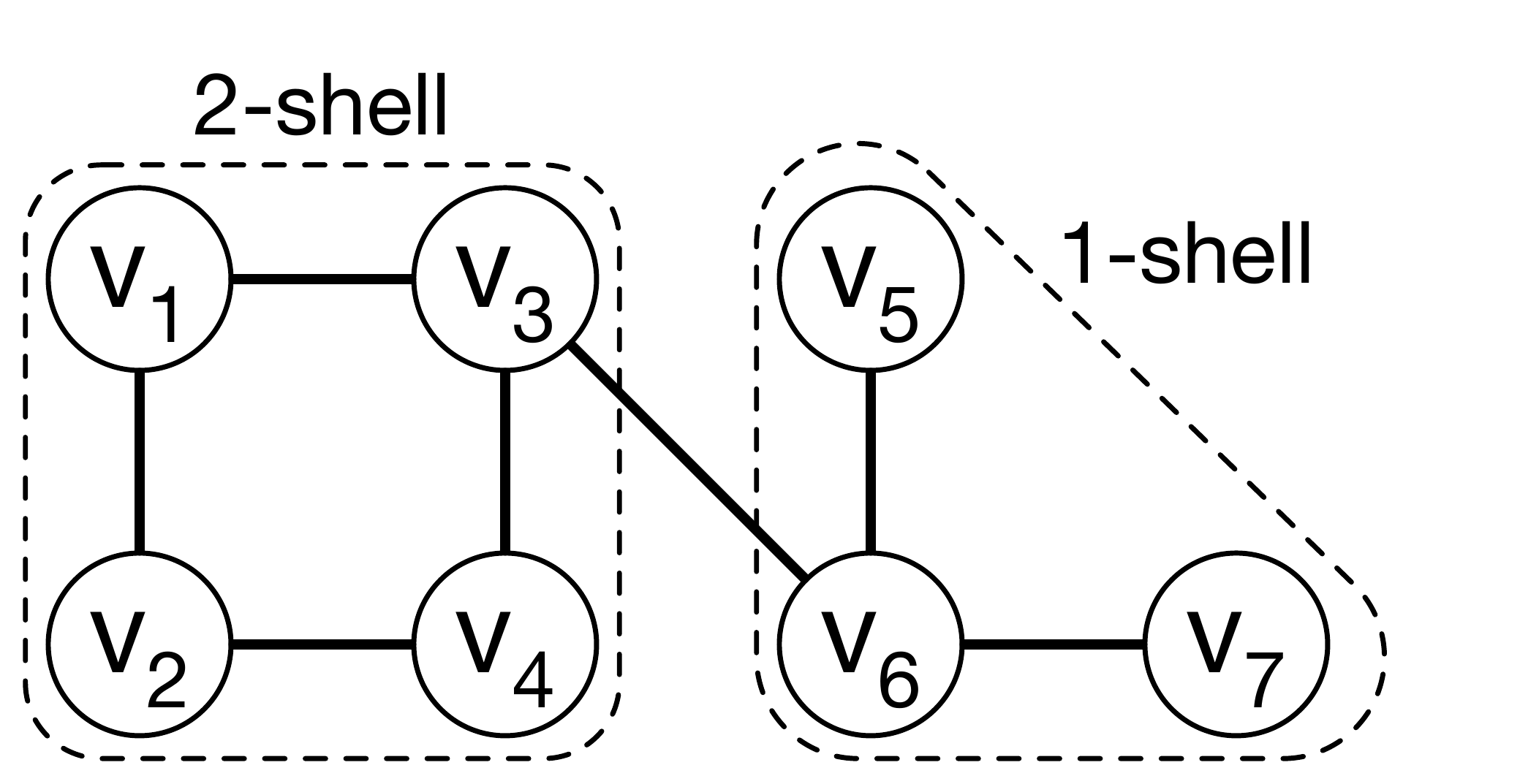}
    \caption{A toy graph $G$}
    \label{fig:itoy1}
\end{subfigure}
\hspace{-1ex}
\begin{subfigure}[t]{.33\linewidth}
    \centering
        \includegraphics[width=\linewidth]{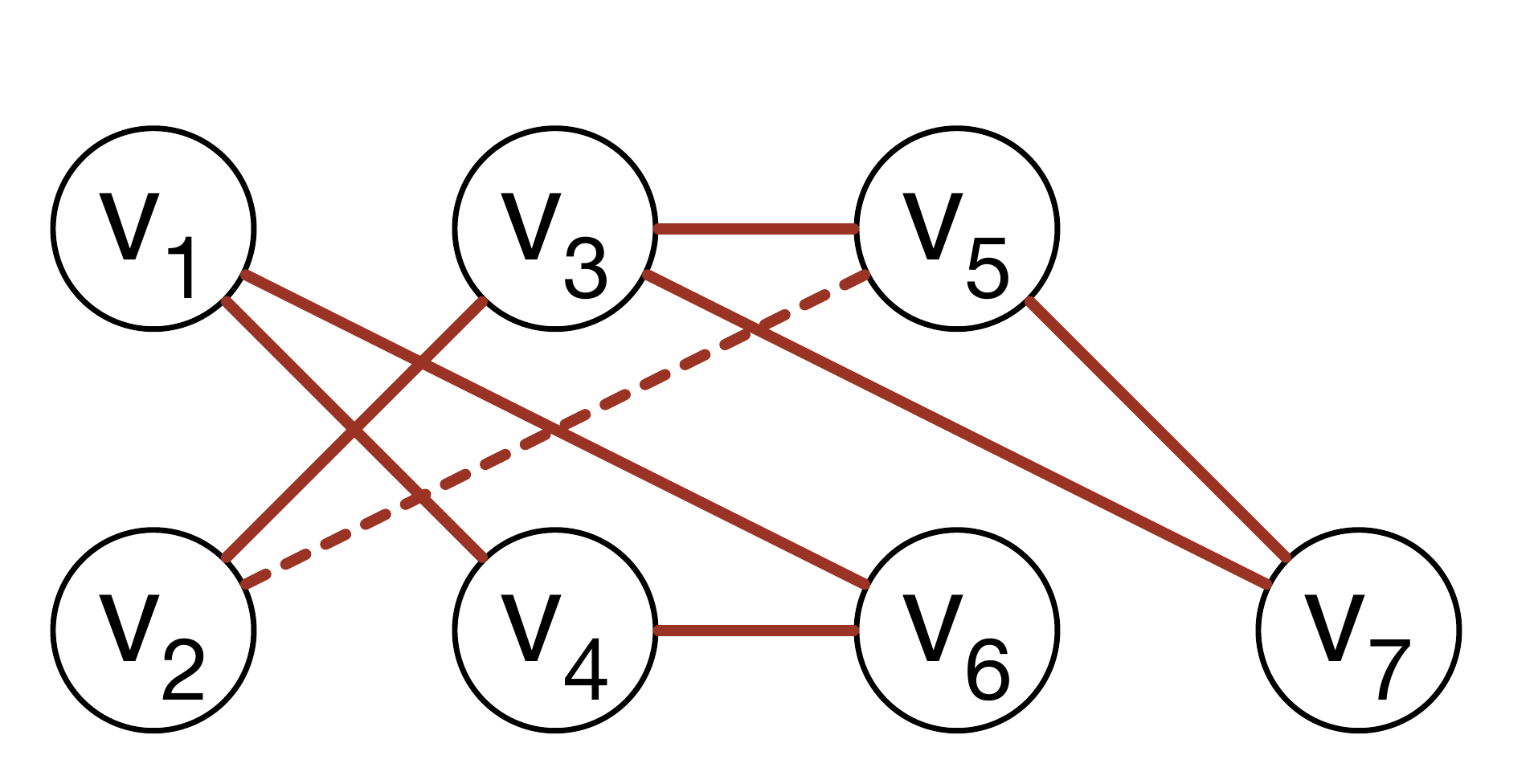}
    \caption{Ins. candidate graph}
    \label{fig:itoy2}
  \end{subfigure}
\hspace{-1ex}
  \begin{subfigure}[t]{.33\linewidth}
    \centering
        \includegraphics[width=\linewidth]{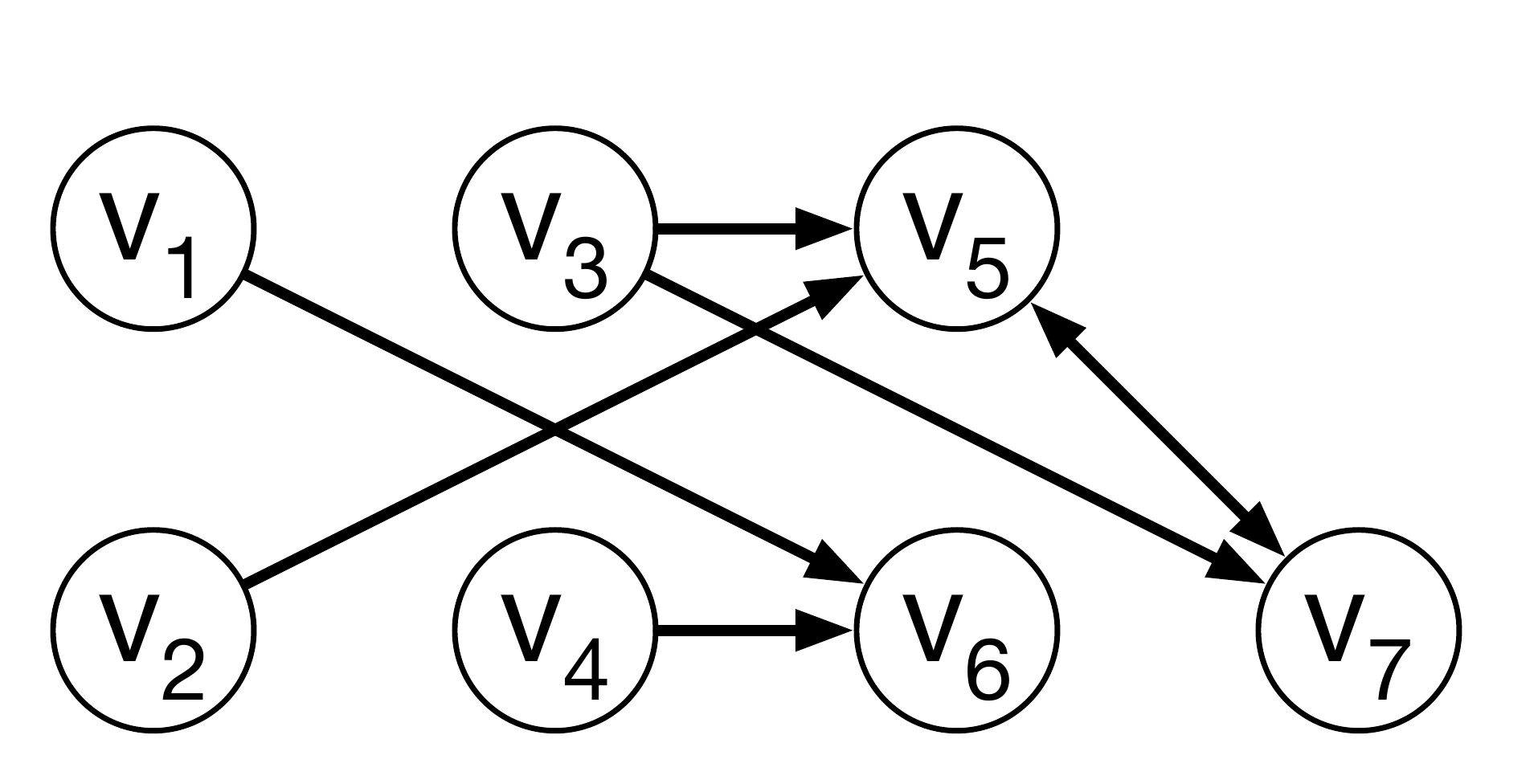}
    \caption{Ins. dependency graph}
    \label{fig:itoy3}
 \end{subfigure}    
 \end{minipage}
\vspace{-1ex}
    \caption{\footnotesize {\bf Examples for insertion candidate ($b=2$) and dependency graphs.}}
    \label{fig:insertion_dependency}
\vspace{-5ex}
\end{figure}

We define the dependency relationships between nodes by using the insertion candidate graph, $G^{ic}$.
For each edge $(u, v) \in G^{ic}$, we check how the core numbers of $u$ and $v$ change when $(u, v)$ is inserted to $G$.
$u$ is said to be dependent on node $v$, denoted as a relationship $v \rightarrow u$, if $K(u)$ increases after inserting the edge $(u,v)$.
For a given graph $G=(V, E)$ and $G^{ic}=(V, E^{ic})$, we define {\it insertion dependency graph}, $G^{id}=(V, E^{id})$, as a directed graph such that an edge $(u, v) \in E^{id}$ if $(u, v) \in E^{ic}$ and $K(v, G \cup (u, v)) > K (v, G)$.
Here, $G^{id}$ is always a subgraph of $G^{ic}$.
We give an example in~\cref{fig:insertion_dependency}.
For the toy graph on the left, corresponding insertion candidate graph is given in the middle (for $b=2$).
All the nodes except $v_2$ has at least two distance-2 neighbors.
To ensure $v_2$ has two edges, we randomly select a node, $v_5$, and put an edge between them.
Straight edges in the candidate graph are the edges due to the distance-2 neighborhood (the if condition above) and the dashed edge is the random edge (from the else condition).
The corresponding insertion dependency graph is shown on the right.
For example, inserting $(v_3, v_5)$ edge would increase $K(v_5)$ and does not impact $K(v_3)$, hence $(v_3 \rightarrow v_5)$ is put.

A node with a large in-degree is dependent on many of its distance-2 (or random) neighbors, implying a lower core resilience.
We define {\it In-Degree Insertion Strength} of a node $u$, $IS_{ID}$, to measure the node's ability to preserve its core number after edge insertion: $IS_{ID}(u)= \frac {1} {deg^{-}(u, G^{id})}$. A node with a large-out degree implies the ability to increase the core numbers of other nodes.
We define {\it Out-Degree Insertion Strength} of a node $u$, $IS_{OD}$, to measure the strength of a node to impact the nodes around it: $IS_{OD}(u)= {deg^{+}(u, G^{id})}$.

\subsubsection{Insertion Strength Computation} \label{sec:insertion}

A naive computation of the insertion dependency graph is to run incremental core decomposition algorithm for every single edge insertion~\cite{sariyuce2013streaming,sariyuce2016incremental}, which is costly.
Here we consider four lemmas that help to determine the core number changes without running the incremental algorithm.

\begin{lemma} \label{lem:insert_HCD}
For a node $u$ such that $K(u, G)=|\Delta_>(u,G)|$, adding a new edge $(u,v)$ s.t. $K(v, G) > K(u, G)$ will increment $K(u)$, i.e., $K(u, G \cup (u, v))=K(u, G)+1$.
\end{lemma}

\longversion{
\begin{proof}
Say $K(u, G)=|\Delta_>(u,G)|=k$. After adding a new edge $(u,v)$, where $K(v, G) > K(u, G)$, the node $u$ will have $k + 1$ neighbors whose core numbers are greater than $k$. As a result, by~\cref{def:core_number_h_index}, $K(u, G')$ will be $k+1$, where $G' = G \cup (u, v)$.
\end{proof}}

\begin{lemma} \label{lem:insert_same_shell1}
For two non-neighbor nodes $u$ and $v$, $(u, v) \notin E$, such that $K(u, G) = K(v, G) = k$, if $|\Delta_>(u,G)|=K(u,G)$ and $|\Delta_>(v,G)|=K(v,G)$, then adding a new edge $(u, v)$ will increment $K(u)$ and $K(v)$.
\end{lemma}

\longversion{
\begin{proof}
Consider the core decomposition algorithm by~\cite{batagelj2011fast_k-shell_decomposition}, where in each iteration $i$, the vertices with degree $i$ are removed from the graph and their core numbers are assigned as $i$.
In $G$, since $K(u, G)=K(v, G)=k$, both nodes $u$ and $v$ are removed in the $k$-th iteration. After adding $(u, v)$, at the $k$-th iteration, both $u$ and $v$ have $k$ neighbors whose core number is greater than $k$ because
$|\Delta_>(u,G)|=| \Delta_>(v,G)|=k$.
As $u$ and $v$ are neighbors, both nodes have an additional neighbor in the same iteration. This means that they both have $k+1$ neighbors in the $k$-th iteration, hence they will not be deleted in the $k$-th iteration. As a result, they will be removed in the next iteration, and their core numbers will be $k+1$.
\end{proof}}

\begin{lemma} \label{lem:insert_diff_shell}
Consider a node $u \in G$ such that $|\Delta_>(u,G)|=K(u,G)-1$. Say $u$ has a neighbor $w$ for which $K(u, G) = K(w, G)$ and $|\Delta_>(w,G)|=K(w,G)$. Adding a new edge $(u, v)$ such that $K(v, G) > K(u, G)$ will increment $K(u)$ and $K(w)$.
\end{lemma}

\longversion{
\begin{proof}
Define the new graph $G'=G\cup(u, v)$. $|\Delta_>(u,G)|$ in $G'$ will be $K(u, G)$ due to the new edge. Now, $|\Delta_>(u,G)|=K(u)$, $|\Delta_>(w,G)|=K(w)$, $K(u, G) = K(w, G)$, and $w$ is a neighbor of $u$. Here, node $u$ and $w$  satisfy the conditions of~\cref{lem:insert_same_shell1} in the new graph, thus the core number of both nodes $u$ and $w$ will increment.
\end{proof}}

\longversion{
\begin{algorithm}[!t]
\caption{\textbf{ISC: Insertion Strength Computation} {$(G(V, E))$} } \label{alg:dependency_graph_insertion}  \DontPrintSemicolon
{\bf Input:} $G~(V, E)$: graph; $G^{ic}$: insertion candidate graph\;
{\bf Output:} $IS_{ID}$, $IS_{OD}$: in-degree and out-degree insertion strength, respectively\;
$G^{id}~(V,E^{'}) \leftarrow $ empty graph \tcp*{insertion dependency graph}
Compute the core numbers of all the nodes in $G$, $K_{.}(G)$\;
\ForEach{edge $(u, v) \in G^{ic}$} {
	\tcp{Assume w.l.o.g that $K(u)\leq K(v)$}
	\If{$K(u) = K(v)$} {
		\If {$|\Delta_>(u,G)|=K(u)$ and $|\Delta_>(v,G)|=K(v)$} {
			$E^{'}$.push($(v,u)$), $E^{'}$.push($(u,v)$)\tcp*{inserting $(u,v)$ will increment both $K(u), K(v)$, by~\cref{lem:insert_same_shell1}}
		}
		\Else {
			$K'(u), K'(v) \leftarrow IncreCoreDec (G, K, (u, v))$\tcp*{by using~\cite{sariyuce2016incremental}}
			\lIf {$K(u) < K'(u)$}{$E^{'}$.push($(v,u)$)}
			\lIf {$K(v) < K'(v)$}{$E^{'}$.push($(u, v)$)} 
		}
	}
	\Else {
		\If {$|\Delta_>(u,G)|=K(u)$}{
			$E^{'}$.push($(v,u)$)\tcp*{$K(u)$ will increment, by~\cref{lem:insert_HCD}}
		}
            	\ElseIf {$|\Delta_>(u,G)|=K(u)-1$}{
			\If {$\exists w \in N(u, G)$ s.t. $K(u) = K(w)$ and $|\Delta_>(w,G)| = K(w)$}{
				$E^{'}$.push($(v,u)$)\tcp*{$K(u)$ will increment, by~\cref{lem:insert_diff_shell}}
			}
		}
		\Else {
			$K'(u), K'(v) \leftarrow IncreCoreDec (G, K, (u, v))$\tcp*{by using~\cite{sariyuce2016incremental}}
			\lIf {$K(u) < K'(u)$}{$E^{'}$.push($(v,u)$)}
		}
	}
} 
\ForEach {node $u$ in $V$} {
	$IS_{ID}(u) \leftarrow  \frac {1}{deg^{-}(u, G^{id})}$, $IS_{OD}(u) \leftarrow  {deg^{+}(u, G^{id})}$
}

Return $IS_{ID}$, $IS_{OD}$
\end{algorithm}
}

\subsubsection{Insertion Strength Algorithm}\label{sec:insalg}

We use the above lemmas (\cref{lem:insert_HCD} to~\cref{lem:insert_diff_shell}) to design ISC (Insertion Strength Computation) algorithm which creates the insertion dependency graph by determining the changes in the core numbers
\ifthenelse{\boolean{extended}}{
~\cref{alg:dependency_graph_insertion} outlines ISC.
}
{
(pseudocode is in \inappendix).
}
We start by computing the $k$-cores by~\cite{batagelj2011fast_k-shell_decomposition}.

We consider the edges of $G^{ic}$, which is given, to build the dependency graph of insertion.
To construct the $G^{id}$, we check whether each edge $e \in G^{ic}$ can be handled by the lemmas given in~\cref{sec:insertion}.
If the conditions in any of the lemmas are satisfied, we can readily determine the dependency graph and tell if inserting the new edge $(u,v)$ will change the $K(u)$ and/or $K(v)$.
If the edge does not fit to any of the lemmas, we use the incremental core decomposition algorithm~\cite{sariyuce2016incremental} to determine the new core numbers.
For each of the cases, if there is any increase in $K(u)$ and/or $K(v)$, we update the $G^{id}$ by inserting directed edges based on the core number changes.
At the end, we calculate and return the insertion strength measures of each node $u \in G$ by using the $G^{id}$.

\noindent {\bf Time and space complexity.} In the worst case, incremental core decomposition algorithm~\cite{sariyuce2016incremental}, which takes $O(|E|)$, is run for each edge in $G^{ic}$. There is $O(b\cdot|V|)$ edges in $G^{ic}$ where $b$ is a constant. In total, the time complexity is $O(|V|\cdot|E|)$. However, this is a loose bound as we show runtime results in~\cref{sec:experimental_evaluation}. In addition to graph ($O(|E|)$), we store $|\Delta_{>} (u, G)|$ values ($O(|V|)$) and $IS_{ID}, IS_{OD}$ values ($O(|V|)$). Overall space complexity is $O(|E|)$.

\begin{center}
\begin{table}[!t]
\centering
\scriptsize
\resizebox{\columnwidth}{!}{
\begin{tabular}{|c|r|r||r|r|r|r||r|r|r|r|}
\hline
\multicolumn{3}{|c||}{\textbf{}} &
  \multicolumn{4}{c||}{\textbf{Removal}} &
  \multicolumn{4}{c|}{\textbf{Insertion}} \\ \hline
\textbf{Graph} &
  \multicolumn{1}{c|}{\textbf{\begin{tabular}[c]{@{}c@{}}$|V|$\end{tabular}}} &
    \multicolumn{1}{c||}{\textbf{\begin{tabular}[c]{@{}c@{}}$|E|$\end{tabular}}} &
  \multicolumn{1}{r|}{\textbf{\begin{tabular}[r]{@{}c@{}}\% Gain\end{tabular}}} &
  \multicolumn{1}{r|}{\textbf{\begin{tabular}[r]{@{}c@{}}Naive\\ (s)\end{tabular}}} &
  \multicolumn{1}{r|}{\textbf{\begin{tabular}[r]{@{}c@{}}RSC\\ (s)\end{tabular}}} &
  \multicolumn{1}{r||}{\textbf{Sp.}} &
    \multicolumn{1}{r|}{\textbf{\begin{tabular}[r]{@{}c@{}}$\frac{|E^{ic}|}{|E|}$\end{tabular}}} &
  \multicolumn{1}{r|}{\textbf{\begin{tabular}[r]{@{}c@{}}Naive\\ (s)\end{tabular}}} &
    \multicolumn{1}{r|}{\textbf{\begin{tabular}[r]{@{}c@{}}ISC\\ (s)\end{tabular}}} &
  \multicolumn{1}{r|}{\textbf{Sp.}} \\ \hline
as19971108	&	3015	&	5156	&	50.2	\%	&	4.88	&	2.93	&	1.67x	&	2.79	&	1.10	&	0.75	&	1.46x	\\
as19990309	&	4759	&	8896	&	54.4	\%	&	12.38	&	6.35	&	1.95x	&	2.58	&	1.70	&	1.31	&	1.30x	\\
bio-dmela	&	7393	&	25569	&	79.4	\%	&	56.96	&	12.85	&	4.43x	&	1.40	&	38.26	&	26.60	&	1.44x	\\
ca-CondMat	&	21363	&	91286	&	89.0	\%	&	575.82	&	67.19	&	8.57x	&	1.11	&	475.15	&	377.32	&	1.26x	\\
ca-Erdos992	&	5094	&	7515	&	39.1	\%	&	10.93	&	7.47	&	1.46x	&	3.11	&	7.48	&	5.23	&	1.43x	\\
ca-GrQc	&	4158	&	13422	&	84.4	\%	&	17.82	&	3.47	&	5.14x	&	1.39	&	32.41	&	26.66	&	1.22x	\\
inf-openflights	&	2939	&	15677	&	86.8	\%	&	15.67	&	2.49	&	6.29x	&	0.90	&	2.64	&	2.28	&	1.16x	\\
inf-power	&	4941	&	6594	&	63.8	\%	&	9.36	&	3.96	&	2.36x	&	2.76	&	504.87	&	486.43	&	1.04x	\\
jazz	&	198	&	2742	&	97.8	\%	&	0.43	&	0.06	&	7.17x	&	0.35	&	1.00	&	0.94	&	1.06x	\\
p2p-Gnutella08	&	6301	&	20777	&	80.3	\%	&	40.83	&	8.99	&	4.54x	&	1.45	&	951.50	&	918.05	&	1.04x	\\
p2p-Gnutella09	&	8114	&	26013	&	78.8	\%	&	63.81	&	14.51	&	4.40x	&	1.49	&	769.12	&	713.63	&	1.08x	\\
soc-hamsterster	&	2426	&	16630	&	93.6	\%	&	13.03	&	1.27	&	10.26x	&	0.72	&	4.82	&	4.22	&	1.14x	\\
soc-wiki-Vote	&	889	&	2914	&	81.2	\%	&	0.88	&	0.29	&	3.03x	&	1.46	&	1.45	&	1.09	&	1.33x	\\
tech-routers-rf	&	2113	&	6632	&	77.8	\%	&	4.26	&	1.30	&	3.28x	&	1.50	&	2.46	&	1.98	&	1.25x	\\
tech-WHOIS	&	7476	&	56943	&	89.9	\%	&	128.82	&	14.51	&	8.88x	&	0.65	&	6.59	&	5.60	&	1.18x	\\
USAir97	&	332	&	2461	&	91.2	\%	&	0.28	&	0.08	&	3.50x	&	0.65	&	0.12	&	0.09	&	1.46x	\\
web-spam	&	4767	&	37375	&	91.5	\%	&	52.48	&	5.28	&	9.94x	&	0.63	&	7.05	&	5.17	&	1.36x	\\ \hline
\end{tabular}
}
\vspace{1ex}
\caption{\footnotesize {\bf Statistics for the networks (first two columns) and runtime results for edge removal and edge insertion (in seconds). \%Gain denotes the savings how much less edges are processed by our algorithm than the naive approach for edge removal. $\frac{|E^{ic}|}{|E|}$ denotes the ratio of the number of edges in the insertion candidate graph to the actual graph. Sp. is the speedup of our algorithms against the naive approaches.}}
\label{table:runtime}
\vspace{-9ex}
\end{table}
\end{center}

\section{Experimental Evaluation}
\label{sec:experimental_evaluation}

We conduct experiments on real-world networks of various types and sizes to evaluate the efficiency and effectiveness of our node-strength measures.
\cref{table:runtime} (first three columns) shows the statistics of the networks, obtained from SNAP\footnote{\url{http://snap.stanford.edu/}} and Network Repository\footnote{\url{http://networkrepository.com/}}. 
All experiments are performed on a Linux operating system (v. 3.10.0-1127) running on a machine with Intel(R) Xeon(R) Gold 6130 CPU processor at 2.10 GHz with 192 GB memory.
We implemented our algorithms in Python 3.6.8.
{\bf Our implementation is publicly available\footnote{\url{https://github.com/erdemUB/ECMLPKDD23}}.}

Since we consider random edge selections to construct $G^{ic}$ and calculate $IS_{ID}$ and $IS_{OD}$, we repeat insertion experiments 10 times to account for randomness and report the average strength measure for each node.
Note that the standard deviation in those computations is quite low, e.g., in {\tt inf-openflights} graph, the standard deviation is less than .18 for most nodes where more than half of the nodes have zero (or close to zero) standard deviation (details are in \inappendix{fig:stdev_count}).

\subsection{Runtime Results}

We first compare the runtime performances of our RSC and ISC algorithms against the naive strategy which simply runs incremental core decomposition algorithms for each edge removal and edge insertion.
One important note is that the three approaches (Subcore, Purecore, and Traversal) proposed in~\cite{sariyuce2013streaming} give different behaviors in our removal and insertion experiments. 
Although the Traversal algorithm is shown to be the best in~\cite{sariyuce2013streaming} for both single edge removal and insertion, we observe that the Subcore algorithm can be made faster for edge insertion in our experiments.
The key is to precompute all the subcores in each $k$-core and reuse when handling edge insertions.
We use this pre-calculation technique and Subcore algorithm in our edge insertion experiments, whereas the Traversal algorithm is used in our edge removal experiments.

\cref{table:runtime} gives the results. For the edge removal, we are able to remove 78.2\% less edges, on average, when compared to the naive approach (fourth column in~\cref{table:runtime}).
This translates to 5.11x faster runtime on average.
For edge insertion, our algorithm well utilizes the lemmas in~\cref{sec:insertion} and gives 1.25x faster computation when compared to the naive approach.

\subsection{Finding Critical Edges}\label{sec:criticalEdges}

Here, we compare our node resilience measures to several baselines for finding critical edges in edge removal and insertion scenarios. 
We use four baselines: Random, Core Number, Degree, Core Strength.
Each method identifies a limited number ($c$) of critical edges to maximize the impact on the core numbers of affected nodes.
For Random, we repeat experiments 50 times and take the average.
For Core Number, Degree, and Core Strength; the score of each edge is determined by the sum of its end points' values and $c$ edges with the highest score are considered.
We assess each method by the percentage of nodes affected, $F$, (decreased or increased from the initial core number) by the removal or insertion of the budget number of edges. 
For all experiments, we vary the budget ($c$) from 50 to 1000 and evaluate our results. For better visualization, we show the results from budget 600 to 1000 in \cref{fig:critical_edge}(c) and \cref{fig:critical_edge}(d).

\vspace{-2ex}
\subsubsection{Edge Removal Experiments}

\begin{figure}[!t]
\vspace{-2ex}
    \centering
\begin{minipage}{1.0\textwidth}
\begin{subfigure}[t]{.25\linewidth}
    \includegraphics[width=\linewidth]{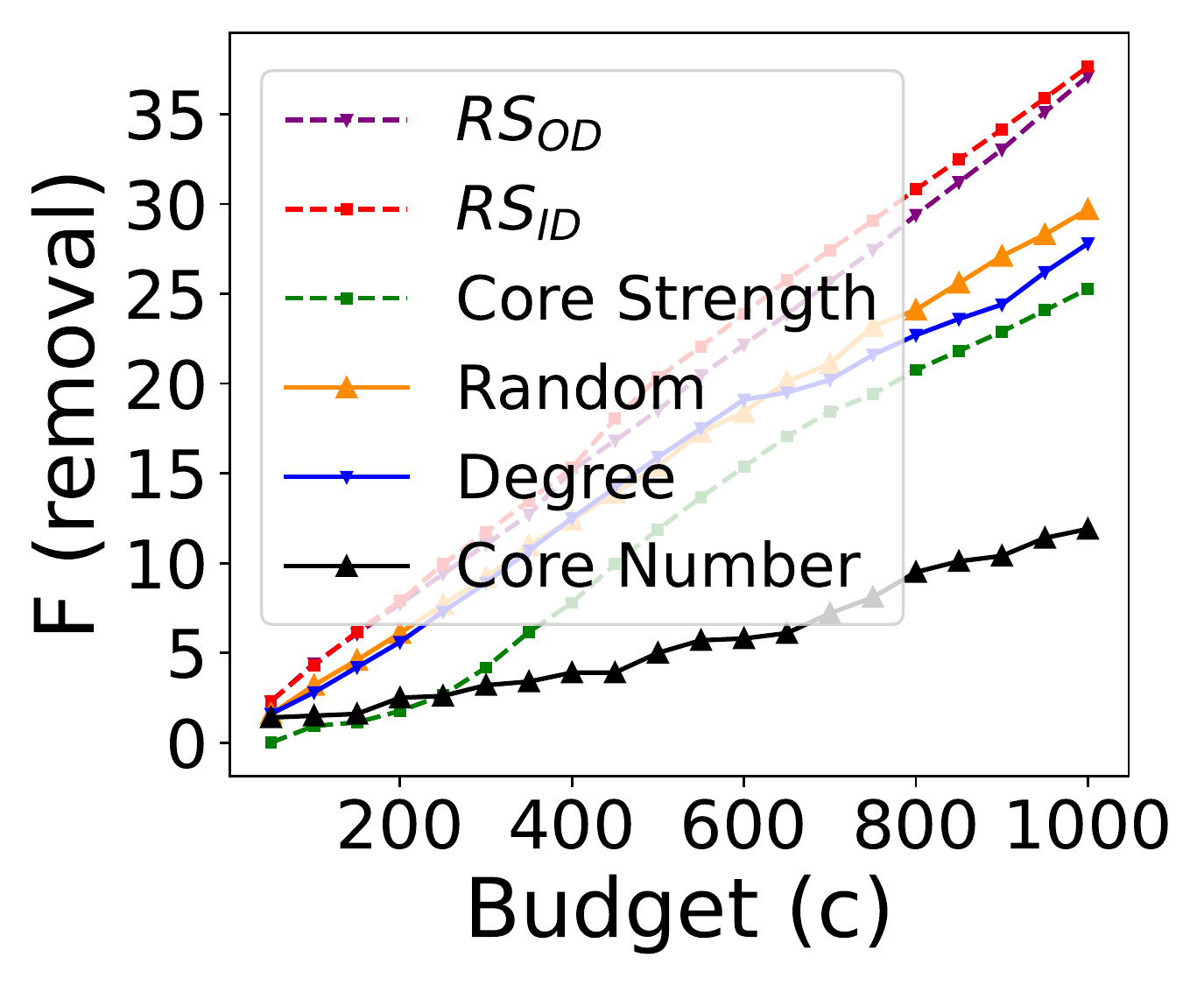}

    \label{fig:rs-as19}
\end{subfigure}
\hspace{-2mm}
\begin{subfigure}[t]{.25\linewidth}
    \includegraphics[width=\linewidth]{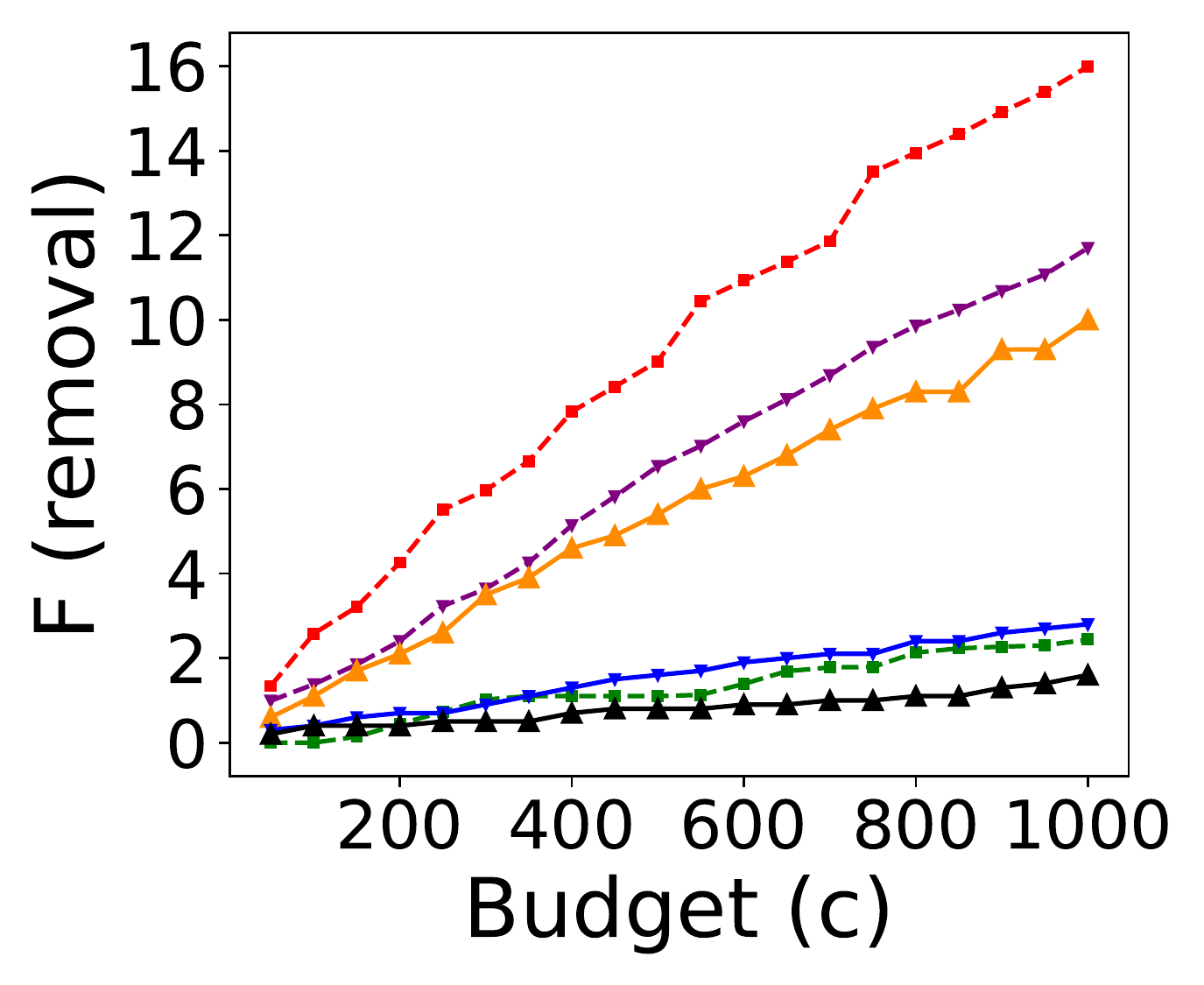}

    \label{fig:rs-cacond}
 \end{subfigure}
 \hspace{-2mm}
\begin{subfigure}[t]{.25\linewidth}
    \includegraphics[width=\linewidth]{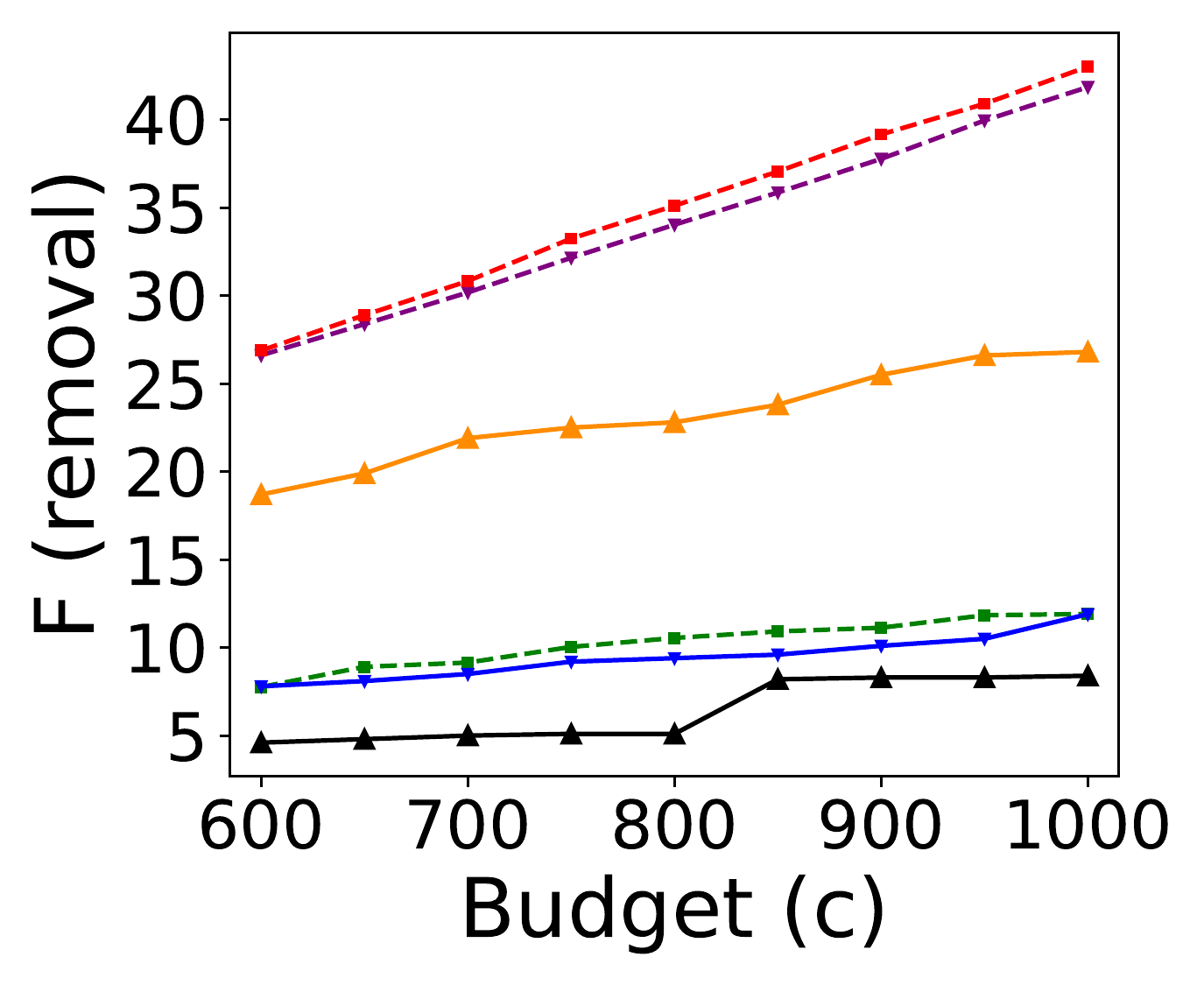}

    \label{fig:rs-infop}
  \end{subfigure}
  \hspace{-2mm}
\begin{subfigure}[t]{.25\linewidth}
    \includegraphics[width=\linewidth]{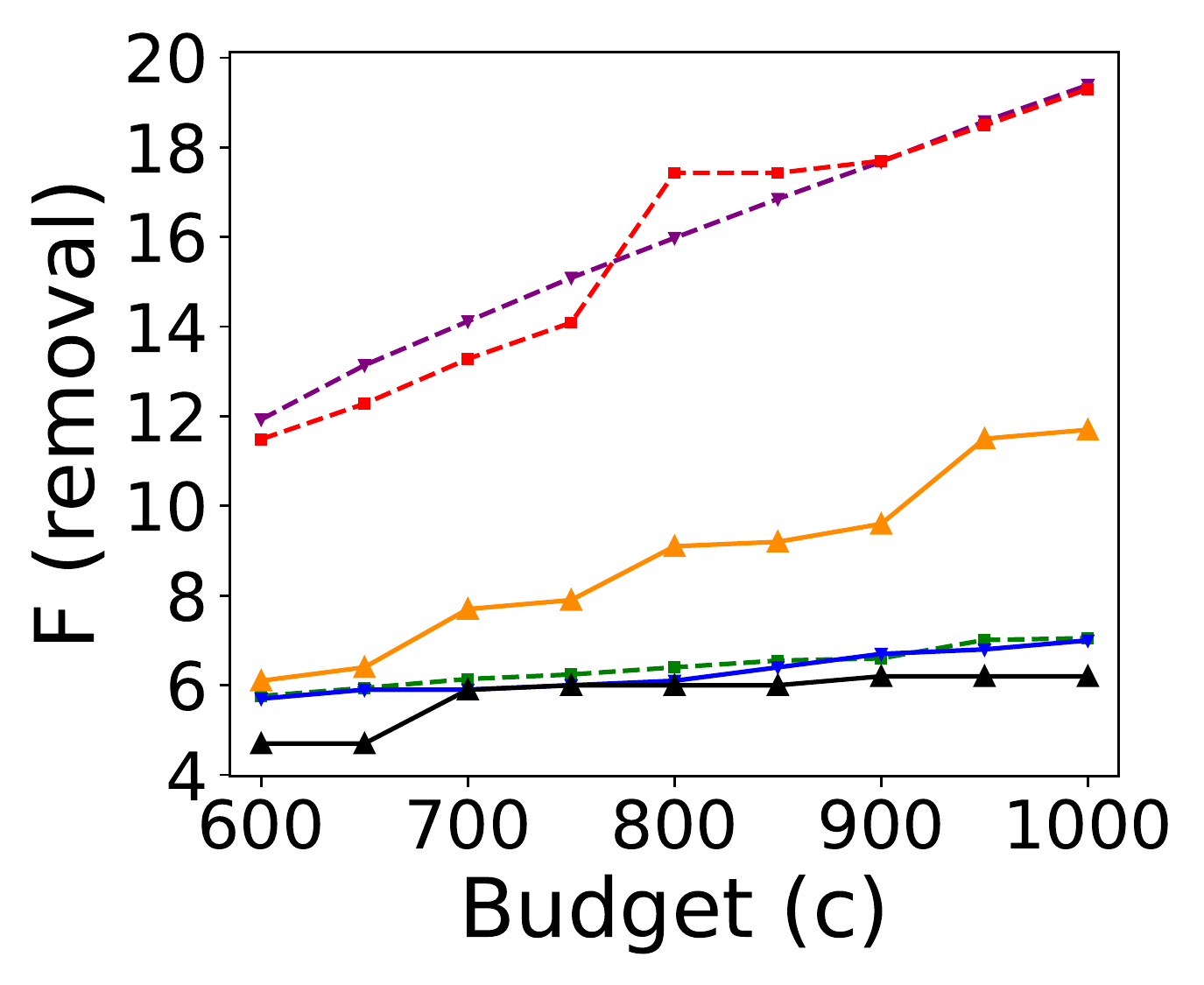}

    \label{fig:rs-p2p}
  \end{subfigure}
\end{minipage}
\vspace{-4ex}

\begin{minipage}{1.0\textwidth}
  \begin{subfigure}[t]{.25\linewidth}
    \includegraphics[width=\linewidth]{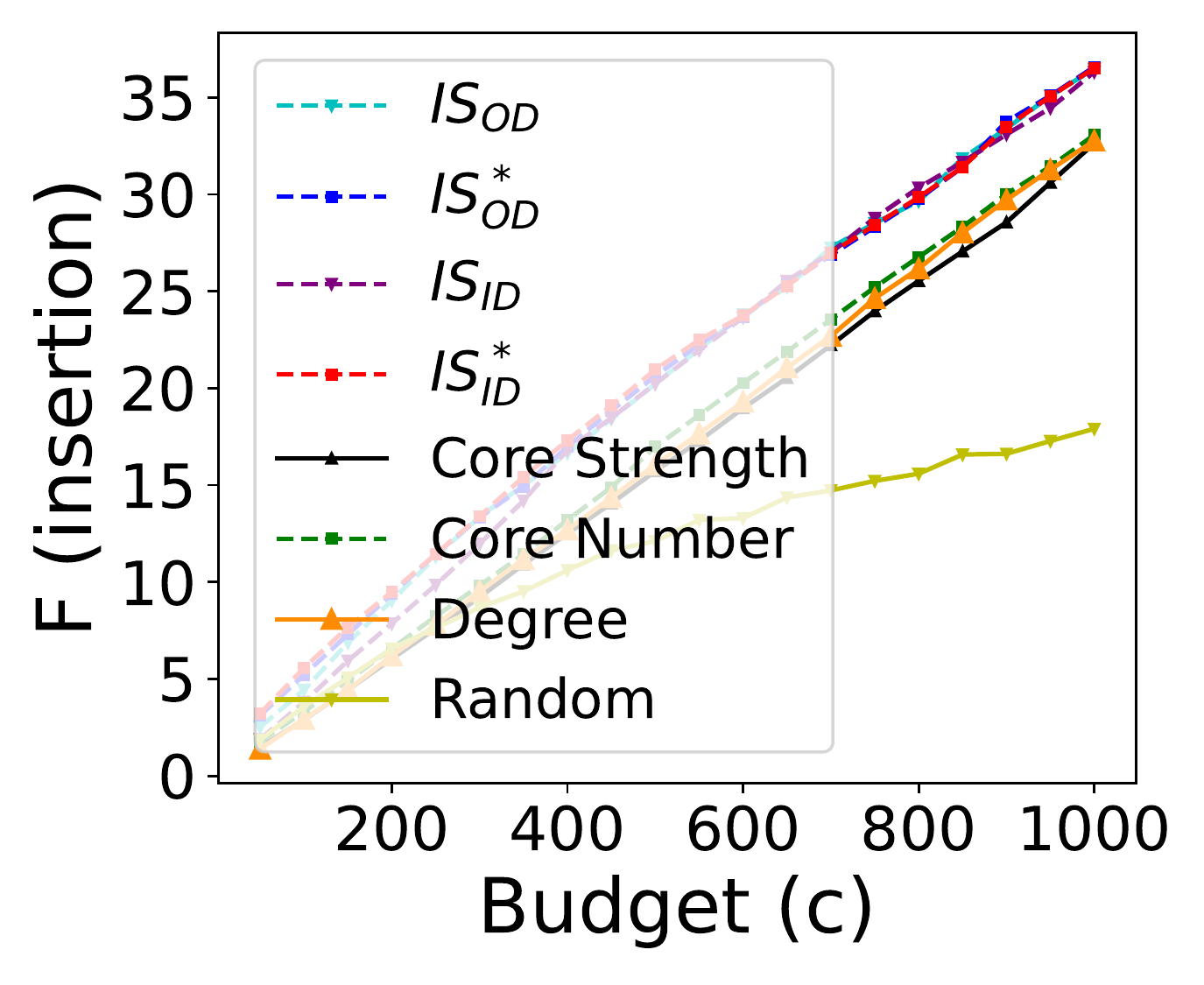}
    \caption{as19971108}
    \label{fig:rs-as19}
\end{subfigure}
\hspace{-2mm}
\begin{subfigure}[t]{.25\linewidth}
    \includegraphics[width=\linewidth]{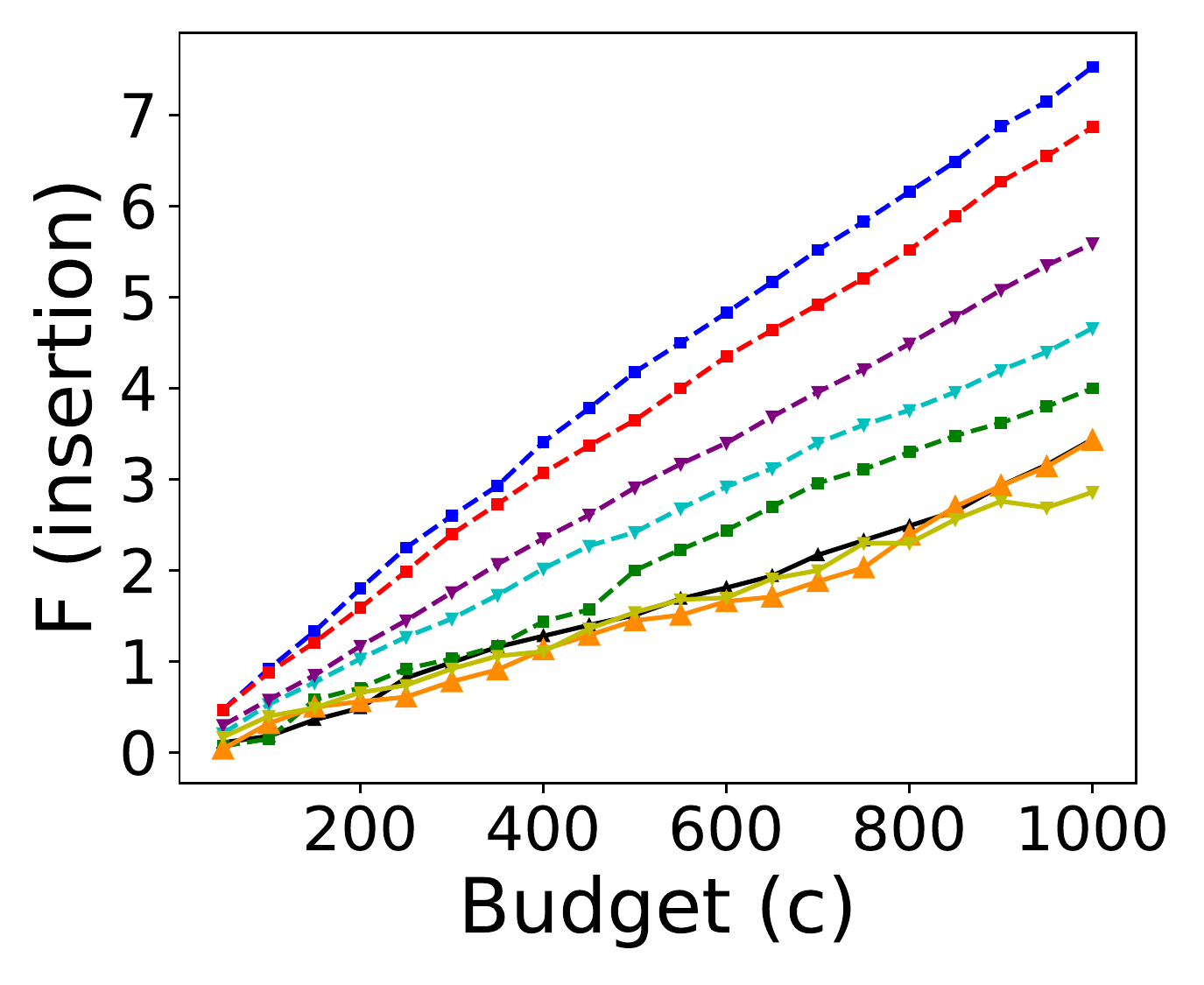}
    \caption{ca-CondMat}
    \label{fig:rs-cacond}
 \end{subfigure}
 \hspace{-2mm}
\begin{subfigure}[t]{.25\linewidth}
    \includegraphics[width=\linewidth]{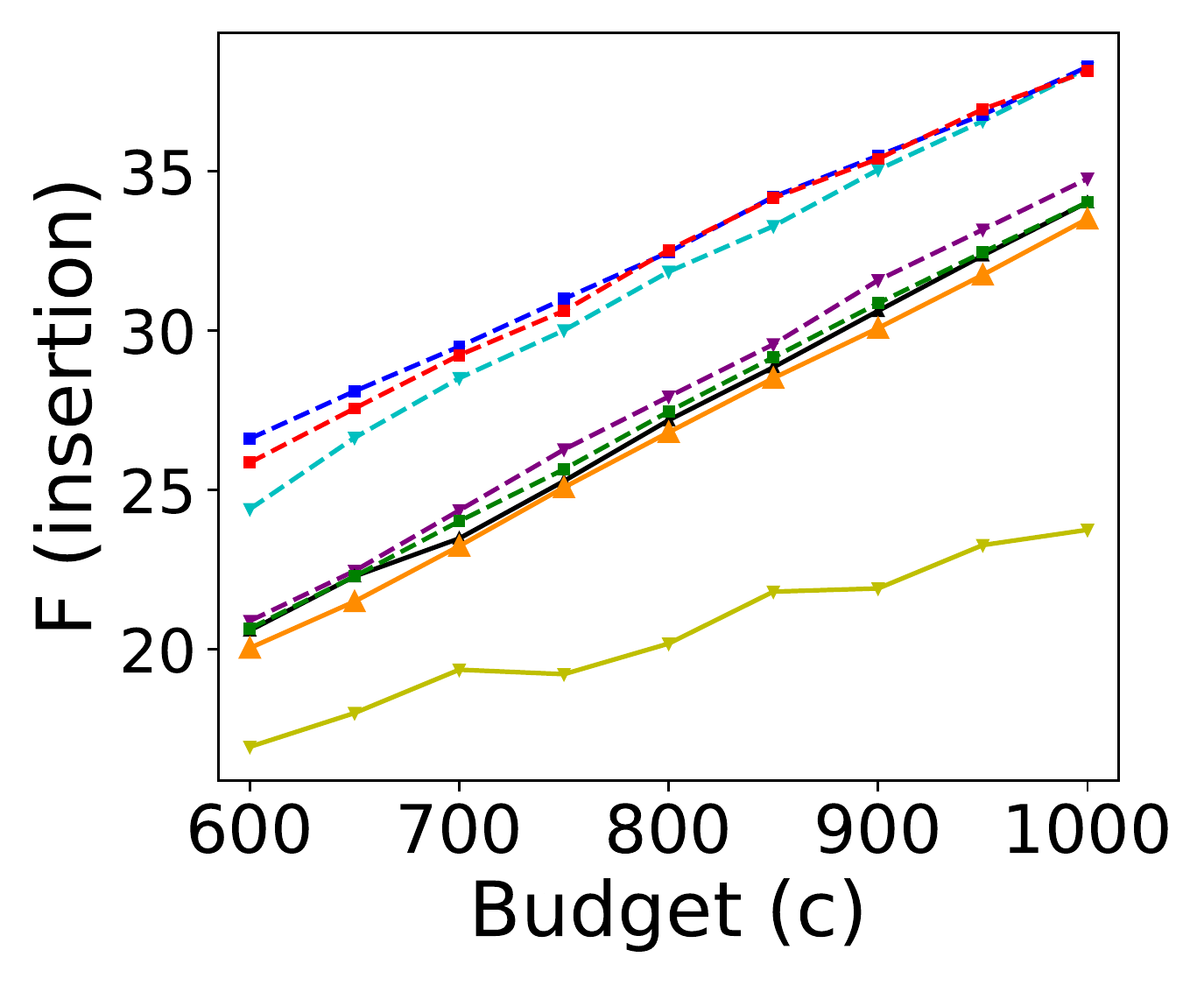}
    \caption{inf-openflights}
    \label{fig:rs-infop}
  \end{subfigure}
  \hspace{-2mm}
\begin{subfigure}[t]{.25\linewidth}
    \includegraphics[width=\linewidth]{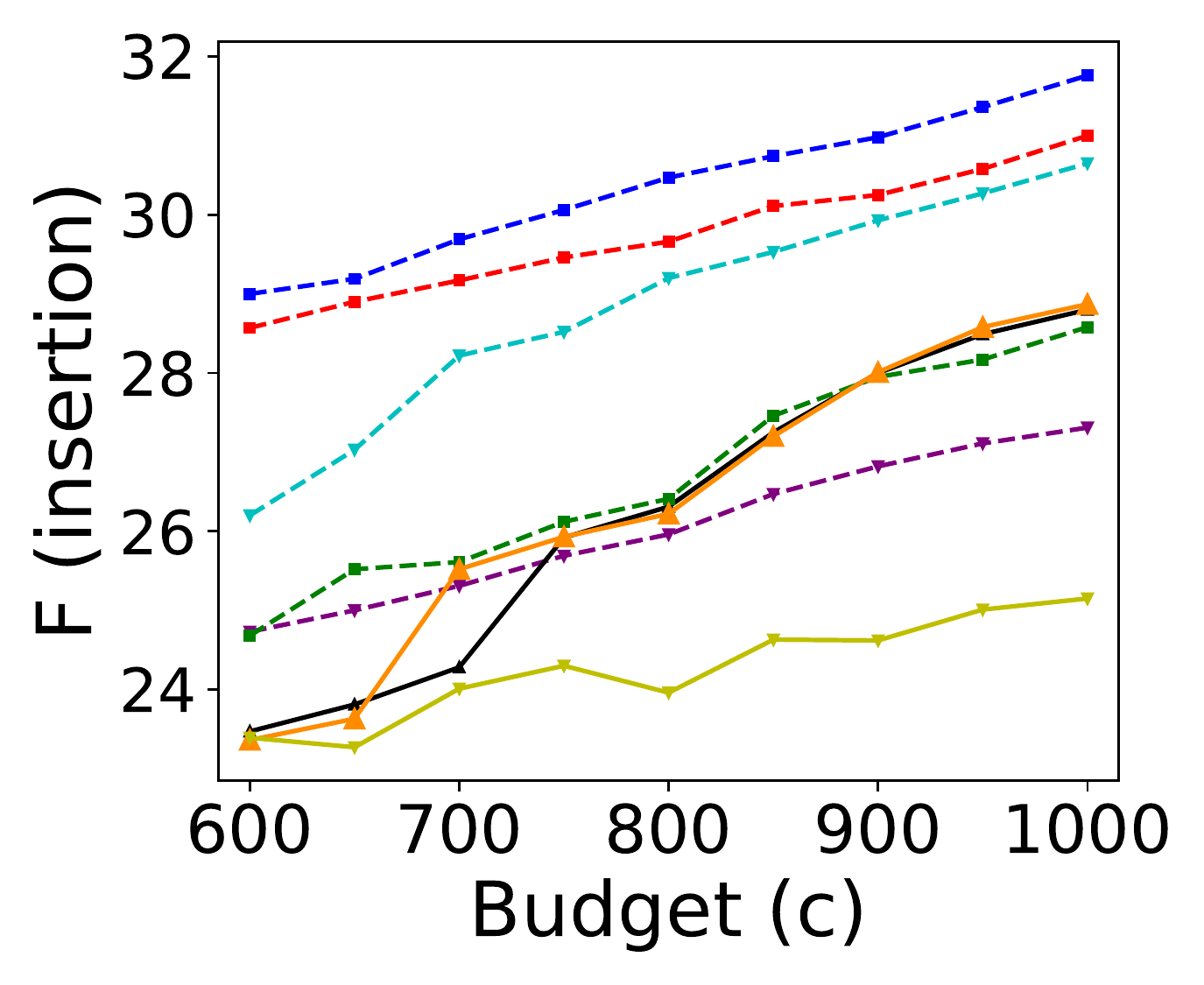}
    \caption{p2p-Gnutella08}
    \label{fig:rs-p2p}
  \end{subfigure}
\end{minipage}
\vspace{-2ex}
\caption{\footnotesize {\bf Finding critical edges by our methods and baselines for edge removal (top row) and edge insertion (bottom row).}}

\label{fig:critical_edge}
\vspace{-5ex}
\end{figure}

We use $RS_{ID}$ and $RS_{OD}$ to select $c$ critical edges to remove from the graph.
For our measures, the score of each edge is set as the sum of its endpoints'  $RS_{ID}$ or $RS_{OD}$ values.
For $RS_{ID}$, we choose $c$ edges with the lowest scores as a node with lower $RS_{ID}$ is more likely to change its core number on edge removal, whereas, for $RS_{OD}$, we select $c$ edges with the highest score as a node with larger $RS_{OD}$ affect other nodes' core numbers more.
We also pay attention to not selecting no more than one edge from any $KAES(S)$, as removing any edges in $KAES(S)$ produces the same core vector for a $k$-corona $S$ by~\cref{lem:same_corevector_kaes}. 
For Random, we choose $c$ random edges from the graph.
\cref{fig:critical_edge} (top row) shows the results for four graphs (results for other graphs are in \inappendix{fig:appendix_edge_deletion}).
Both $RS_{ID}$ and $RS_{OD}$ outperform the baselines.
$RS_{ID}$ is slightly better than $RS_{OD}$ in some graphs and significantly better in a few. 

\vspace{-2ex}

\subsubsection{Edge Insertion Experiments}

We use $IS_{OD}$ and $IS_{ID}$ to select $c$ critical edges to insert to the graph.
We first consider all the non-neighbor node pairs who share at least two common neighbors, called as candidate set, and then select a subset of size $c$ edges of the candidate set by using the baselines or our methods. 
When inserting an edge $(u, v)$, if core number of both $u$ and $v$ increased by a previous insertion, we skip this edge.
For our measures, we define the score of each candidate edge $(u,v)$ as $ \max(IS_{ID}(u), IS_{ID}(v))$ (likewise for $IS_{OD}$), then select the $c$ edges with lowest scores to insert.
Here, we consider maximum endpoint strength as the edge score, unlike the edge removal case where we considered sum, to keep the scores more regularized because the space of edge insertions is larger and can yield very large edge scores if the sum is applied.
We choose the edges with the lowest score as they are the least resilient for incrementing core numbers. 
For Random, we choose $c$ random edges from the candidate set.
\cref{fig:critical_edge} (bottom row) gives the results for four networks (rest are in \inappendix{fig:appendix_edge_insertion}). Overall, $IS_{ID}$ and $IS_{OD}$ consistently outperform the baselines.

We also define a simple variant of our measures to handle the clique-like structures in which core numbers are difficult to increase.
We consider the propagation effect of neighbor nodes by summing up the strength of a node with its neighbors' strengths.
We define Neighbor Sum variants as $IS^*_{ID} (u) = IS_{ID}(u) + \sum_{v\in N(u)}^{} IS_{ID}(v)$ (likewise for $IS^*_{OD}$).
As above we define the score of each candidate edge the maximum strength of its endpoints and then select the $c$ edges with the lowest
scores to insert. As shown in~\cref{fig:critical_edge} (bottom row), $IS^*_{ID}$ and $IS^*_{OD}$ significantly outperform all the other methods in {\tt ca-CondMat}, which is a co-authorship network formed by cliques of authors on a paper.

\vspace{-2ex}
\subsection{Identifying Influential Spreaders}\label{sec:infspreader}

In this section, we consider the problem of identifying influential spreaders in the SIR model.
We use our node resilience measures as well as three baselines to choose 20\% nodes in a given graph as the initially infected node set.
For our measures ($RS_{ID}, RS_{OD}, IS_{ID}, IS_{OD}$), we choose the node with largest strength 
from the highest $k$-shell, then do the same for the next highest shell ($(k-1)$-shell), and so on until the 1-shell. Then we repeat this process until 20\% of the nodes are chosen. Ties are broken randomly. As the highest strength nodes are more resilient upon graph changes, they are more important for influence maximization than others.
Regarding the baselines, we choose the methods that rely on core numbers---the $k$-shell strategy~\cite{kitsak2010identification_influential_kshell}, the $IKS$ method~\cite{wang2020identifying_improved_kshell}, and the Core Strength measure (the nodes with the largest values)---to select the 20\% initially infected node set.
To ensure a smooth transmission in the SIR model, we fix S$\rightarrow$I probability $\mu = 0.01$ and set the value of I$\rightarrow$R probability $\mathbf{\beta}$ to be a little bit bigger than ${\beta_{min}} = \langle k \rangle/
\langle k^2 \rangle$ \cite{castellano2010thresholds_transmission_rate}, where $\langle k \rangle$ and $\langle k^2 \rangle$ are the first and the second moment of the degree distribution, as done in~\cite{wang2020identifying_improved_kshell} (exact $\beta$ values are in \inappendix{table:dataset_statistics}).
For each method, we run the model 50 times and take the average.
We consider the percentage of affected nodes at time $t$, denoted as $S(t)$, to evaluate the spreading effect of the initially infected node set.

\begin{figure}[!t]
\vspace{-2ex}
    \centering

    \begin{minipage}{1.0\textwidth}
\begin{subfigure}[t]{.25\linewidth}
    \includegraphics[width=\linewidth]{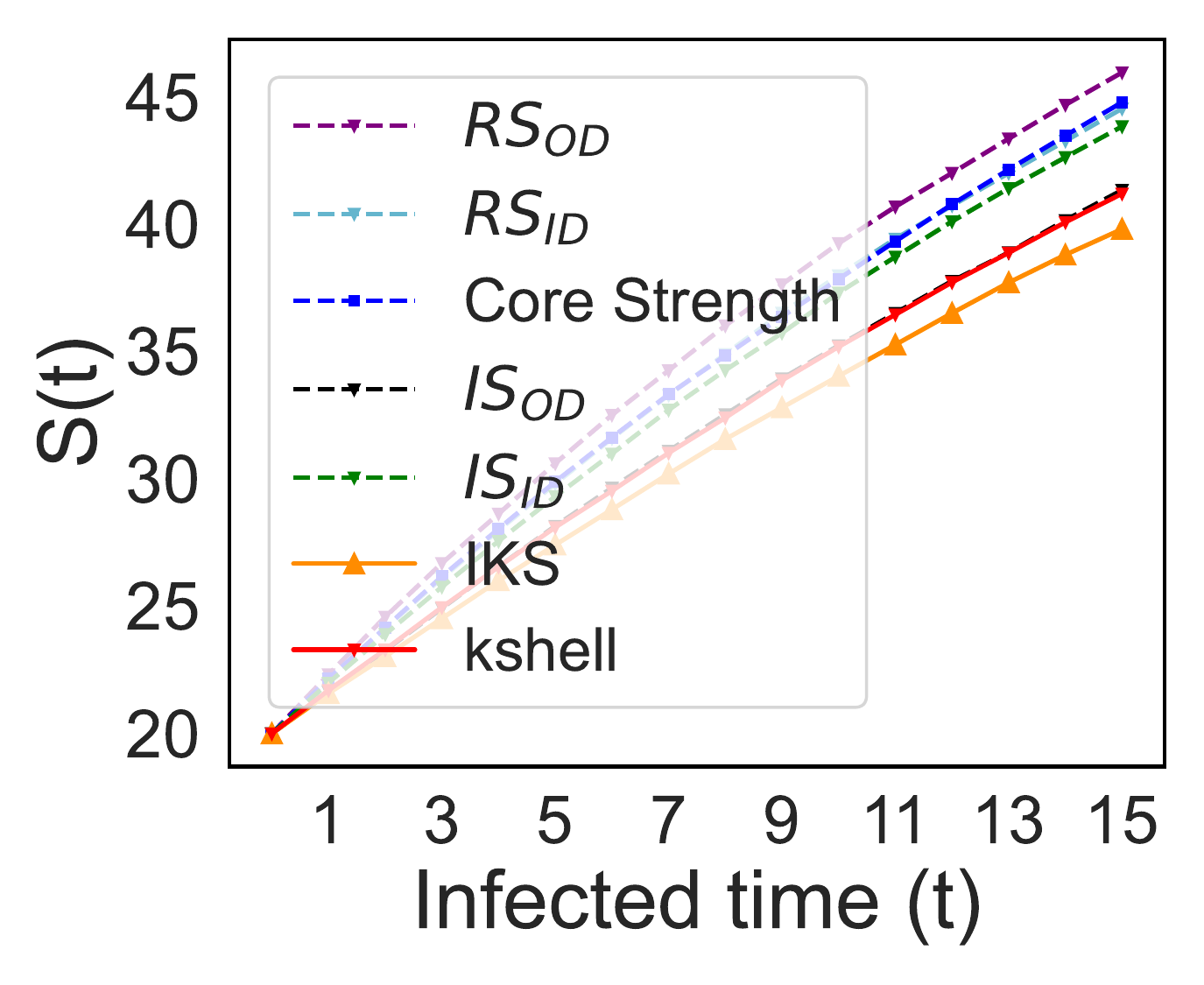}
    \vspace{-3ex}
    \caption{as19971108}
    \label{fig:rs-as19}
\end{subfigure}
\hspace{-2mm}
\begin{subfigure}[t]{.25\linewidth}
    \includegraphics[width=\linewidth]{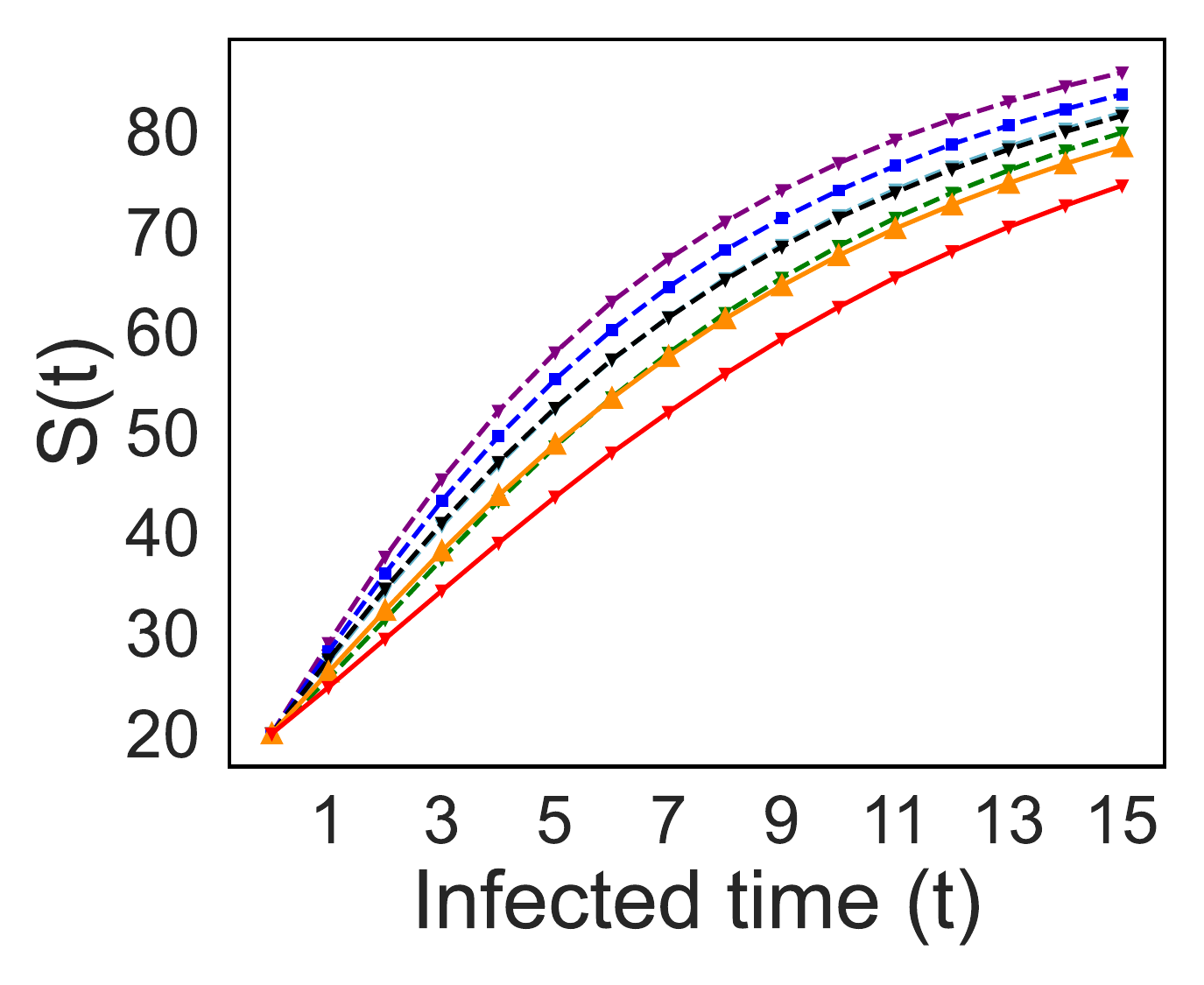}
    \vspace{-3ex}
    \caption{ca-CondMat}
    \label{fig:rs-cacond}
 \end{subfigure}
 \hspace{-2mm}
\begin{subfigure}[t]{.25\linewidth}
    \includegraphics[width=\linewidth]{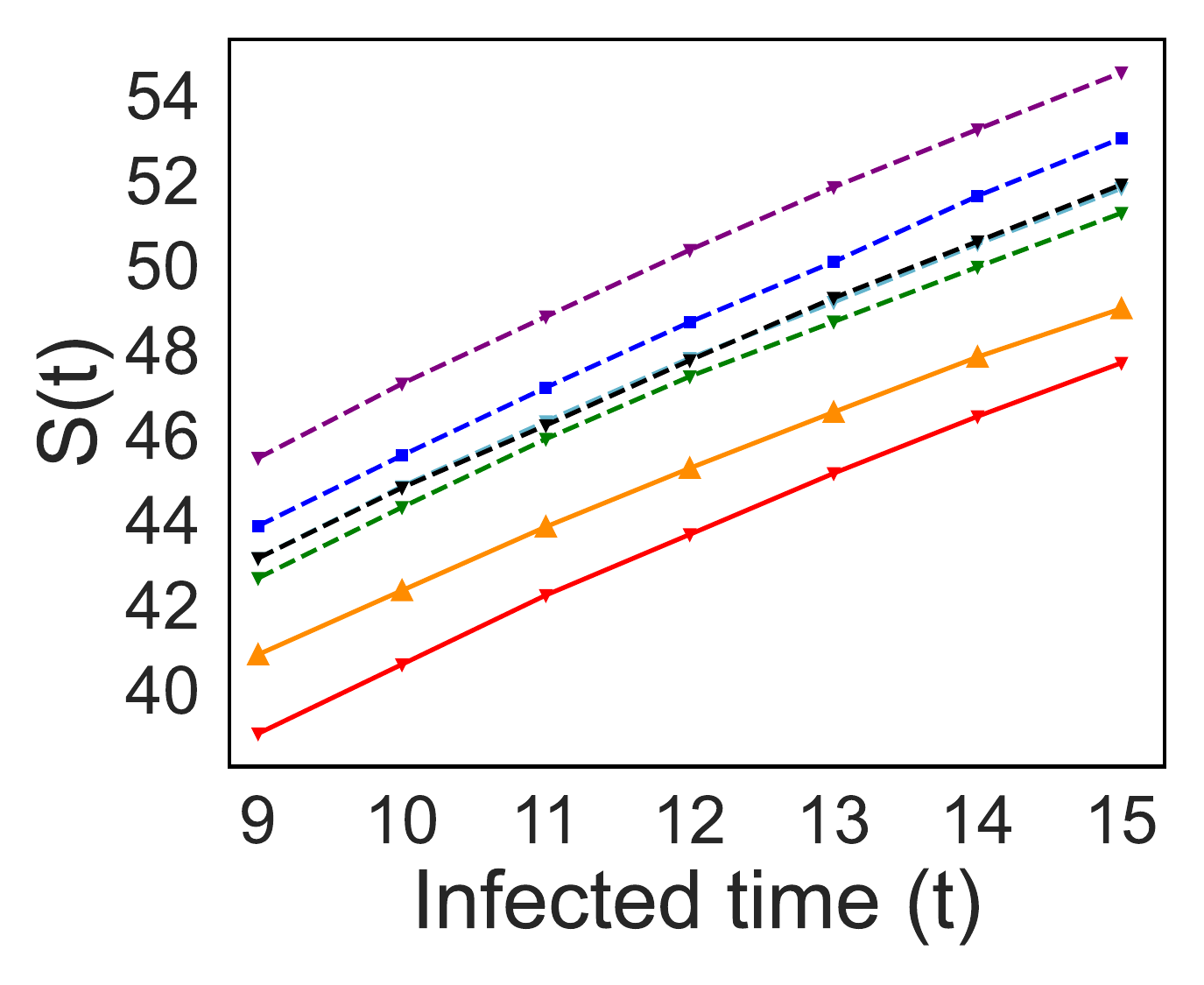}
    \vspace{-3ex}
    \caption{inf-openflights}
    \label{fig:rs-infop}
  \end{subfigure}
  \hspace{-2mm}
\begin{subfigure}[t]{.25\linewidth}
    \includegraphics[width=\linewidth]{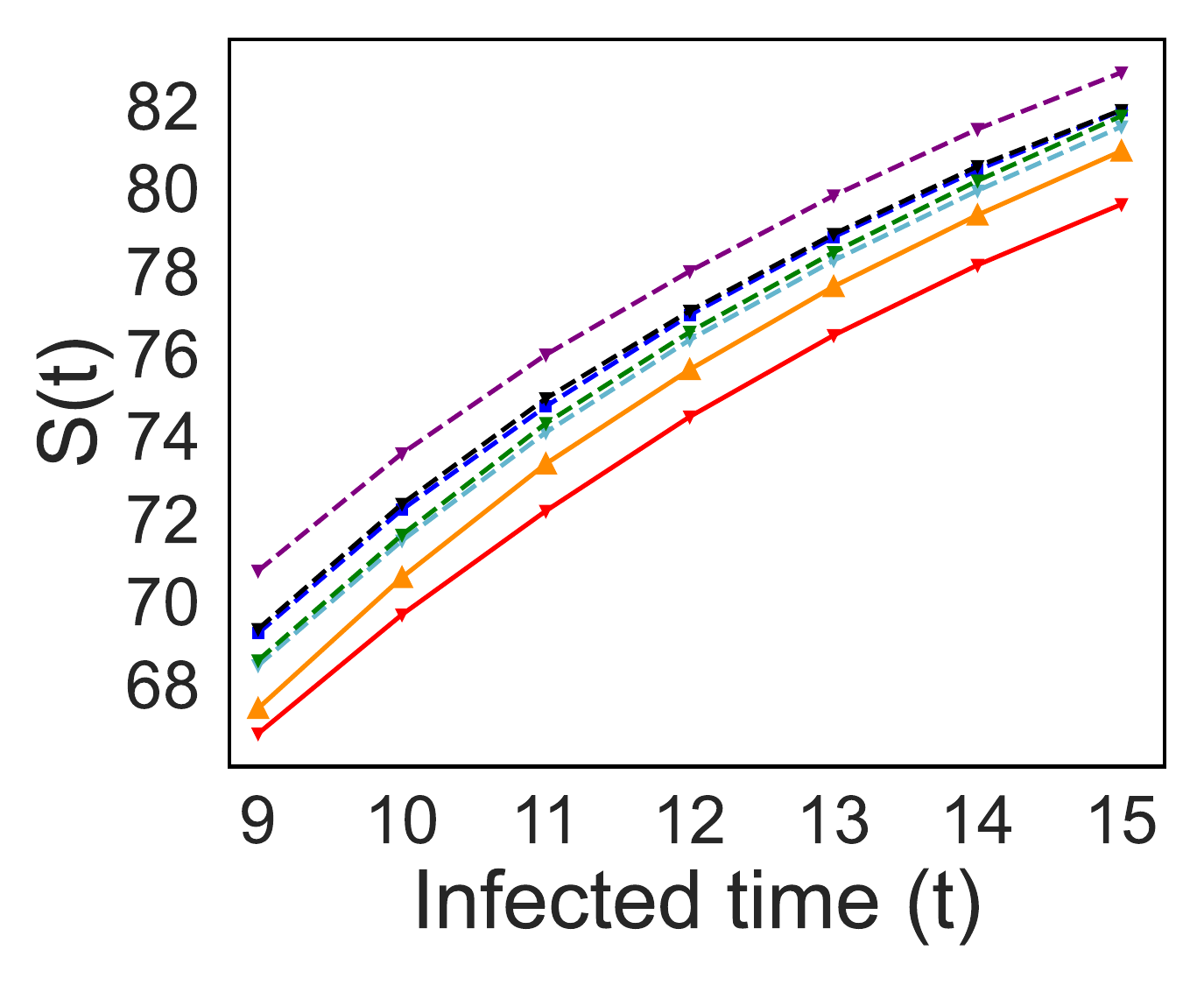}
    \vspace{-3ex}
    \caption{p2p-Gnutella08}
    \label{fig:rs-p2p}
  \end{subfigure}   
\end{minipage}
\vspace{-2ex}
   \caption{\footnotesize {\bf Identifying influential spreaders by our measures and baselines.}}

    \label{fig:influential_figure}
 \vspace{-5ex}
  
\end{figure}

\cref{fig:influential_figure} shows $S(t)$ as a function of $t \in [0, 15]$ for four networks (results for other graphs are in \inappendix{fig:appendix_influential_figure}).
As $t$ increases, $S(t)$ rises and eventually reaches a steady value.
Overall, our node strength measures outperform the $k$-shell and $IKS$ strategies.
Core Strength measure shows superior performance than some of our measures but $RS_{OD}$ consistently outperforms all the methods.
The reason for this behavior is that the nodes with large $RS_{ID}$ do not always have large core numbers whereas the nodes with large $RS_{OD}$ are consistently in highest $k$-cores.

\section{Conclusions and Future Work}

In this paper, we studied the problem of node-based core resilience upon edge removals and edge insertions.
We first showed that the Core Strength~\cite{www18} does not correctly capture the core resilience of a node upon edge removals.
Then we introduced the concept of dependency graph to capture the impact of neighbor nodes (for removal) and probable future neighbor nodes (for insertion) on the core number of a given node. 
We defined node strengths in dependency graphs based on in- and out-degrees and introduced efficient heuristics to compute those.
Experiments show that our heuristics are faster than the naive approaches and our strength measures outperform the existing baselines on two key applications, finding critical edges and identifying influential spreaders. For future work, we plan to speed up the computation of insertion strength measures and also consider more realistic scenarios to construct the $G^{ic}$.\\ 

\noindent {\bf Acknowledgments.} Hossain and Sar{\i}y\"{u}ce were supported by NSF Award\\
\noindent \#1910063 and used resources of the Center for Computational Research at the University at Buffalo\jakir{\footnote{\url{https://ubir.buffalo.edu/xmlui/handle/10477/79221}}}. Soundarajan was supported by NSF Award \#1908048.

\section{Ethical Statement}

Our contribution is algorithmic in nature, building on previously proposed concepts. We work on public datasets. We do not foresee any ethical implications of our work.

\bibliography{reference}

\ifthenelse{\boolean{extended}}
{
\newpage 
\clearpage 
\section{Appendix}

\begin{center}

\begin{table}[]
\centering

\begin{tabular}{|c|c|c|c|c|c|c|}
\hline
    \textbf{Type}     & \textbf{$\mathbf{Graph Name}$}         & \textbf{$\mathbf{\lvert V\lvert }$} & \textbf{$\mathbf{\lvert E\lvert }$}   & $\boldsymbol{{\beta}_{min}}$   & $\boldsymbol{\beta}$ & $\boldsymbol{d_{avg}}$  \\ \hline \hline
\multirow{2}{*}{AS}   & as19971108      & 3015  & 5156  & 0.011 & 0.02 & 1.71  \\ \cline{2-7} 
                      & as19990309      & 4759  & 8896  & 0.008 & 0.02 & 1.87  \\ \hline
BIO                   & bio-dmela       & 7393  & 25569 & 0.042 & 0.05 & 3.46  \\ \hline
\multirow{3}{*}{CA}   & ca-CondMat      & 21363 & 91286 & 0.045 & 0.05 & 4.27  \\ \cline{2-7} 
                      & ca-Erdos992     & 5094  & 7515  & 0.061 & 0.07 & 1.48  \\ \cline{2-7} 
                      & ca-GrQc         & 4158  & 13422 & 0.056 & 0.06 & 3.23  \\ \hline
\multirow{2}{*}{INF}  & inf-openflights & 2939  & 15677 & 0.018 & 0.02 & 5.33  \\ \cline{2-7} 
                      & inf-power       & 4941  & 6594  & 0.258 & 0.26 & 1.33  \\ \hline
Jazz                  & Jazz            & 198   & 2742  & 0.026 & 0.03 & 13.85 \\ \hline
\multirow{2}{*}{P2P}  & p2p-Gnutella08  & 6301  & 20777 & 0.057 & 0.06 & 3.30  \\ \cline{2-7} 
                      & p2p-Gnutella09  & 8114  & 26013 & 0.059 & 0.06 & 3.21  \\ \hline
\multirow{2}{*}{SOC}  & soc-hamsterster & 2426  & 16630 & 0.024 & 0.03 & 6.85  \\ \cline{2-7} 
                      & soc-wiki-Vote   & 889   & 2914  & 0.055 & 0.06 & 3.28  \\ \hline
\multirow{2}{*}{TECH} & tech-routers-rf & 2113  & 6632  & 0.046 & 0.05 & 3.14  \\ \cline{2-7} 
                      & tech-WHOIS      & 7476  & 56943 & 0.007 & 0.02 & 7.62  \\ \hline
Air                   & USAir           & 332   & 2461  & 0.023 & 0.03 & 7.41  \\ \hline
WEB                   & web-spam        & 4767  & 37375 & 0.014 & 0.02 & 7.84  \\ \hline
\end{tabular}
\vspace{1ex}
\caption{\bf {\footnotesize Statistics of networks.  In this table, $\boldsymbol{\lvert V\lvert }$ is the number of nodes, $\boldsymbol{\lvert E\lvert }$ is the number of edges, $\boldsymbol{d_{avg}}$ is the average degree, $\boldsymbol{\beta_{min}}$ is the transmission rate and $\boldsymbol{\beta}$ is chosen slightly larger than $\boldsymbol{\beta_{min}}$. Here, we include the networks from 10 different domains (Type). The full forms are, AS-Autonomous System, BIO-Biological,  CA-Collaboration, INF-Infrastructure, Jazz-jazz musicians, P2P-peer to peer, SOC-Social, TECH-Technological, Air-USAir97, and WEB-Web spam.}}

\label{table:dataset_statistics}
\end{table}
\end{center}

\begin{figure}[!h]
    \center
    \centering

    \begin{subfigure}[t]{.25\linewidth}
    {{ \includegraphics[width=\linewidth]{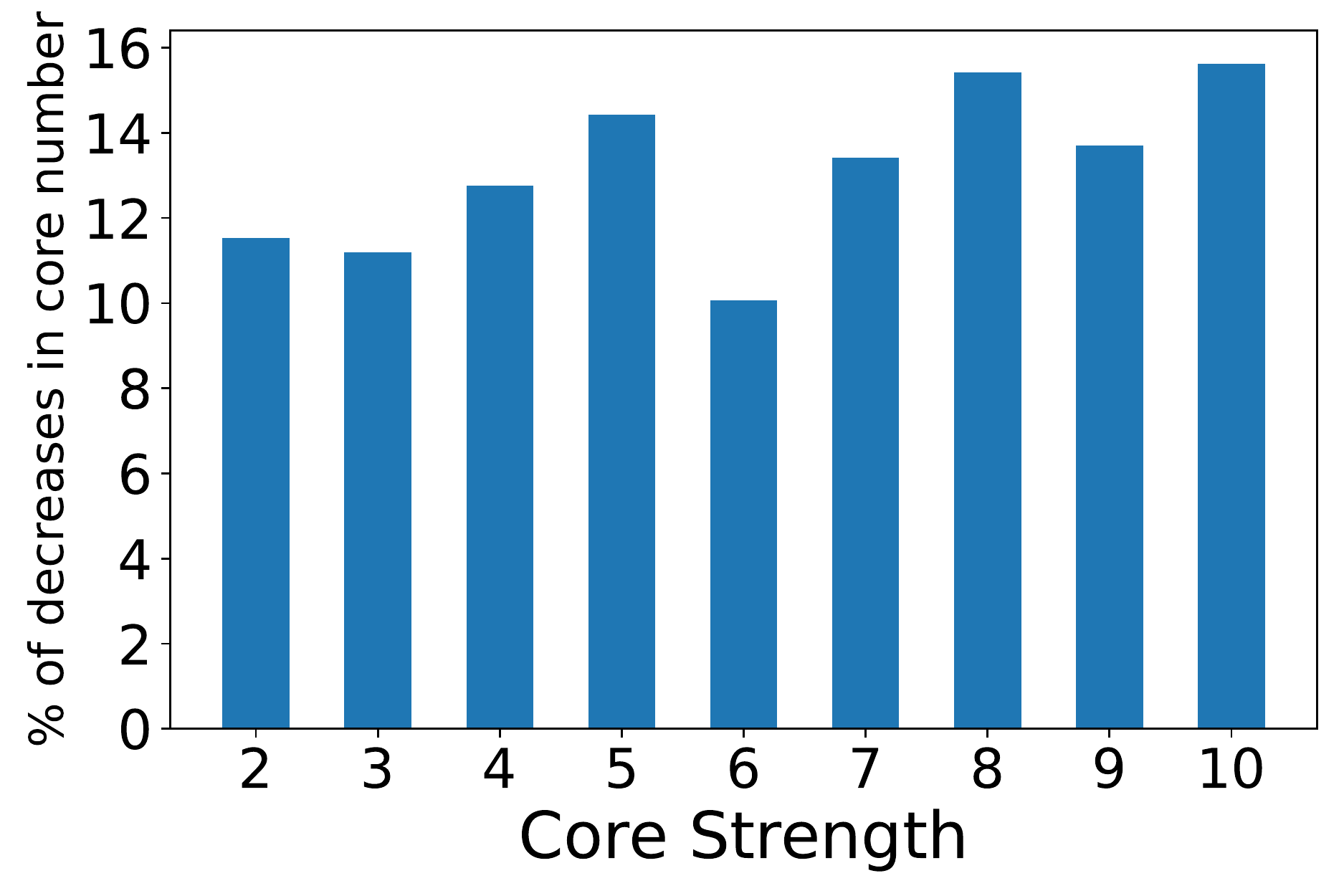}
    \caption{inf-openflights}}}
    \end{subfigure}
    \hspace{-2mm}
    \begin{subfigure}[t]{.25\linewidth}
    {{ \includegraphics[width=\linewidth]{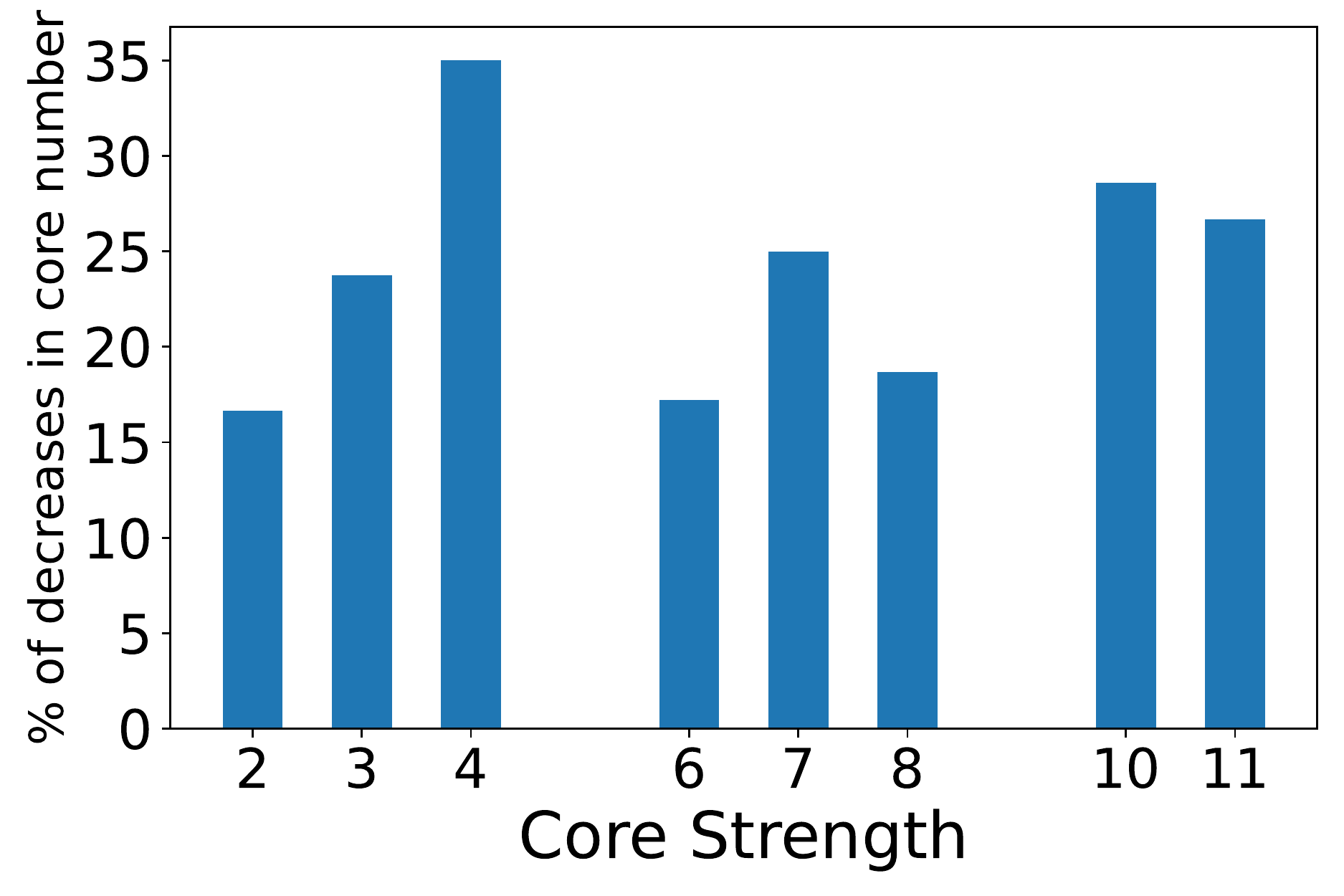}
    \caption{soc-Wiki-vote}}}
    \end{subfigure}
    \hspace{-2mm}
    \begin{subfigure}[t]{.25\linewidth}
    {{ \includegraphics[width=\linewidth]{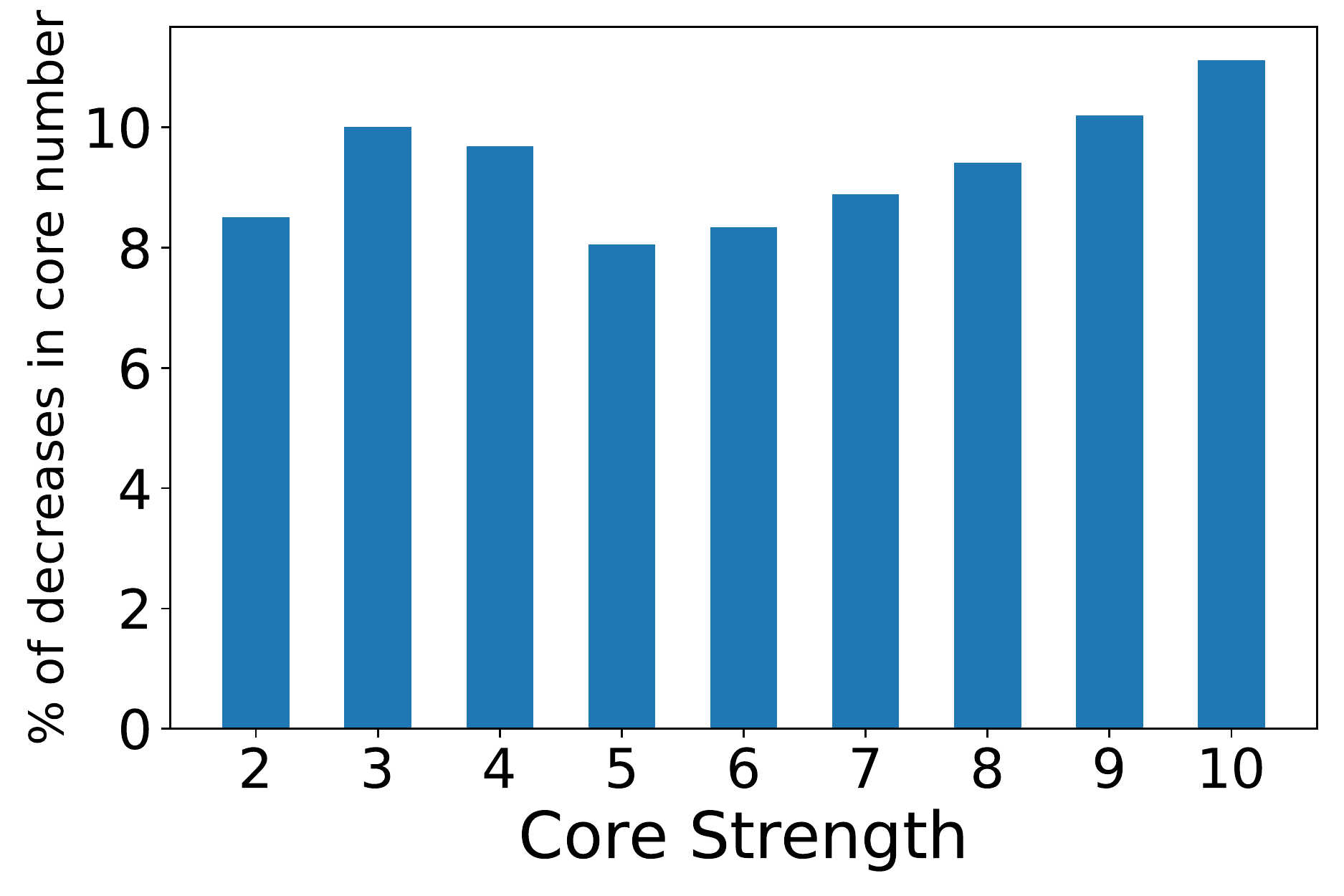}
    \caption{tech-pgp}}}
    \end{subfigure}
    \hspace{-2mm}
    \begin{subfigure}[t]{.25\linewidth}
    {{ \includegraphics[width=\linewidth]{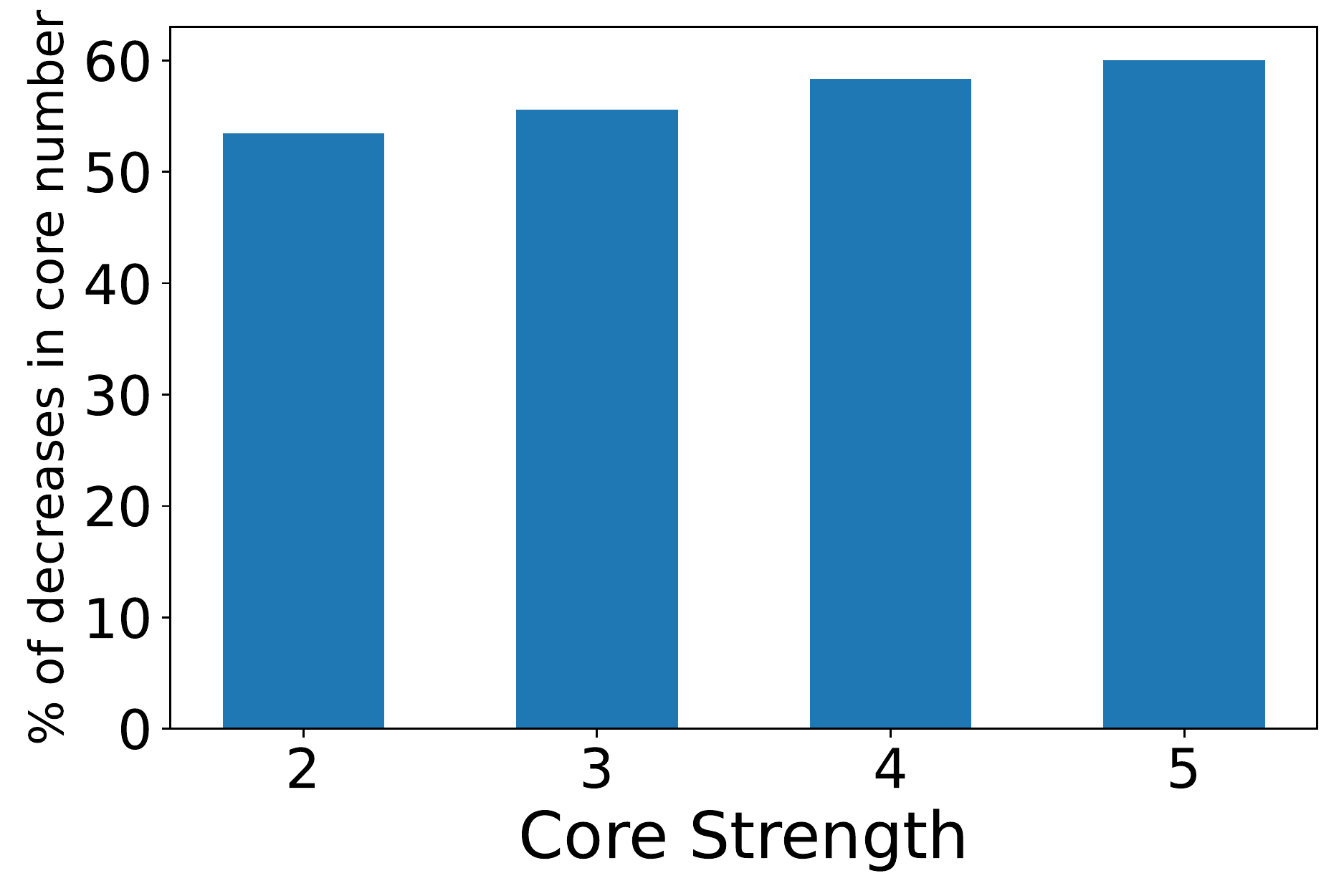} 
    \caption{as19990309}}}
    \end{subfigure}
    \hspace{-2mm}

    \caption{\bf {\footnotesize Statistics for the incorrect definition of Core Strength. The $\boldsymbol{y}$-axis is the percentage of time the core number decrease and the $\boldsymbol{x}$-axis is the Core Strength number.}}
    
    \label{fig:appendix_CS_wrong}

\end{figure}

\begin{figure}
    \centering
\begin{subfigure}[t]{.47\linewidth}
    \centering    
    \includegraphics[width=\linewidth]{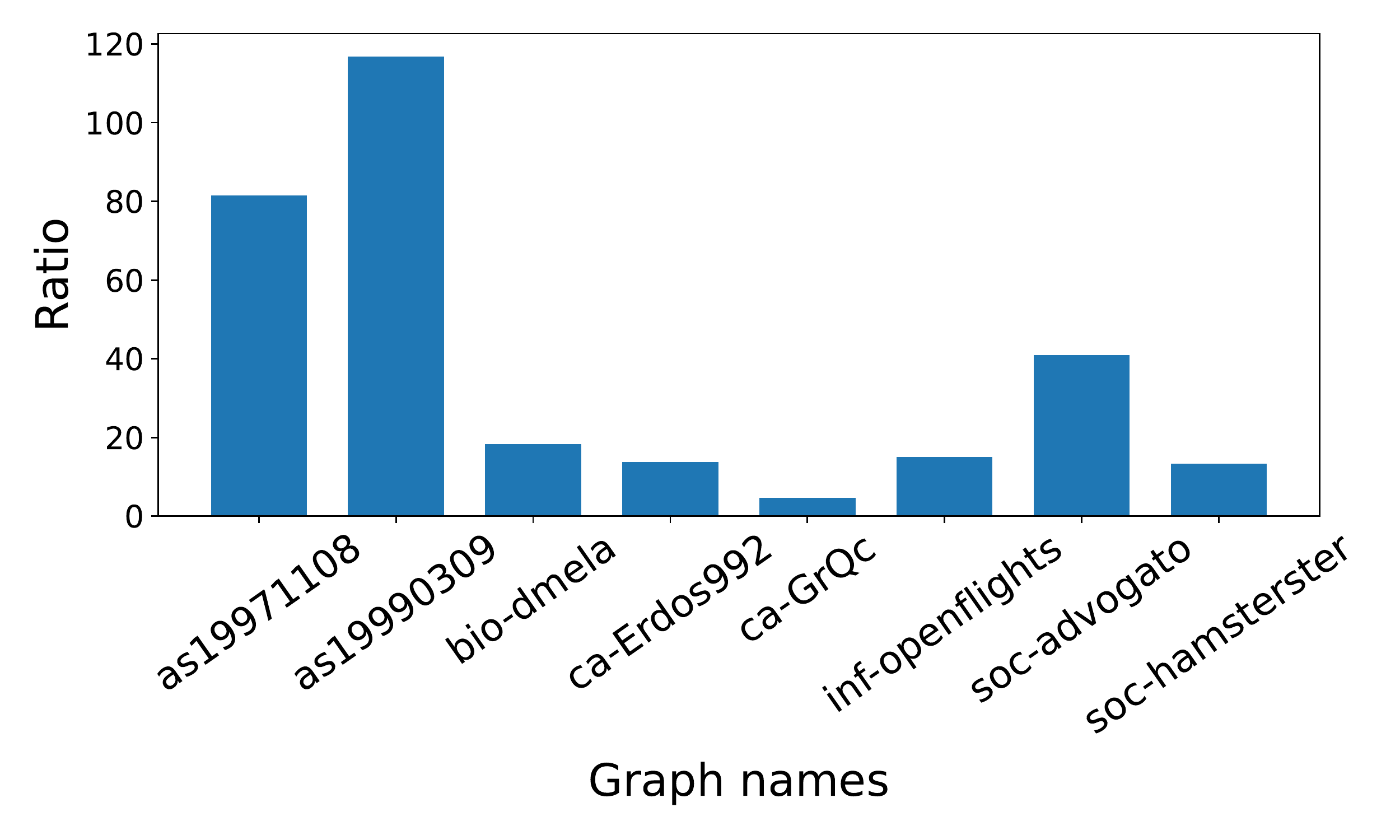}
    \caption{Ratio of the number of unconnected pairs that have at least one common neighbor to the number of edges}
    \label{fig:d2_neighbor_count}
\end{subfigure}
\hspace{2ex}
\begin{subfigure}[t]{.47\linewidth}
    \centering    
    \includegraphics[width=\linewidth]{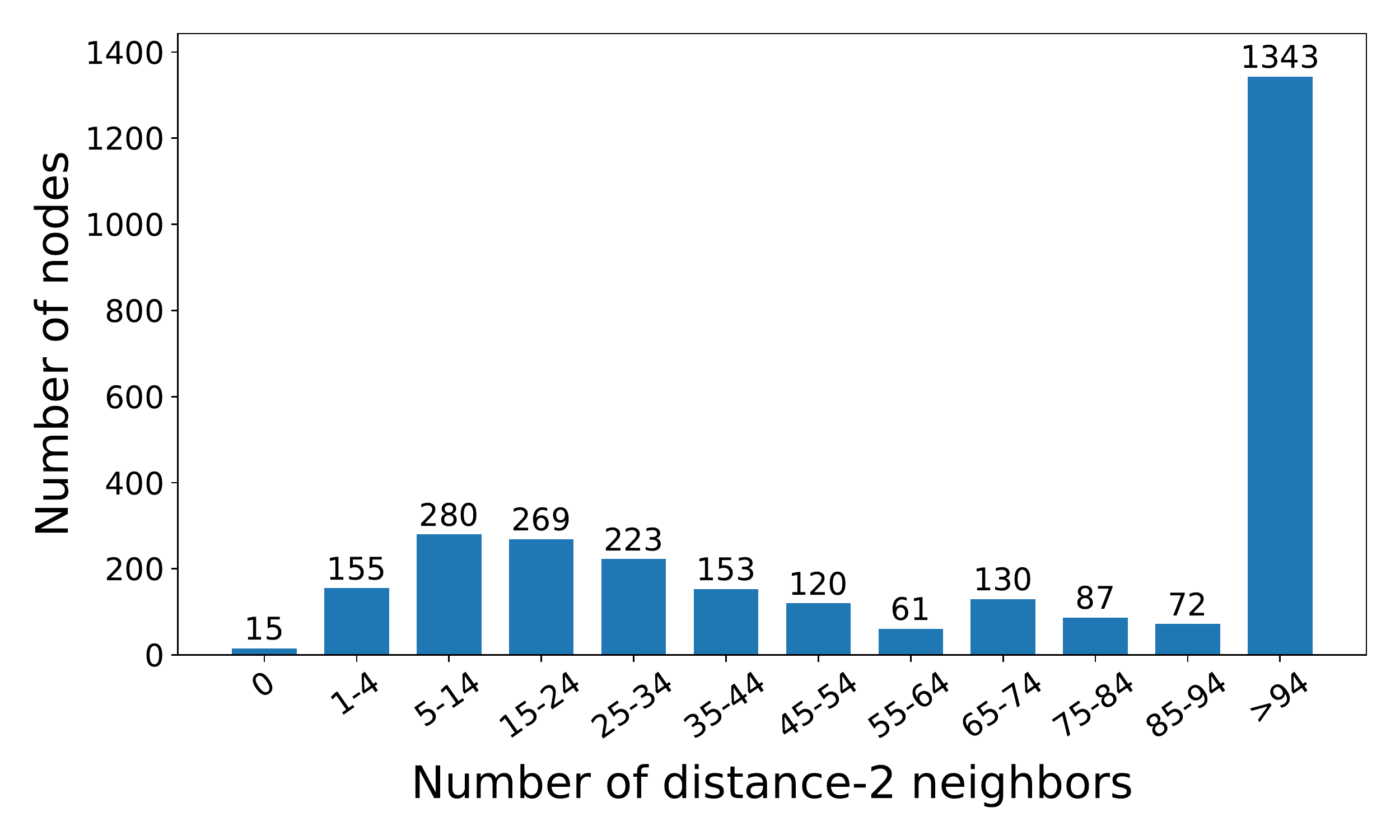}
    \caption{Histogram of distance-2 neighbor counts per nodes. $x$-axis denotes the number of distance-2 neighbors and $y$-axis represents the number of nodes having the distance-2 neighbors of $x$-axis values ({\tt inf-openflights}).}
    \label{fig:d2_dist}
\end{subfigure}
    \caption{\bf {\footnotesize Statistics for unconnected node pairs with at least one common neighbor.}}
    \label{fig:d2-hist}
\end{figure}

\begin{figure}
    \centering
    \includegraphics[width=10cm, height = 6cm]{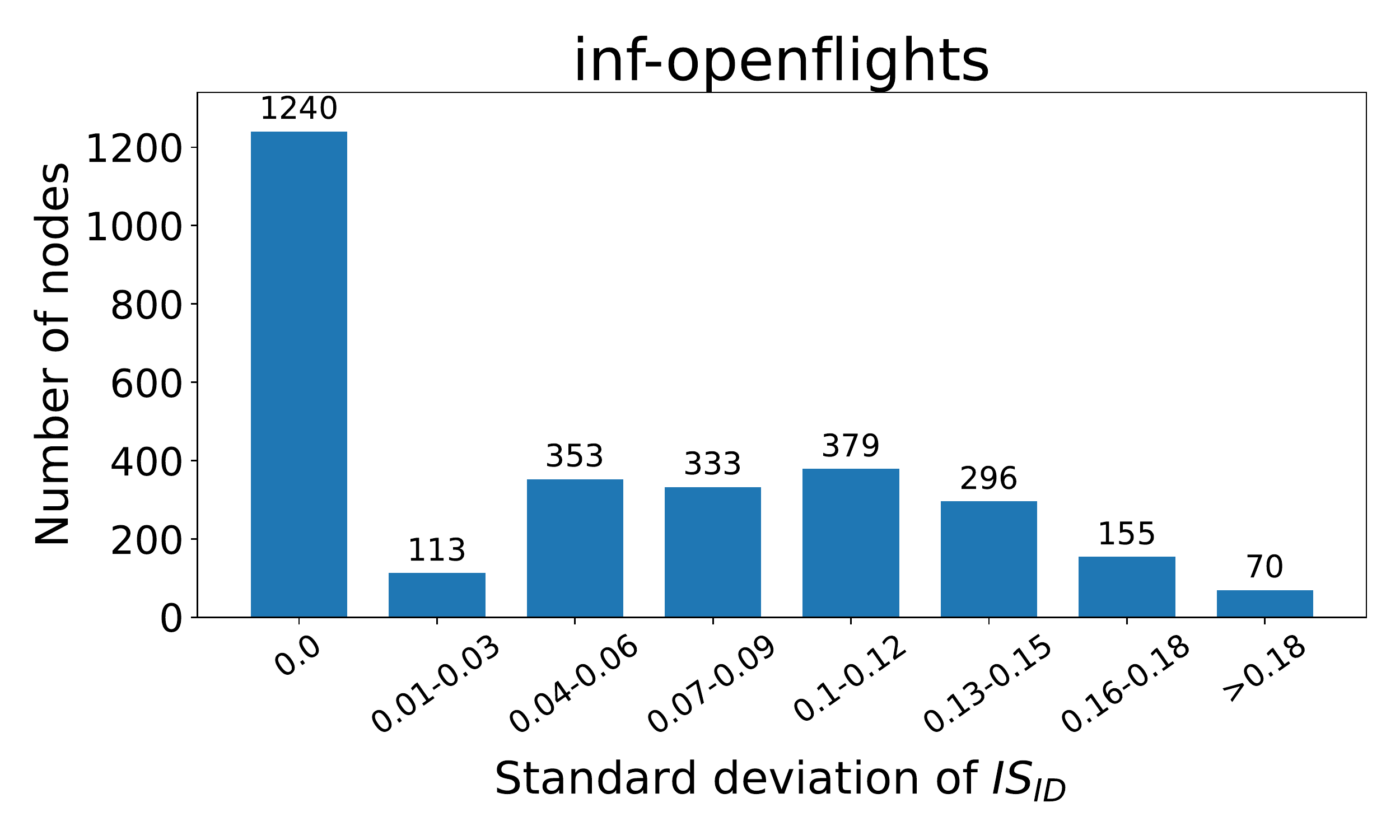}
    \caption{\bf {\footnotesize Histogram of standard deviation for In-Degree Insertion Strength ($IS_{ID}$). The $X$-axis represents the standard deviation range and $Y$-axis represents the number of nodes whose standard deviation falls in that range.}}
    \label{fig:stdev_count}
\end{figure}

\begin{figure}[h]
    \centering
    \begin{subfigure}[t]{.25\linewidth}
    {{ \includegraphics[width=\linewidth]{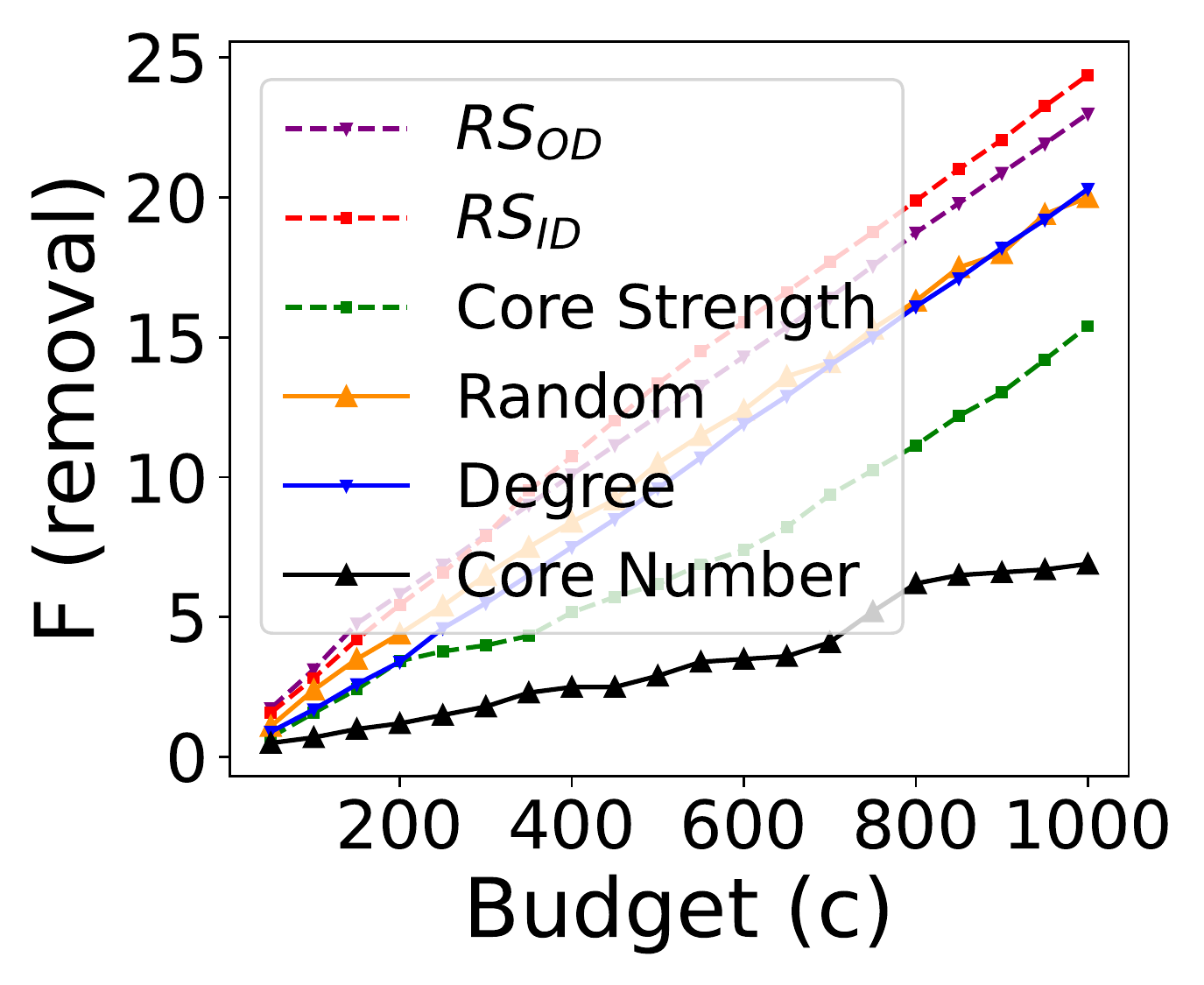} 
    \caption{as19990309}}}
    \end{subfigure}
    \hspace{-2mm}
    \begin{subfigure}[t]{.25\linewidth}
    {{ \includegraphics[width=\linewidth]{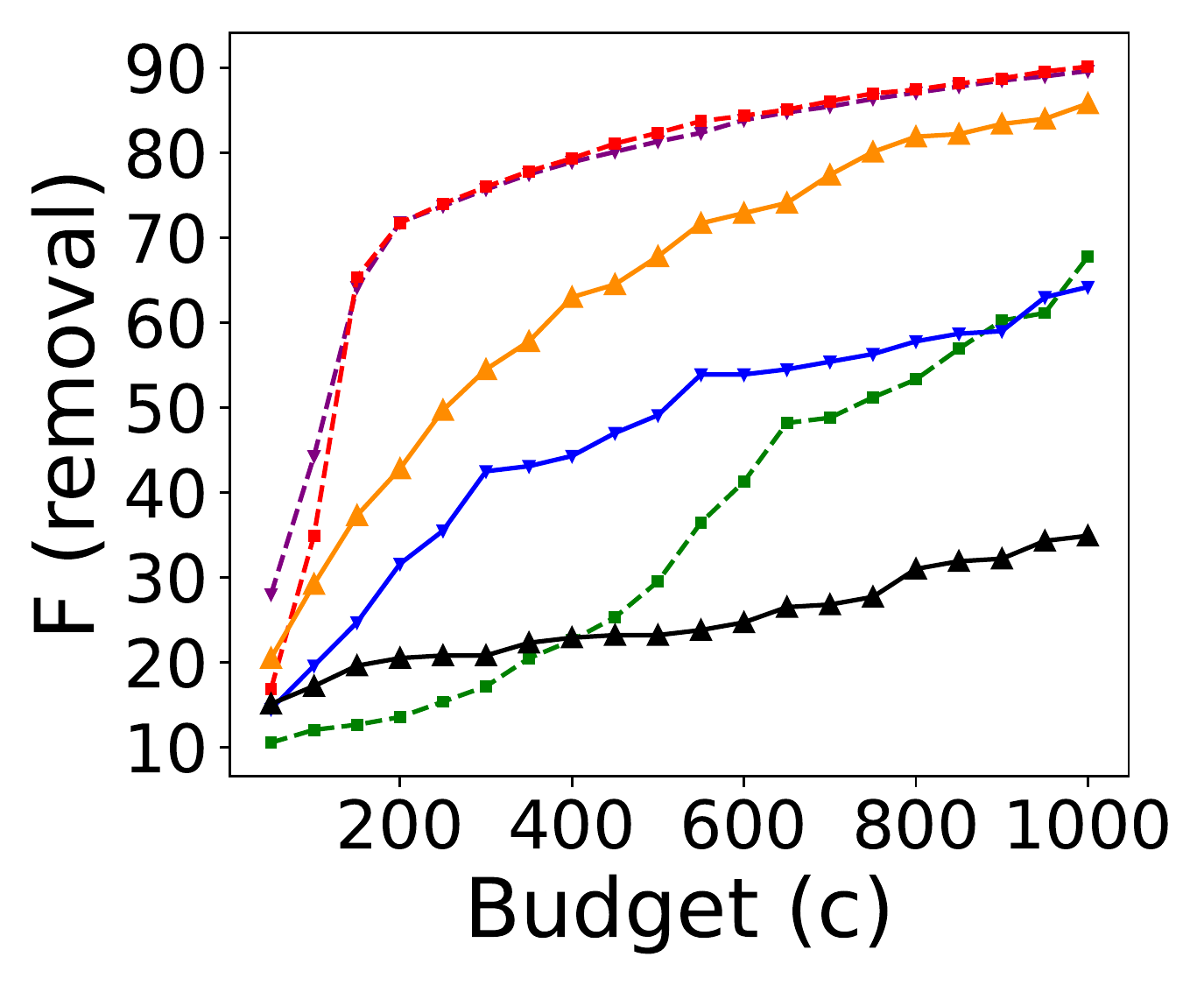}
    \caption{USAir97}}}
    \end{subfigure}
    \hspace{-2mm}
    \begin{subfigure}[t]{.25\linewidth}
    {{ \includegraphics[width=\linewidth]{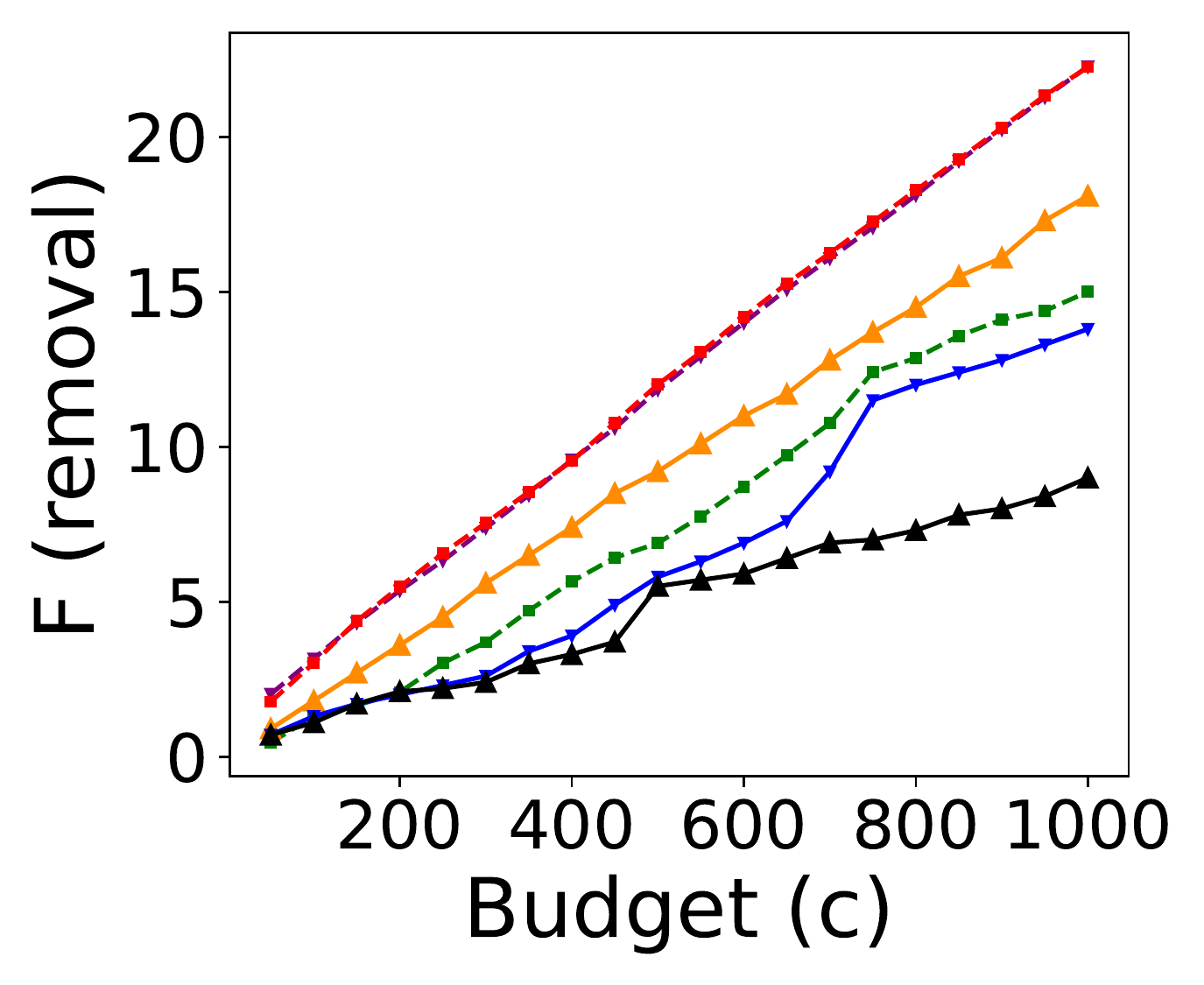} 
    \caption{ca-Erdos992}}}
    \end{subfigure}
    \hspace{-2mm}
    \begin{subfigure}[t]{.25\linewidth}
    {{ \includegraphics[width=\linewidth]{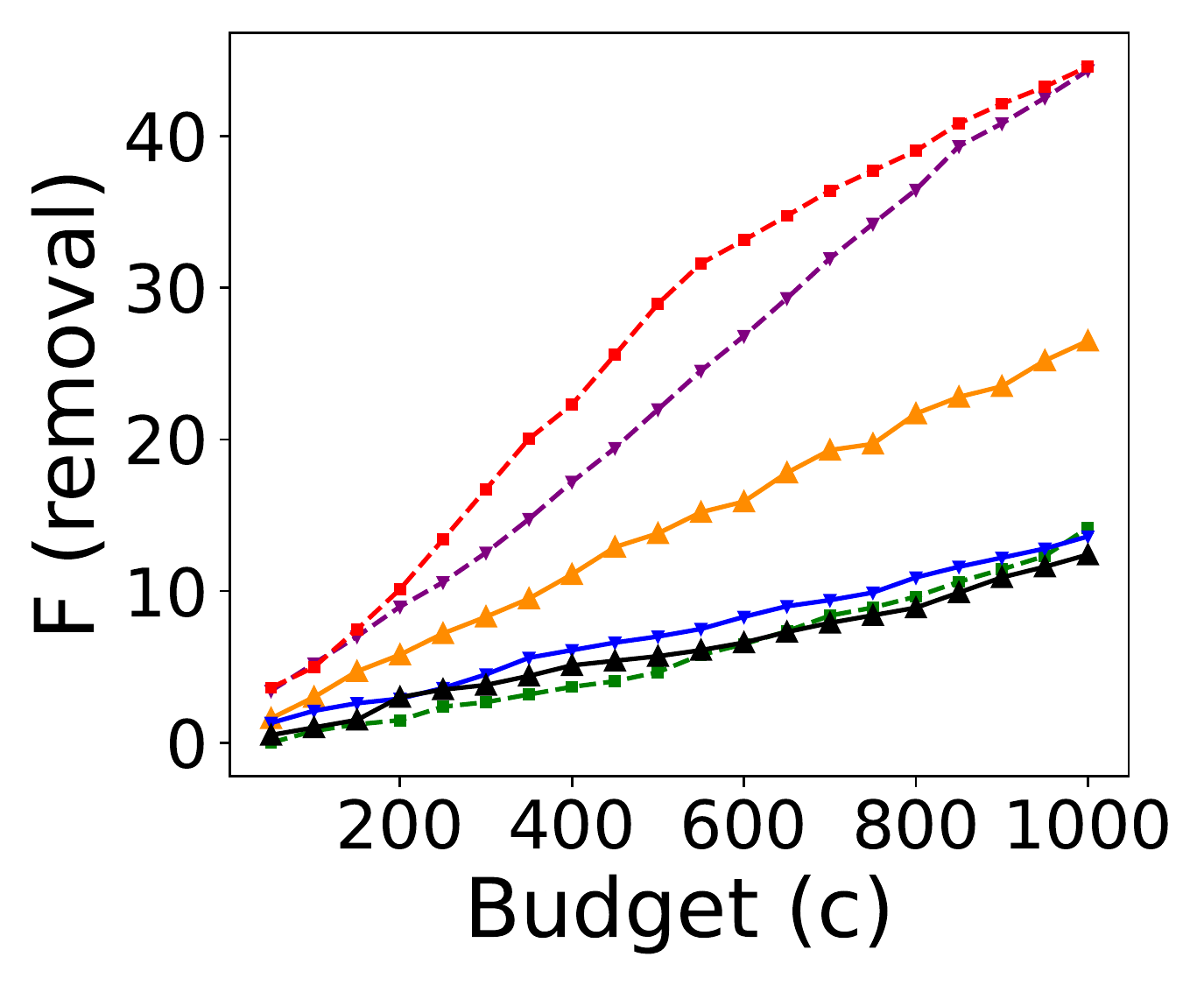} 
    \caption{inf-power}}}
    \end{subfigure}
    \hspace{-2mm}
    \begin{subfigure}[t]{.25\linewidth}
    {{ \includegraphics[width=\linewidth]{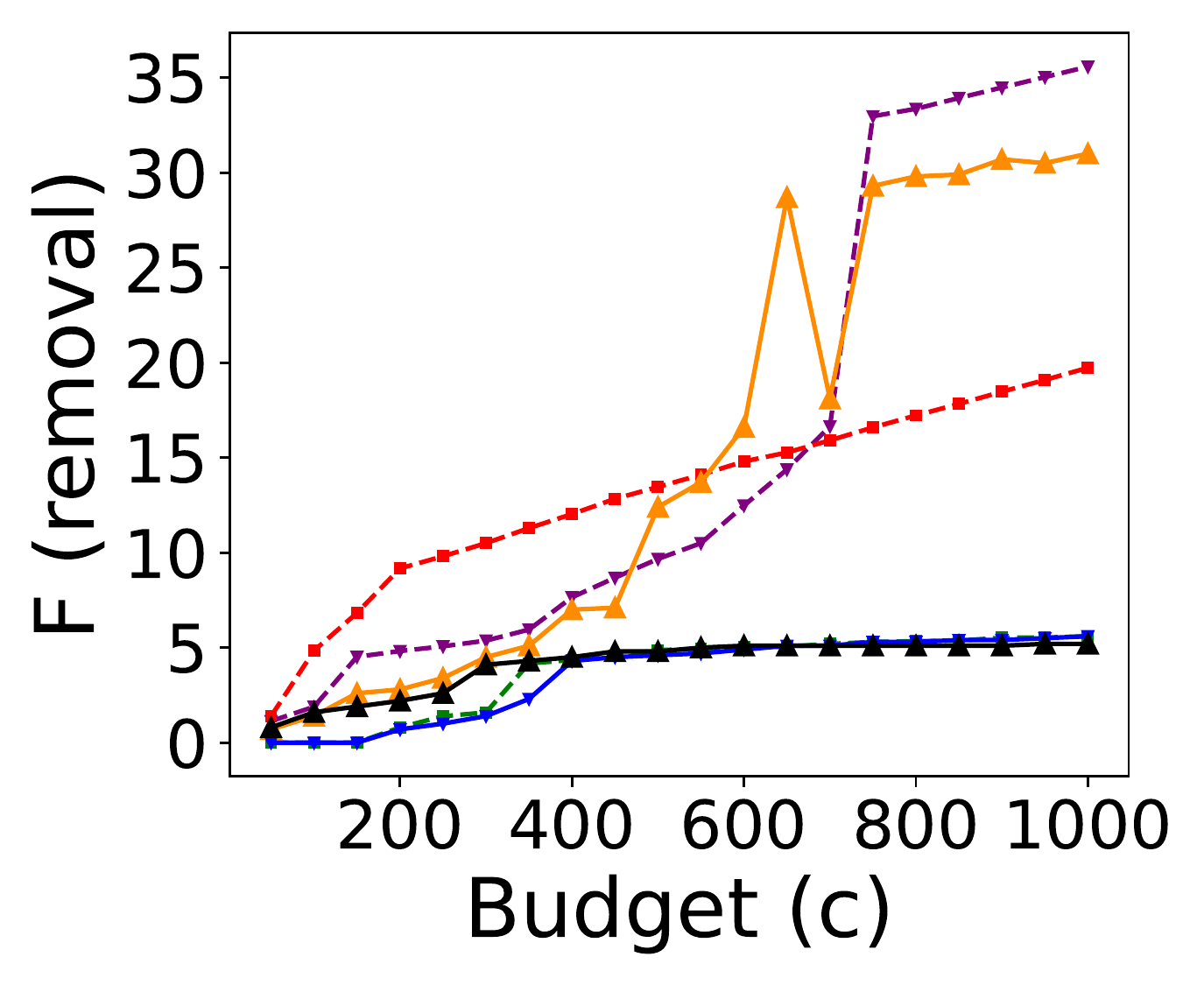} 
    \caption{p2p-Gnutella09}}}
    \end{subfigure}
    \hspace{-2mm}
    \begin{subfigure}[t]{.25\linewidth}
    {{ \includegraphics[width=\linewidth]{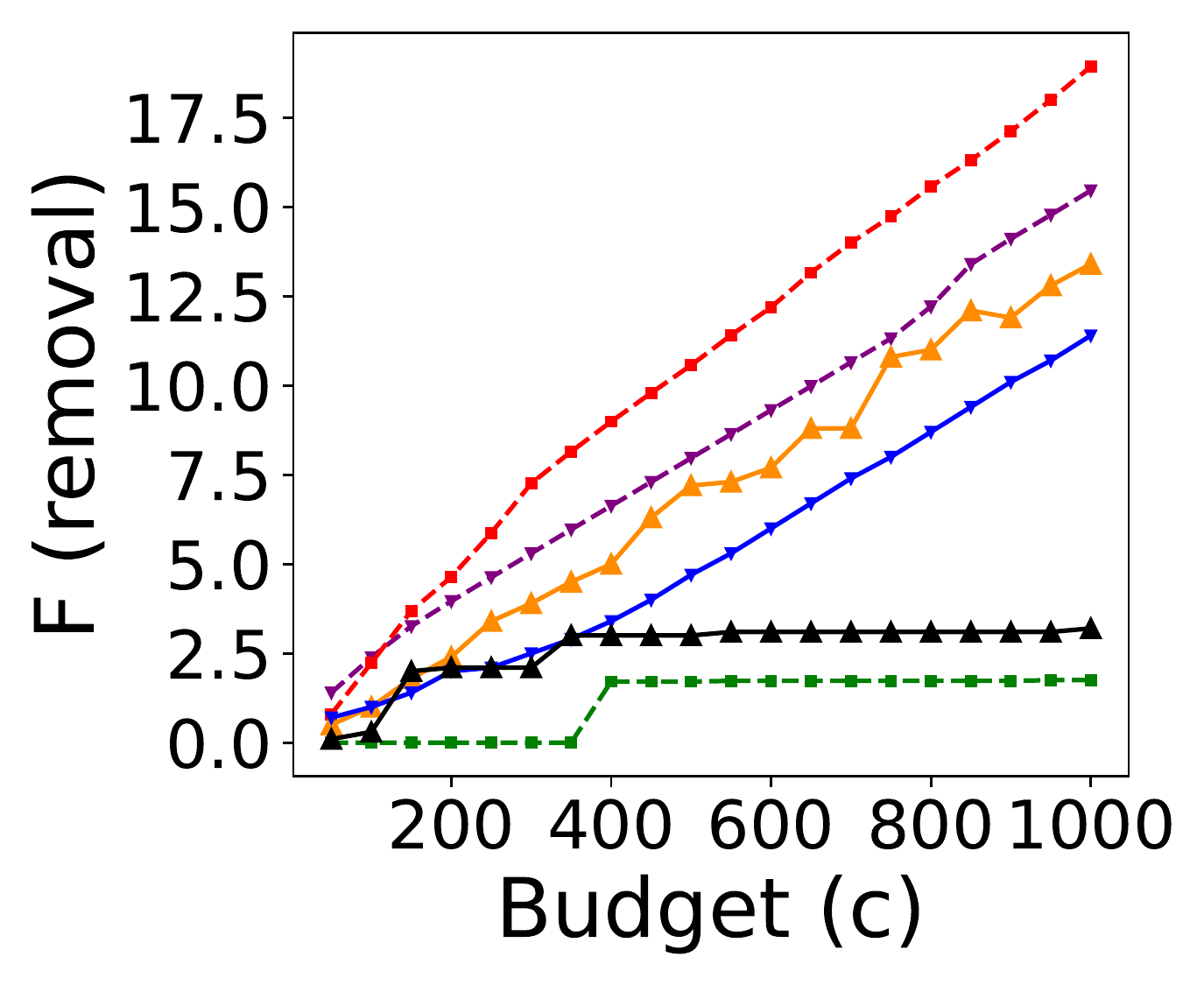} 
    \caption{tech-WHOIS}}}
    \end{subfigure}
    \hspace{-2mm}
    \begin{subfigure}[t]{.25\linewidth}
    {{ \includegraphics[width=\linewidth]{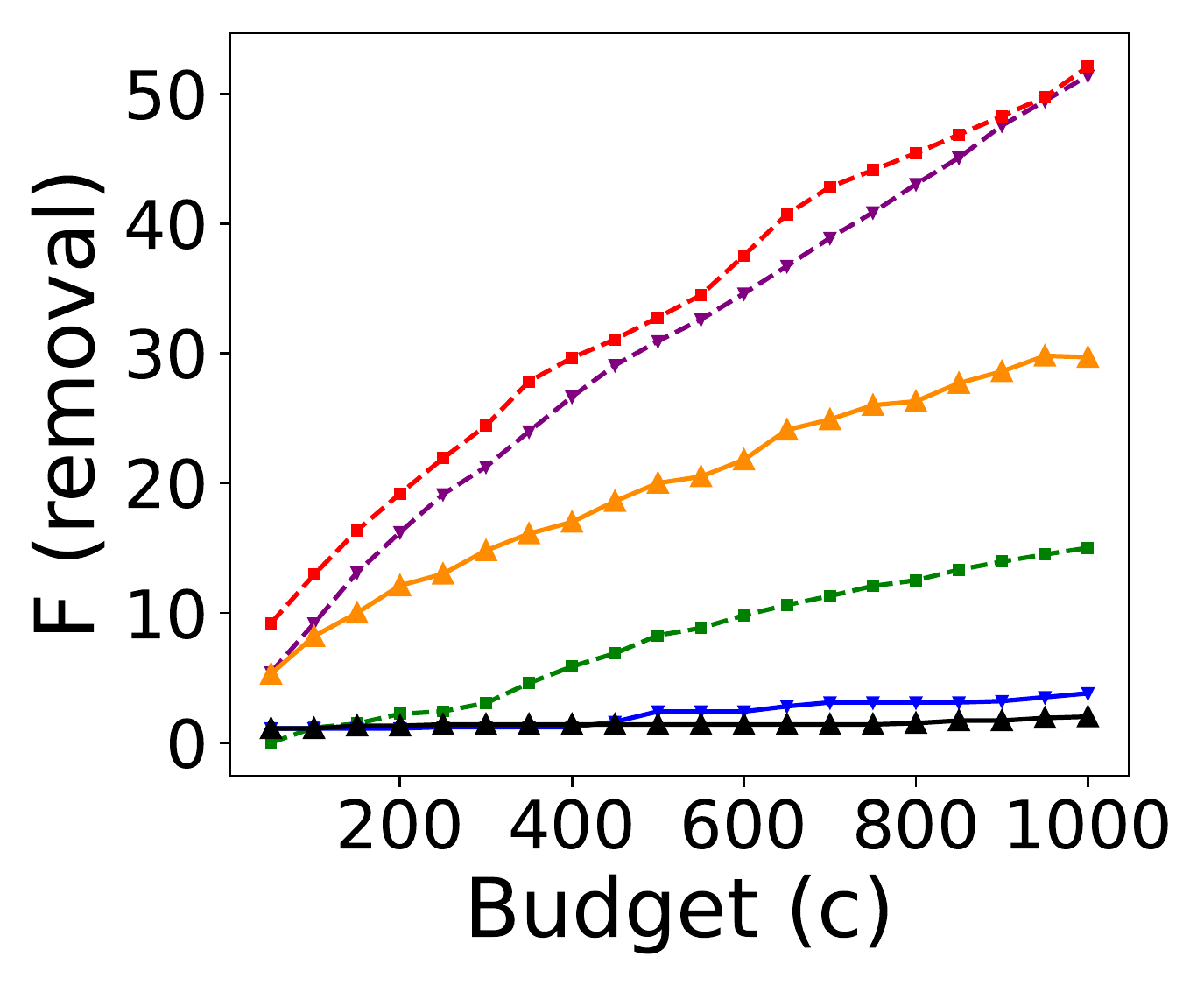}
    \caption{ca-GrQc}}}
    \end{subfigure}
    \hspace{-2mm}
    \begin{subfigure}[t]{.25\linewidth}
    {{ \includegraphics[width=\linewidth]{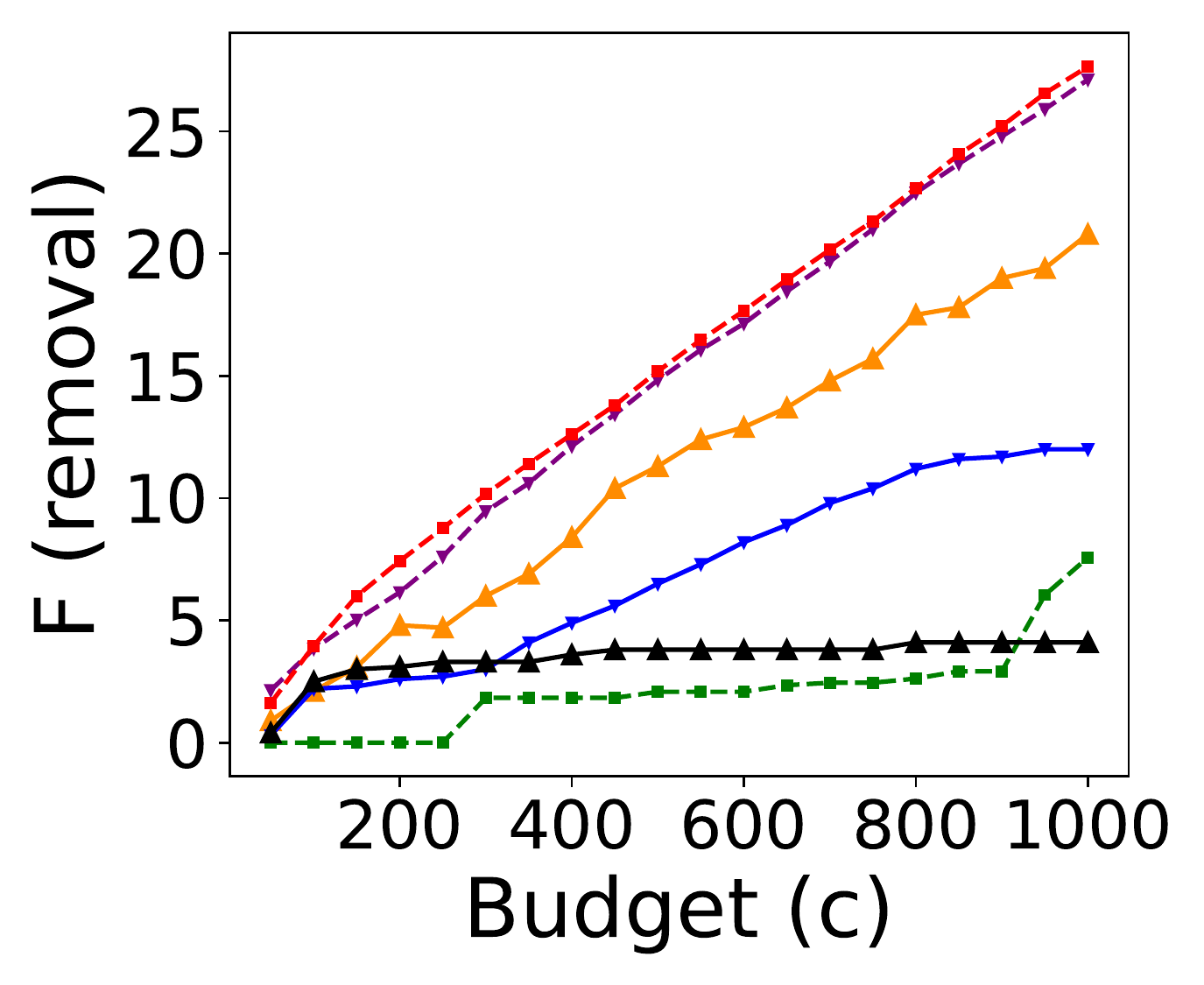}
    \caption{web-spam}}}
    \end{subfigure}
    \hspace{-2mm}

    \caption{\bf {\footnotesize Finding critical edges by our methods and baselines for edge removal.}}
    
    \label{fig:appendix_edge_deletion}
\end{figure}

\begin{figure}[!h]
    \centering
    \begin{subfigure}[t]{.25\linewidth}
    {{ \includegraphics[width=\linewidth]{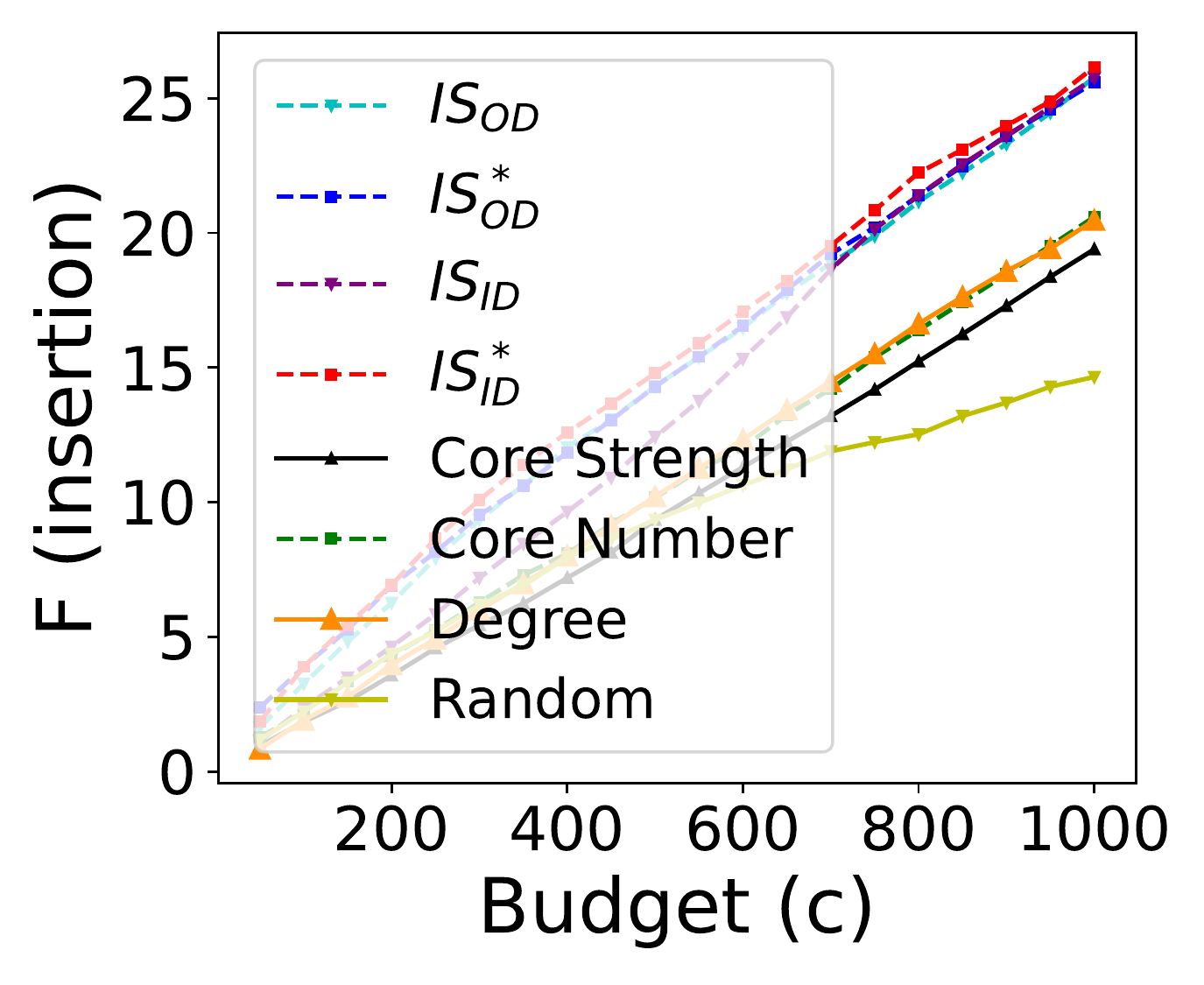} 
    \caption{as19990309}}}
    \end{subfigure}
    \hspace{-2mm}
    \begin{subfigure}[t]{.25\linewidth}
    {{ \includegraphics[width=\linewidth]{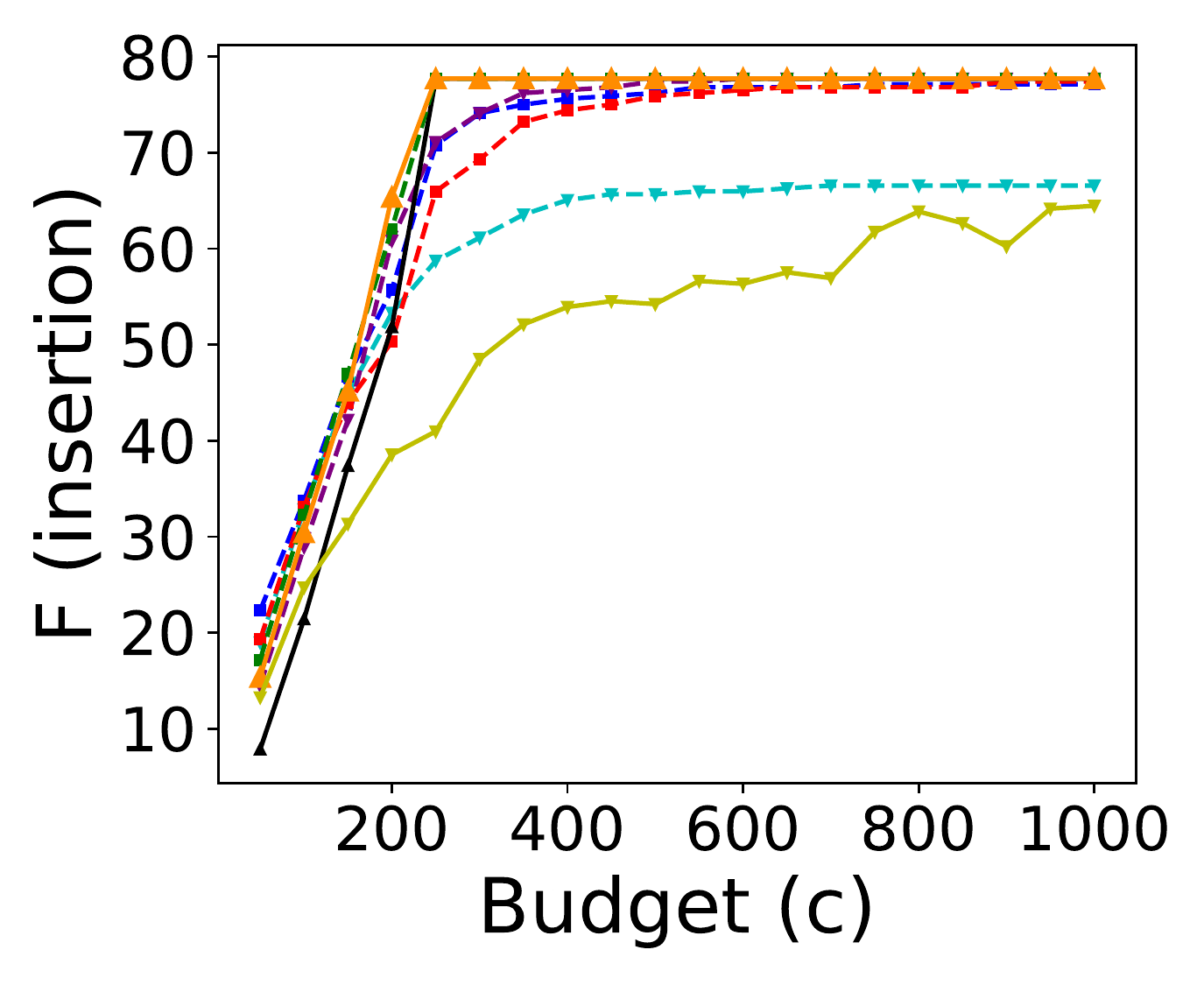}
    \caption{USAir97}}}
    \end{subfigure}
    \hspace{-2mm}
    \begin{subfigure}[t]{.25\linewidth}
    {{ \includegraphics[width=\linewidth]{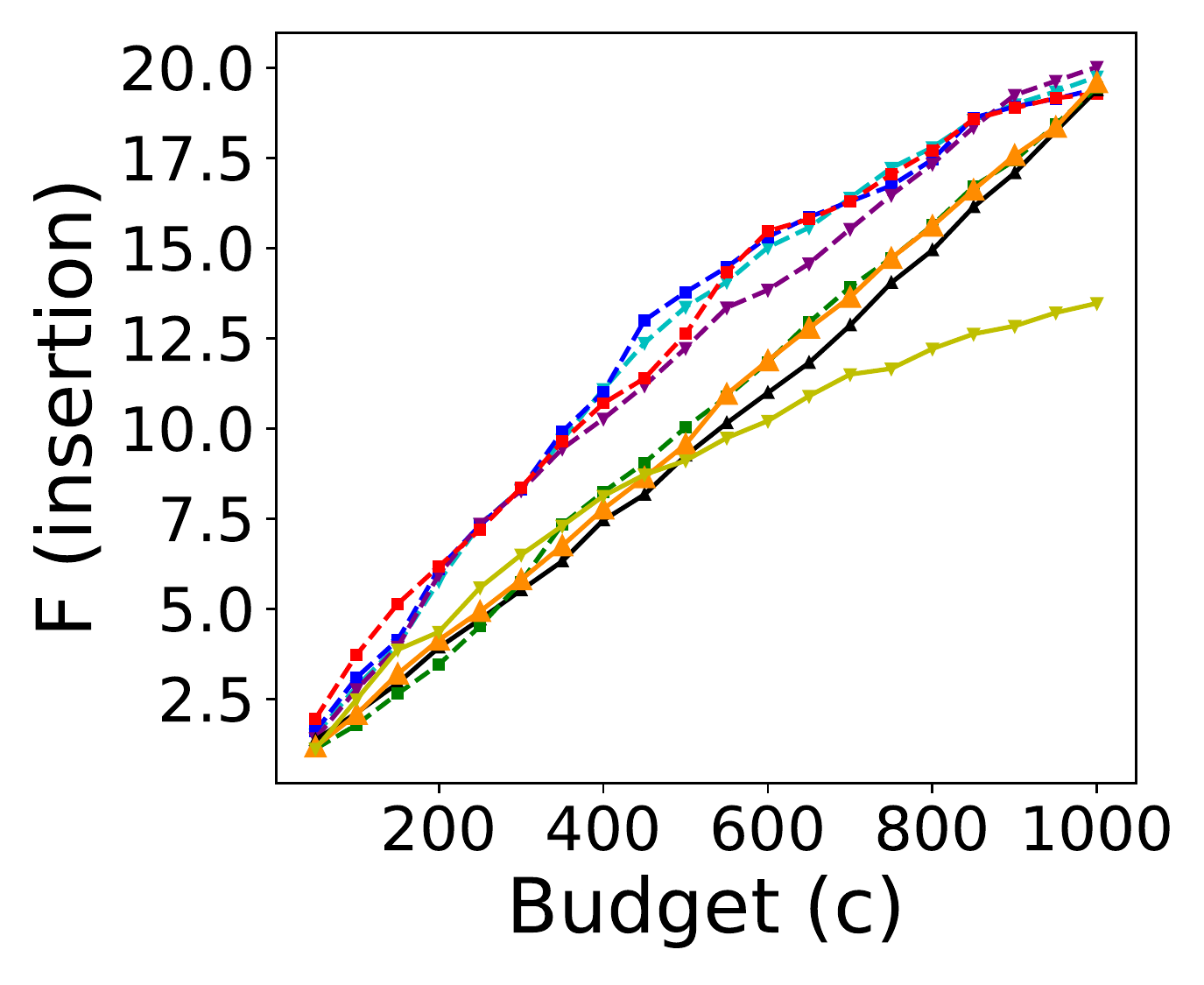} 
    \caption{ca-Erdos992}}}
    \end{subfigure}
    \hspace{-2mm}
    \begin{subfigure}[t]{.25\linewidth}
    {{ \includegraphics[width=\linewidth]{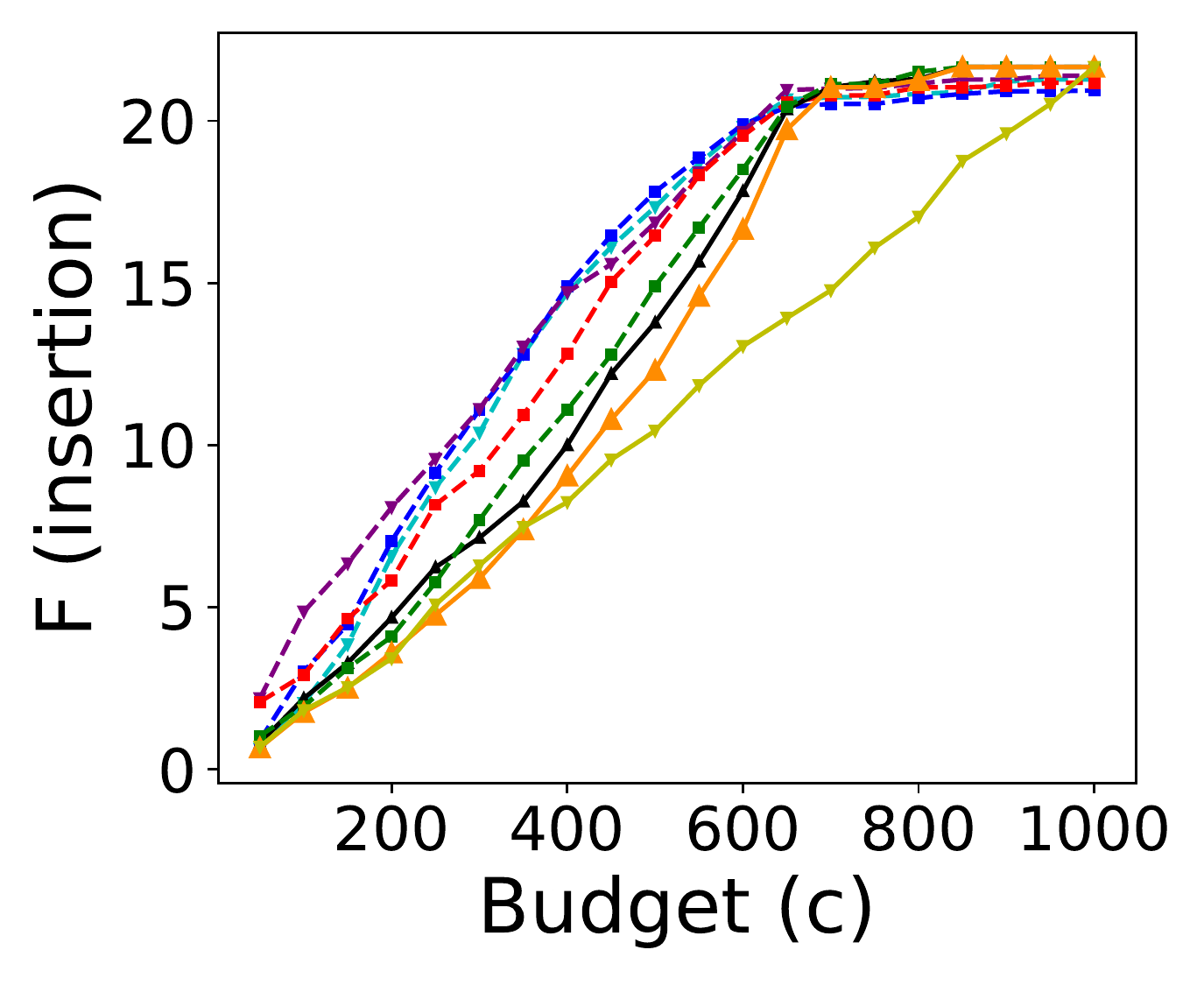} 
    \caption{inf-power}}}
    \end{subfigure}
    \hspace{-2mm}
    \begin{subfigure}[t]{.25\linewidth}
    {{ \includegraphics[width=\linewidth]{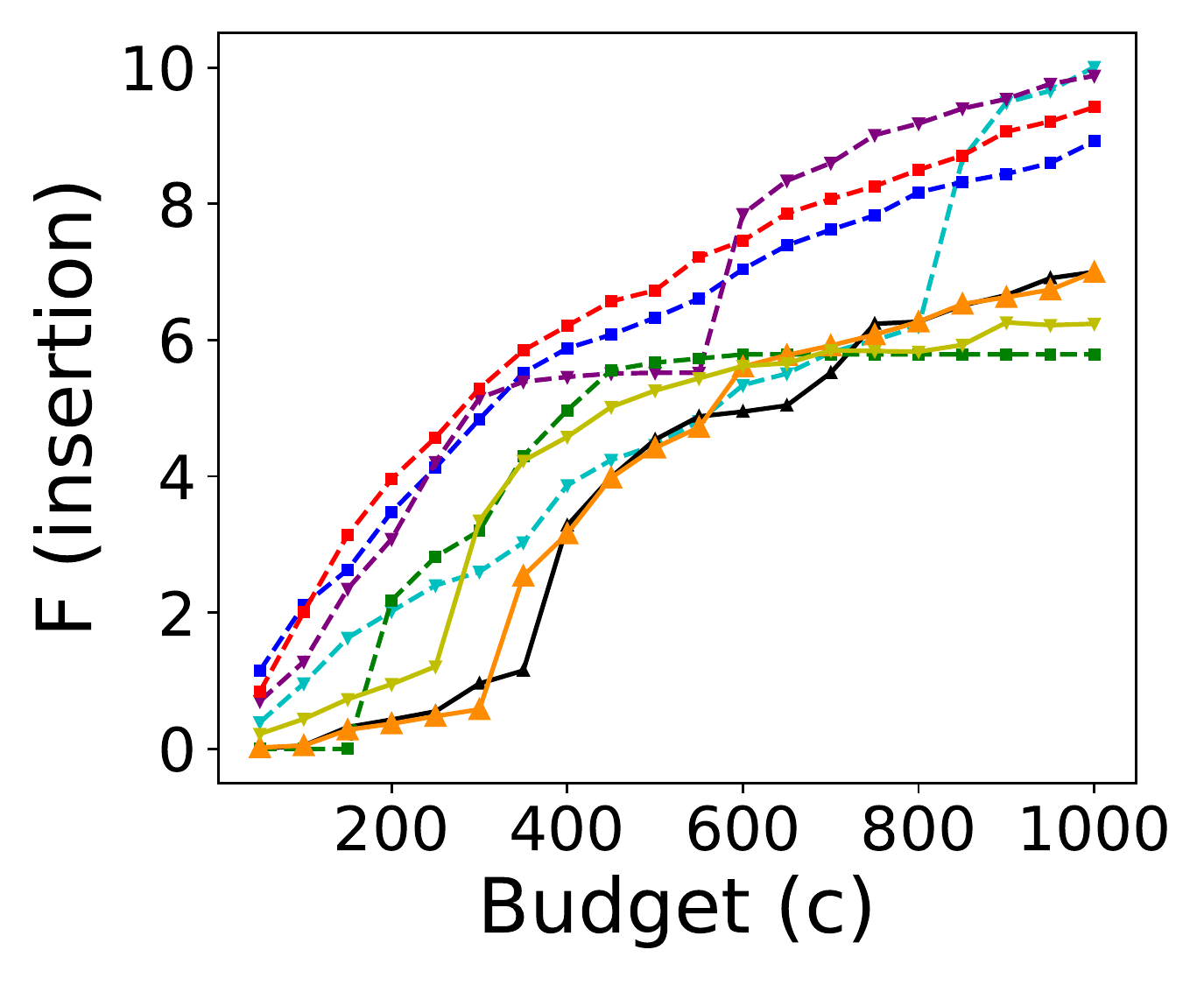} 
    \caption{p2p-Gnutella09}}}
    \end{subfigure}
    \hspace{-2mm}
    \begin{subfigure}[t]{.25\linewidth}
    {{ \includegraphics[width=\linewidth]{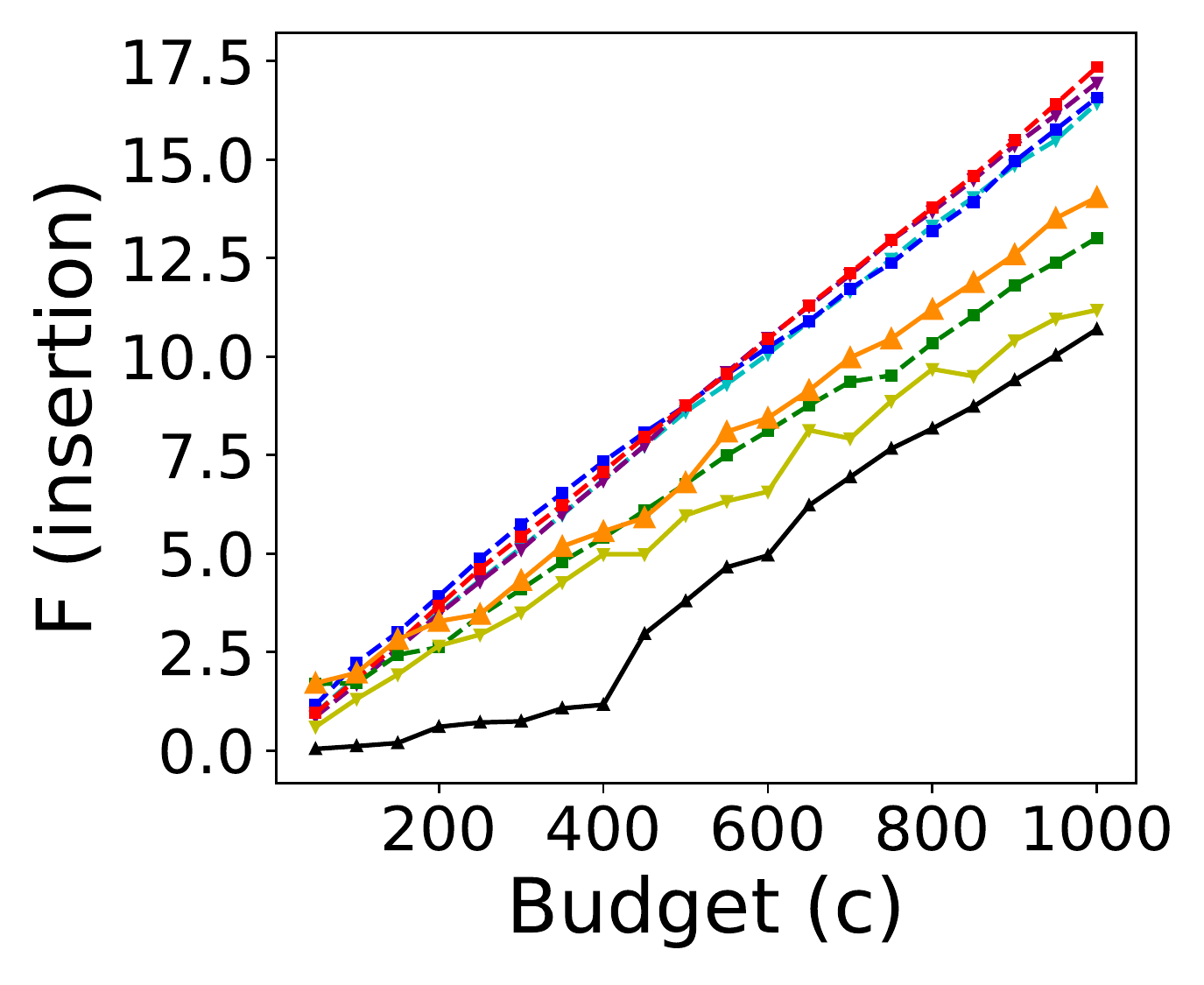} 
    \caption{tech-WHOIS}}}
    \end{subfigure}
    \hspace{-2mm}
    \begin{subfigure}[t]{.25\linewidth}
    {{ \includegraphics[width=\linewidth]{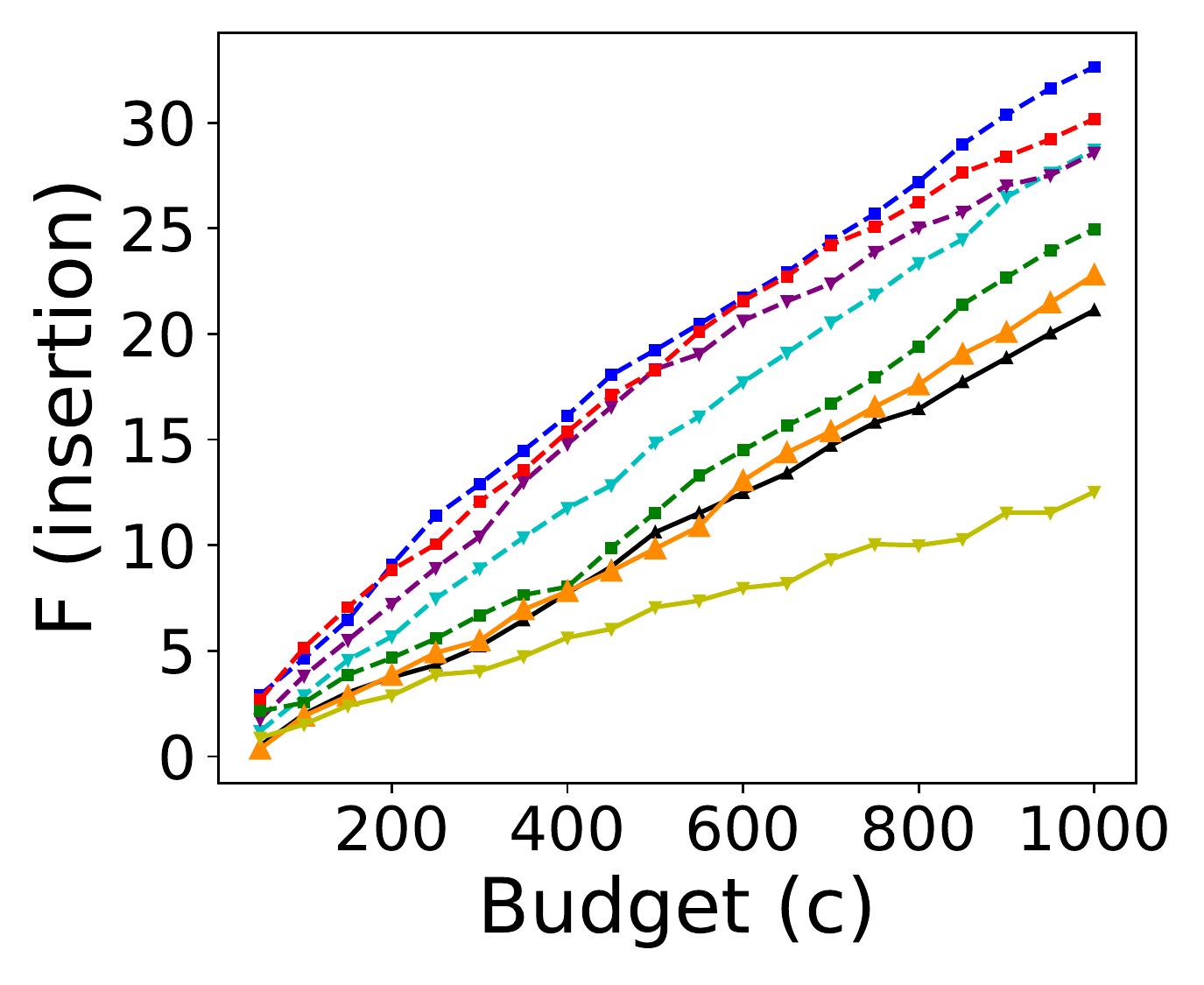}
    \caption{ca-GrQc}}}
    \end{subfigure}
    \hspace{-2mm}
    \begin{subfigure}[t]{.25\linewidth}
    {{ \includegraphics[width=\linewidth]{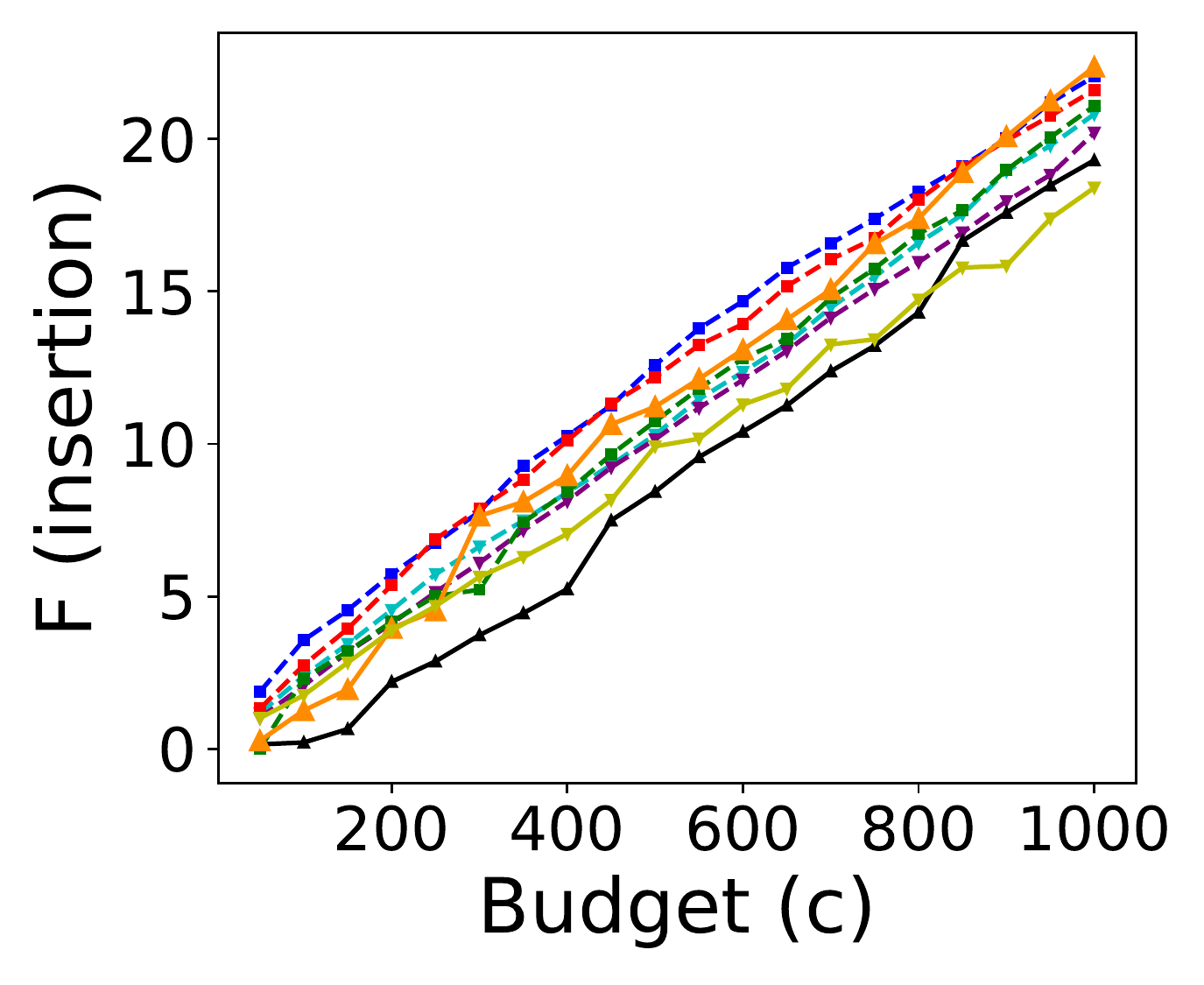}
    \caption{web-spam}}}
    \end{subfigure}
    \hspace{-2mm}
    
    \caption{\bf {\footnotesize Finding critical edges by our methods and baselines for edge insertion.}}
    
    \label{fig:appendix_edge_insertion}
\end{figure}

\begin{figure}[h]
    \centering
    \begin{subfigure}[t]{.25\linewidth}
    {{ \includegraphics[width=\linewidth]{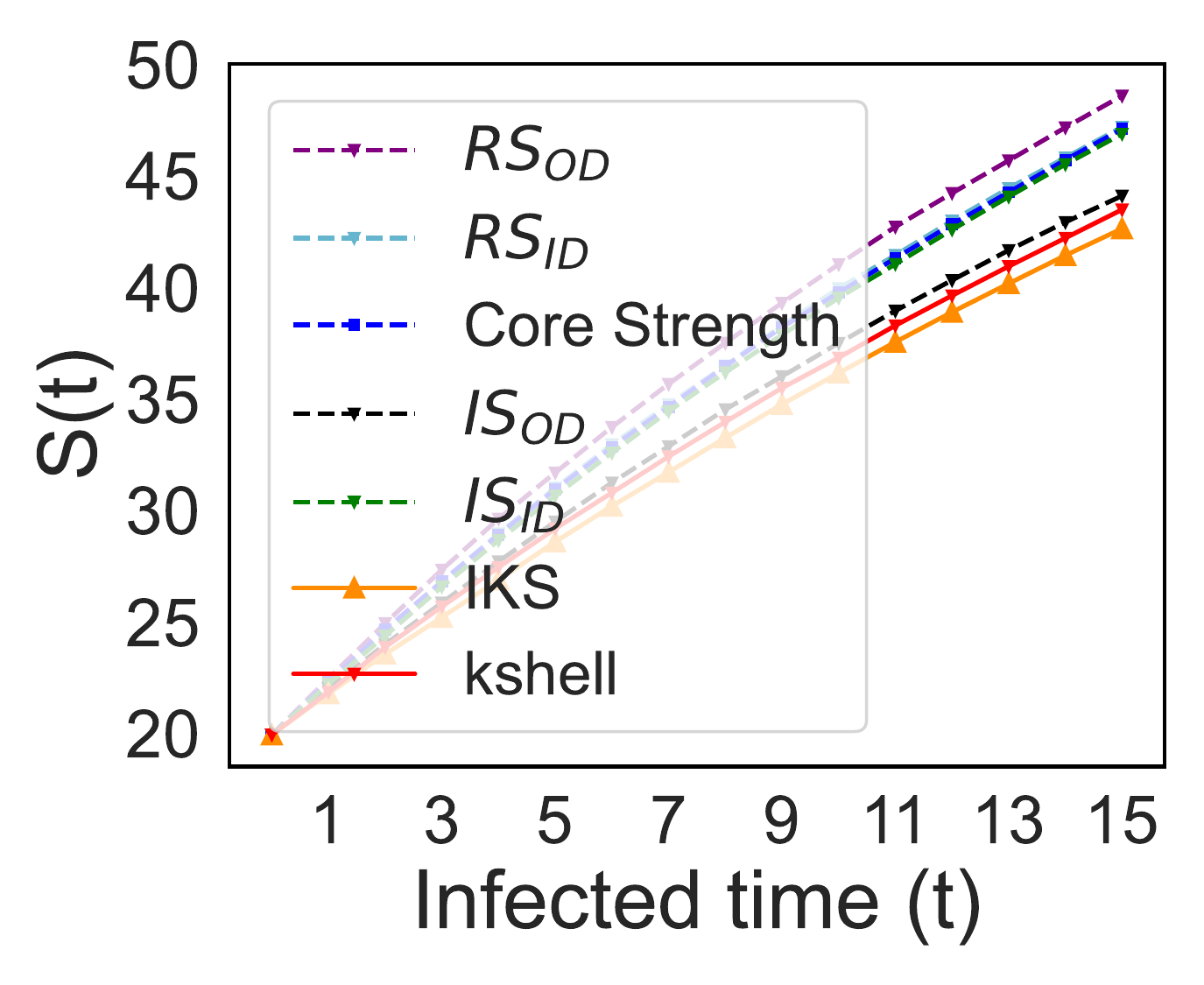} 
    \caption{as19990309}}}
    \end{subfigure}
    \hspace{-2mm}
    \begin{subfigure}[t]{.25\linewidth}
    {{ \includegraphics[width=\linewidth]{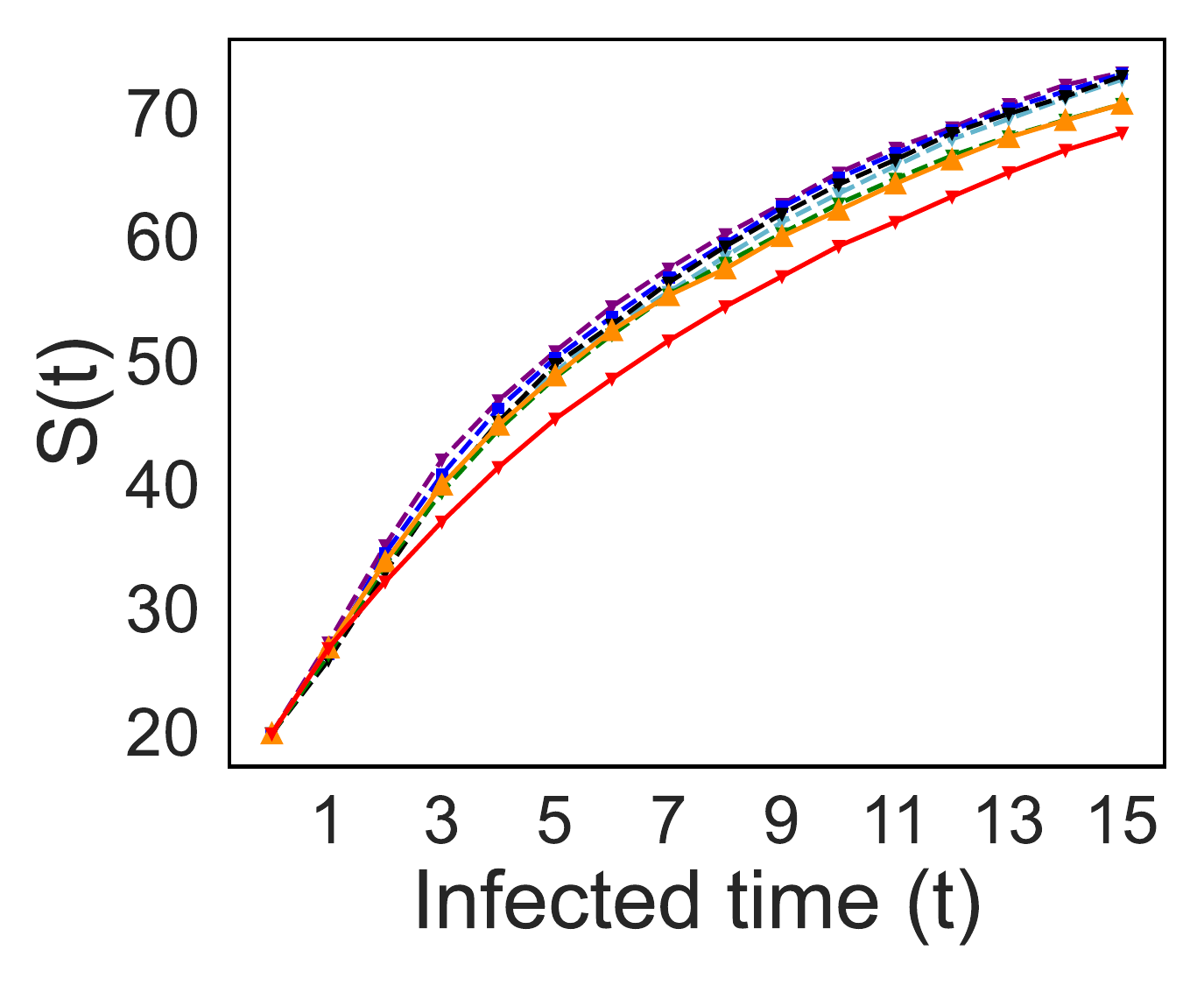}
    \caption{USAir97}}}
    \end{subfigure}
    \hspace{-2mm}
    \begin{subfigure}[t]{.25\linewidth}
    {{ \includegraphics[width=\linewidth]{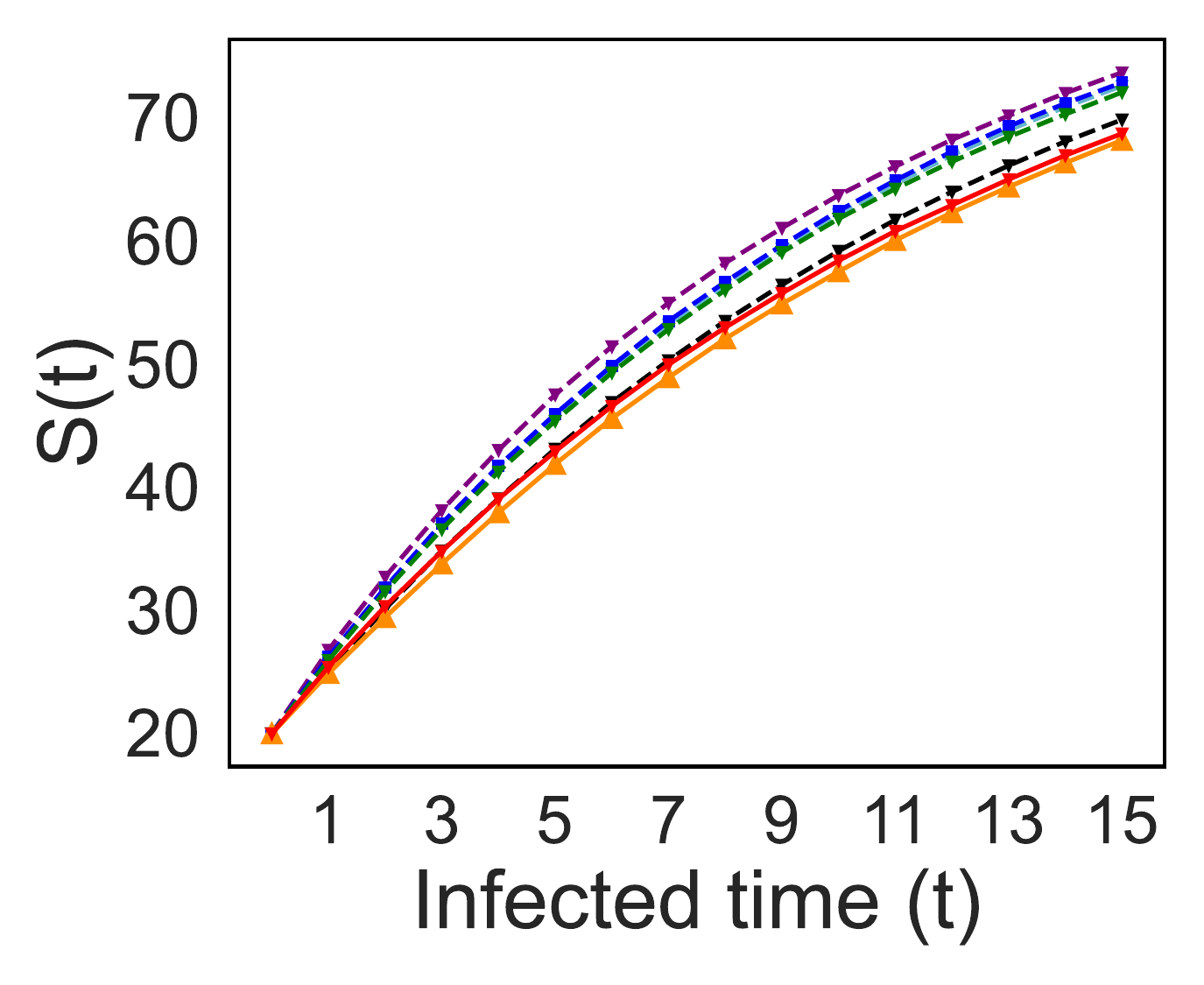} 
    \caption{ca-Erdos992}}}
    \end{subfigure}
    \hspace{-2mm}
    \begin{subfigure}[t]{.25\linewidth}
    {{ \includegraphics[width=\linewidth]{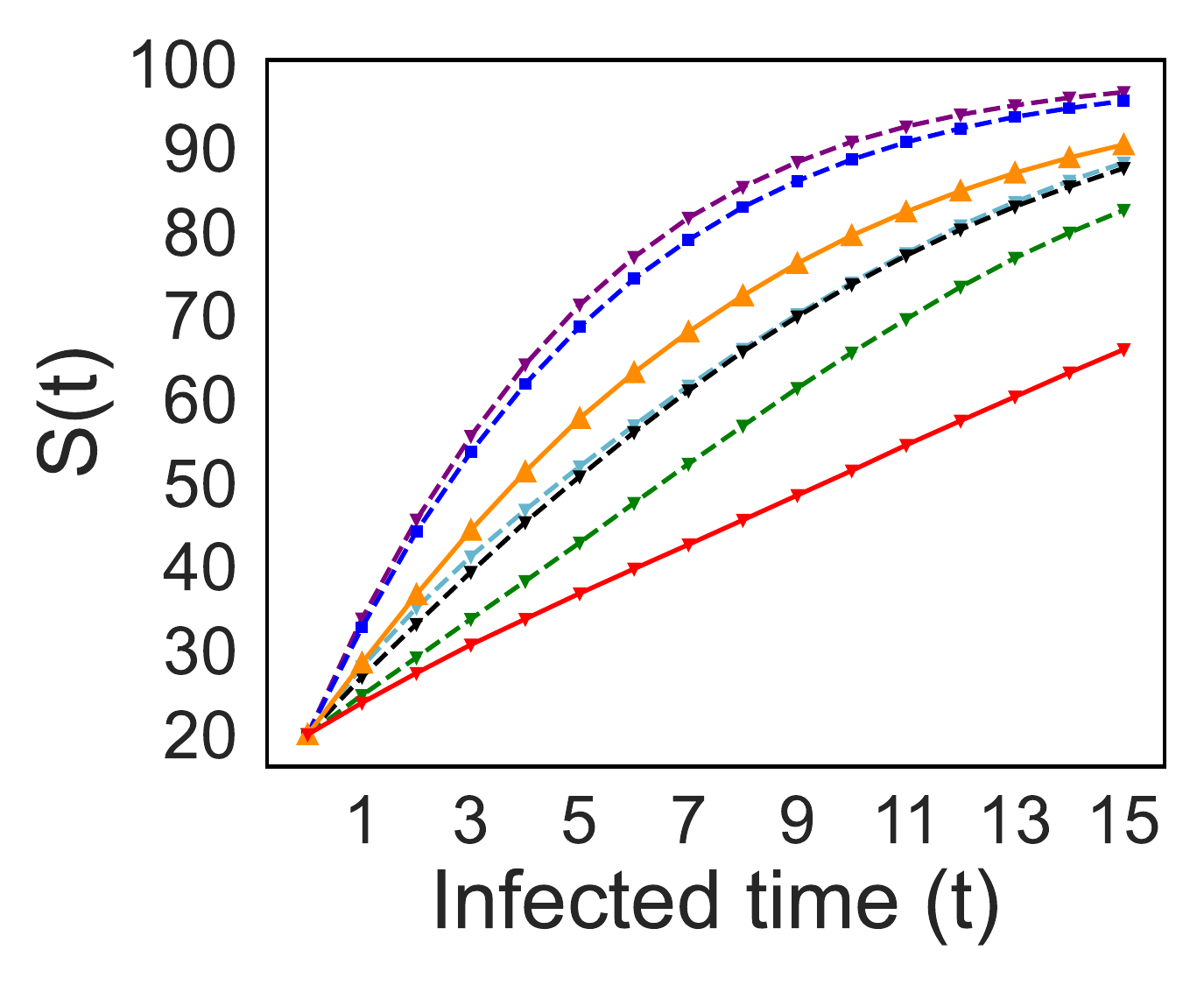} 
    \caption{inf-power}}}
    \end{subfigure}
    \hspace{-2mm}
    \begin{subfigure}[t]{.25\linewidth}
    {{ \includegraphics[width=\linewidth]{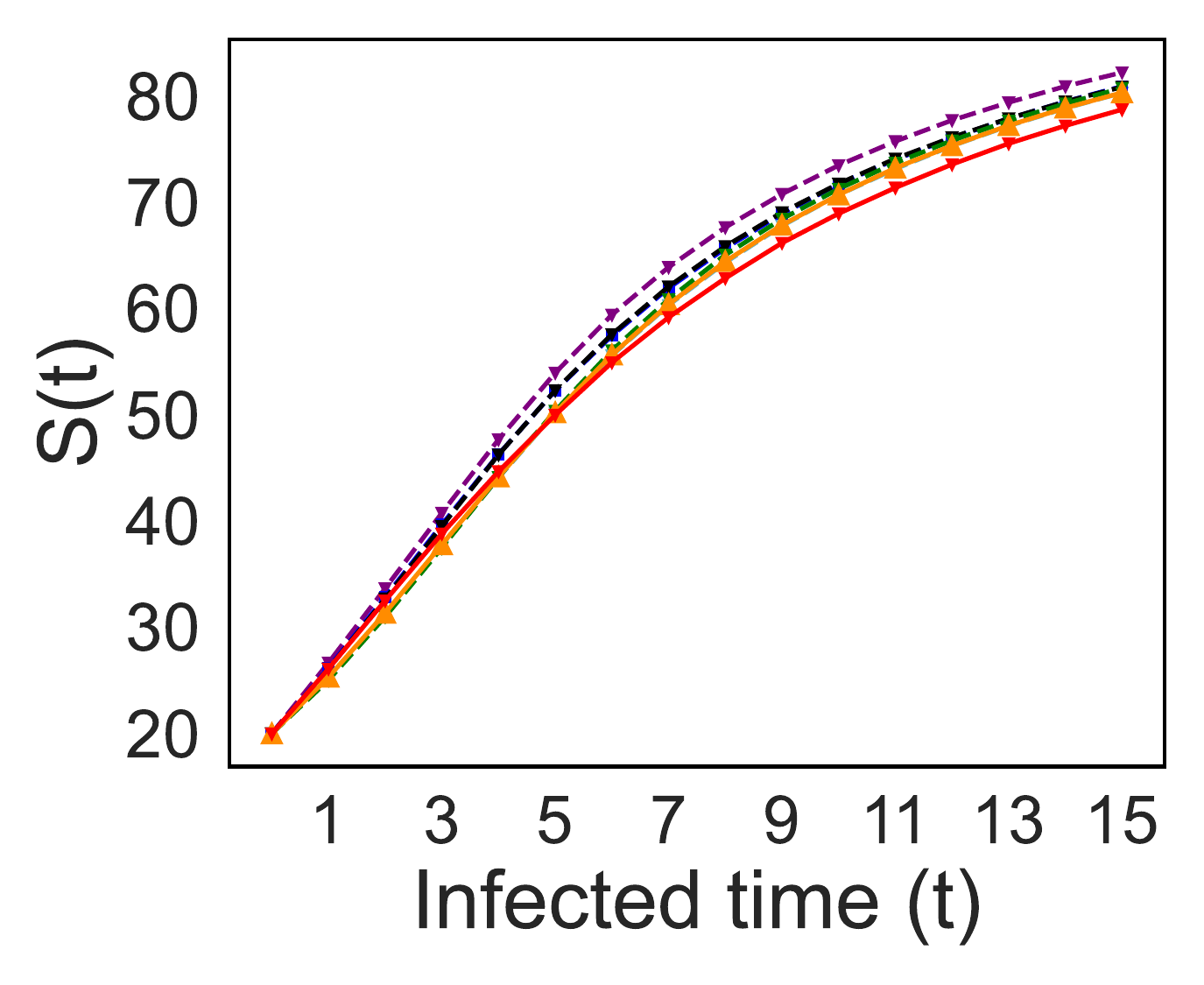} 
    \caption{p2p-Gnutella09}}}
    \end{subfigure}
    \hspace{-2mm}
    \begin{subfigure}[t]{.25\linewidth}
    {{ \includegraphics[width=\linewidth]{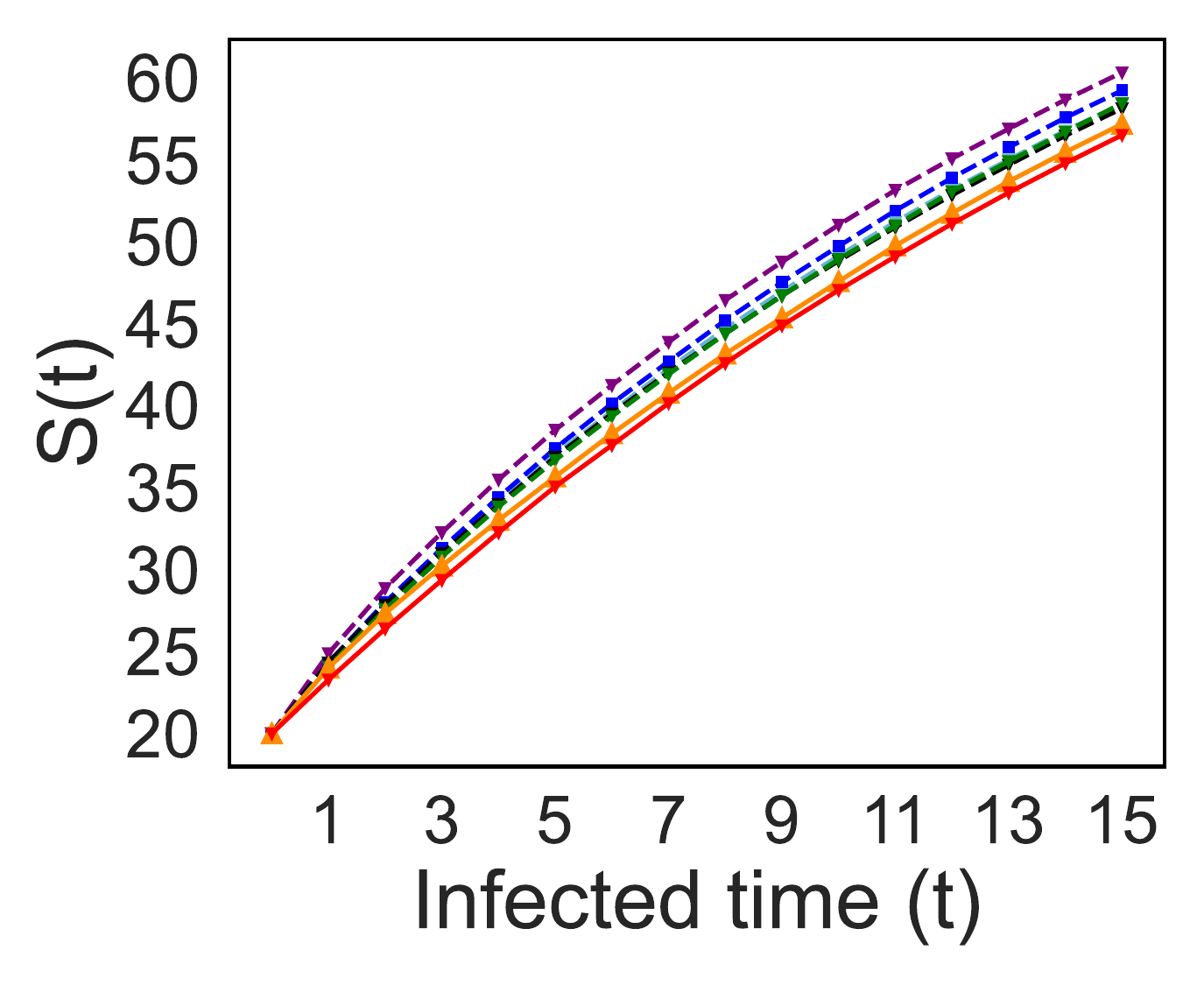} 
    \caption{tech-WHOIS}}}
    \end{subfigure}
    \hspace{-2mm}
    \begin{subfigure}[t]{.25\linewidth}
    {{ \includegraphics[width=\linewidth]{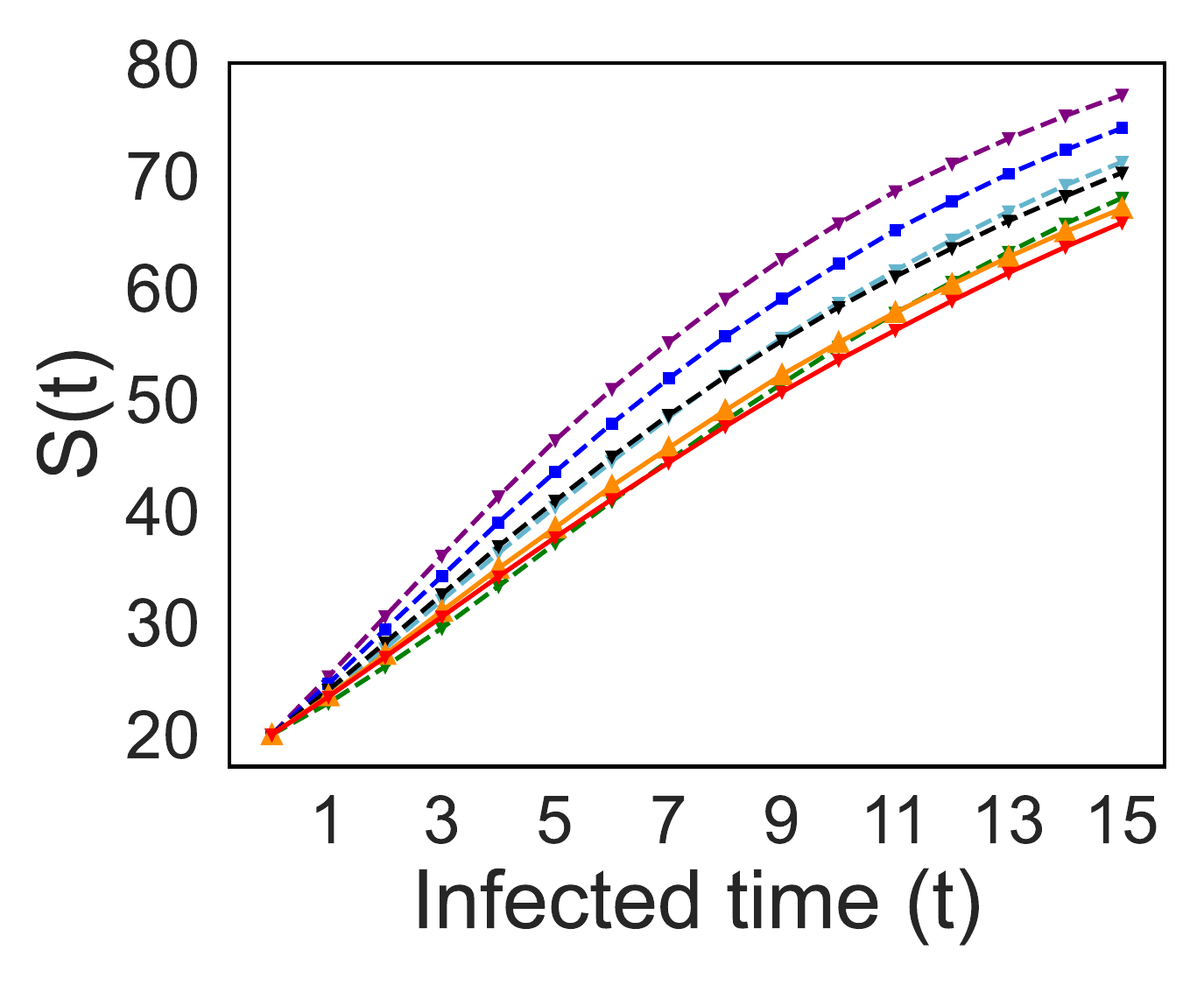}
    \caption{ca-GrQc}}}
    \end{subfigure}
    \hspace{-2mm}
    \begin{subfigure}[t]{.25\linewidth}
    {{ \includegraphics[width=\linewidth]{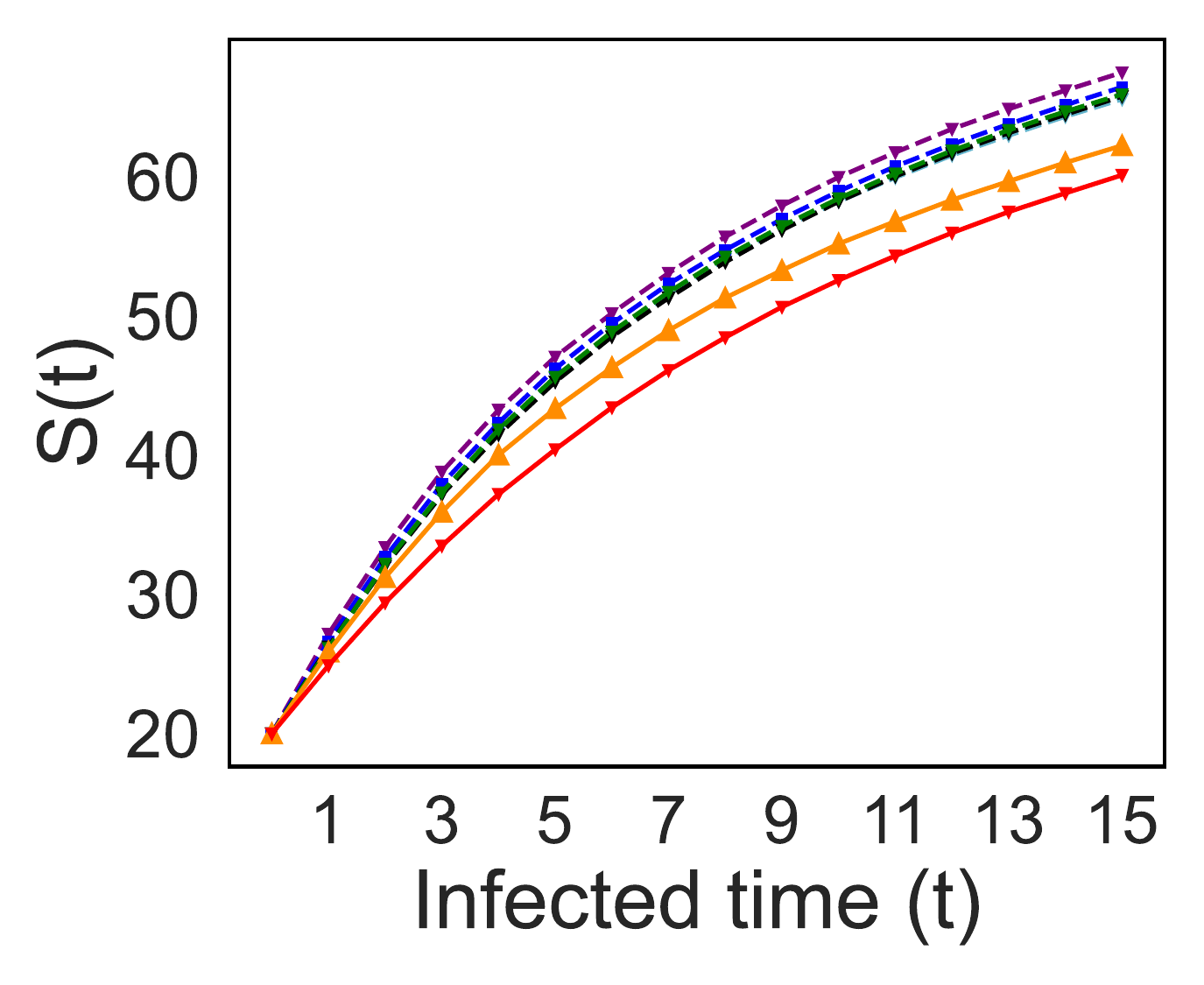}
    \caption{web-spam}}}
    \end{subfigure}
    \hspace{-2mm}
    
    \caption{\bf {\footnotesize Identifying influential spreaders by our measures and baselines.}}
    
    \label{fig:appendix_influential_figure}
    
\end{figure}

}
{}

\end{document}